\documentclass[journal=jacsat,manuscript=article]{achemso}

\usepackage[version=3]{mhchem} 

\author{Dumitru-Claudiu Sergentu}
\altaffiliation{Universitatea Alexandru Ioan Cuza din Ia\textcommabelow{s}i, Laboratorul RA-03 (RECENT AIR), Ia\textcommabelow{s}i, Romania}

\author{Boris Le Guennic}

\author{R\'emi Maurice}
\affiliation[ISCR]{Univ Rennes, CNRS ISCR (Institut des Sciences Chimiques de Rennes) -- UMR 6226, Rennes, France}
\email{remi.maurice@univ-rennes.fr}

\title[model Hamiltonian]{The resolution of the weak-exchange limit made rigorous, simple and general in binuclear complexes}

\abbreviations{}

\newcommand{\hms}{$\hat{H}^{\mathrm{MS}}$}
\newcommand{\hmod}{$\hat{H}^\mathrm{mod}$}
\newcommand{\hgs}{$\hat{H}^{\mathrm{GS}}$}
\newcommand{\hzee}{$\hat{H}^{\mathrm{Zee}}$}
\newcommand{\heff}{$\hat{H}^{\mathrm{eff}}$}
\newcommand{\js}{$J\hat{S}_a\hat{S}_b$}
\newcommand{\da}{$\hat{S}_a \bar{\bar{D}}_a\hat{S}_a$}
\newcommand{\db}{$\hat{S}_b \bar{\bar{D}}_b\hat{S}_b$}
\newcommand{\dab}{$\hat{S}_a \bar{\bar{D}}_{ab}\hat{S}_b$}
\newcommand{\za}{$\mu_B\vec{B}\bar{\bar{g}}_a\hat{S}_a$}
\newcommand{\zb}{$\mu_B\vec{B}\bar{\bar{g}}_b\hat{S}_b$}

\newcommand{\dm}{$\bar{d} \hat{S}_a \times \hat{S}_b$}

\newcommand{\icm}{cm$^{-1}$}

\usepackage{braket}
\usepackage{graphicx, nicefrac, xfrac}
\usepackage{adjustbox}
\usepackage{booktabs}

\usepackage{xr-hyper} 
\usepackage{siunitx}
\usepackage{caption}
\usepackage{hyperref}
\usepackage{lscape}
\usepackage{rotating}

\usepackage{pdfpages}

\usepackage{xr}
\begin{document}

\begin{tocentry}

\includegraphics[width=8.25cm,height=4.45cm]{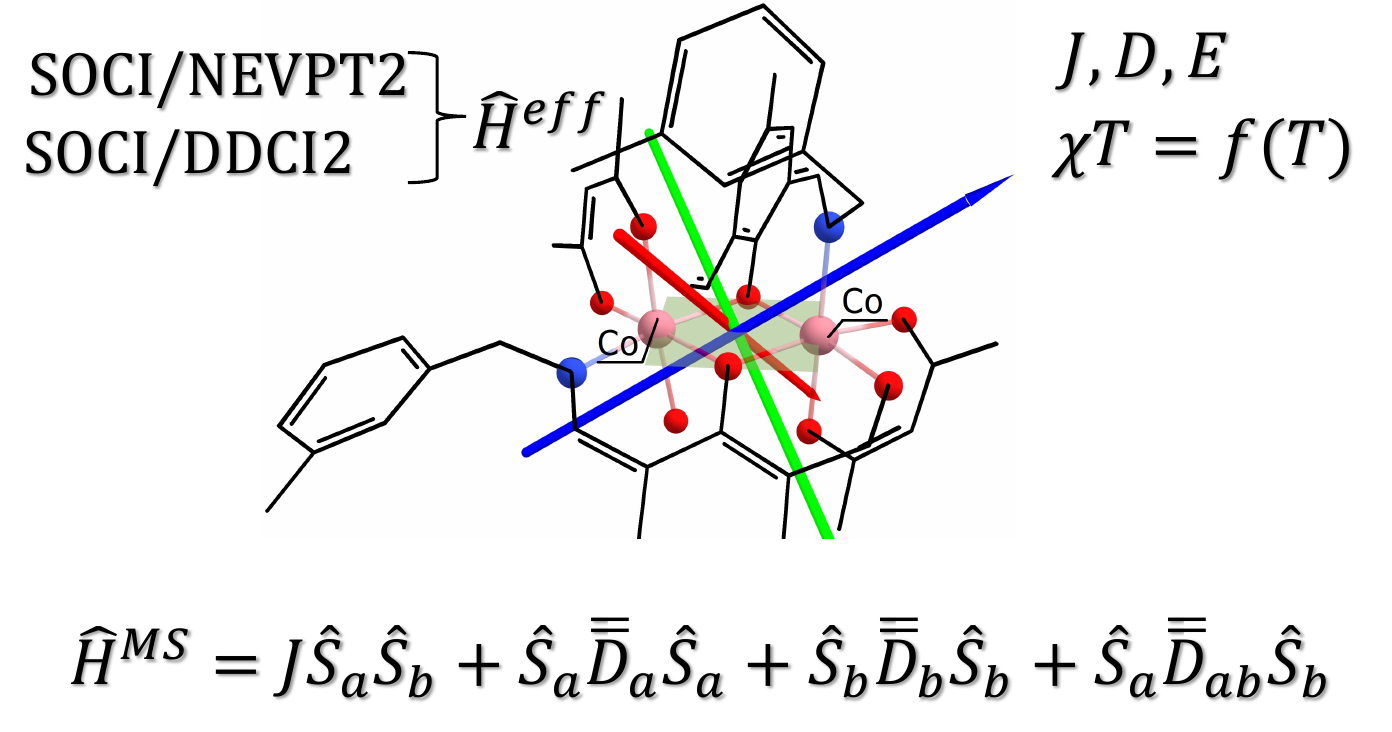}

\textbf{Synopsis:} The standard multispin model Hamiltonian ($\hat{H}^\mathrm{MS}$) is revived for calculating magnetic properties in binuclear complexes with weakly-coupled magnetic centers. Resolution and parameter extraction from $\hat{H}^\mathrm{MS}$ is achieved with a simple yet robust strategy that is applicable regardless of point group symmetry.
\end{tocentry}

\begin{abstract}
The correct interpretation of magnetic properties in the weak-exchange regime has remained a challenging task for several decades. In this regime, the effective exchange interaction between local spins is quite weak, of the same order of magnitude or smaller than the various anisotropic terms, which generates a complex set of levels characterized by spin spin mixing. Although the model multispin Hamiltonian in the absence of local orbital momentum, \hms{} = \js{} + \da{} +\db{} + \dab{}, is considered good enough to map the experimental energies at zero field and in the strong-exchange limit, theoretical works pointed out limitations of this simple model. This work revives the use of \hms{} from a new theoretical perspective, detailing point-by-point a strategy to correctly map the computational energies and wave functions onto \hms{} , thus validating it regardless of the exchange limit. We will distinguish two cases, based on experimentally characterized dicobalt(II) complexes from the literature. If centrosymmetry imposes alignment of the various rank-2 tensors constitutive of \hms{} in the first case, the absence of any symmetry element prevents such alignment in the second case. In such a context, the strategy provided herein becomes a powerful tool to rationalize the experimental magnetic data, since it is capable of fully and rigorously extracting the multispin model without any assumption on the orientation of its constitutive tensors. Furthermore, the strategy allows to question the use of the spin Hamiltonian approach by explicitly controlling the projection norms on the model space, which is showcased in the second complex where local orbital momentum could have occurred (distorted octahedra). Finally, previous theoretical data related to a known dinickel(II) complex is reinterpreted, clarifying initial wanderings regarding the weak exchange limit.
\end{abstract}

\section{Introduction}
\label{sec:introduction}

Recent decades witnessed significant advancements in the engineering of single-molecule magnets (SMMs) with compelling magnetic properties closer and closer to room temperature.\cite{Goodwin2017:a, Guo:2017a, Guo:2018c, Gould:2022a} At stake may be the future of storage devices and quantum information systems,\cite{Bogani:2008a, Coronado:2020a, Corcuera:2023a} but these SMMs also provide a playground to explore complex electronic structures, new quantum-mechanical phenomena,\cite{Gatteschi:2003a, Wernsdorfer:2005a, Schlegel:2008a} and develop novel strategies for evaluating them. 

SMMs typically exhibit magnetic anisotropy through the intertwined effect of spin-orbit coupling (SOC) and anisotropic crystal fields (CFs) and retain magnetic bistability with an energy barrier for magnetization reversal below a certain blocking temperature.\cite{Coronado:2020a} To go beyond the blocking temperatures observed in d-element polynuclear complexes, research shifted to the field of f-element single ion magnets after the discovery of the properties of the bis(phthalocyaninato)terbium anion.\cite{Ishikawa:2005a} Although f-element molecules make the current state-of-the-art SMM prototypes,\cite{Woodruff:2013a, Ishikawa:2003a, Guo:2018c, Zhu:2019a, Gould:2022a} with well-known recipes designed to improve their magnetic properties,\cite{Liddle:2015a, Chilton:2015a} potential d- and mixed f/d-element SMMs are also investigated at a pace faster than before.\cite{Wang:2011b, Swain:2021a, Chandrasekhar:2022a, Magott:2022a} 

In the laboratory, information on magnetic anisotropy is commonly evidenced through electron paramagnetic resonance (EPR) and magnetic susceptibility ($\chi$) studies. The outcome is interpreted through the language of model Hamiltonians,\cite{Boca:2004a, Maurice:2016a, Moreira:2002a} dressed with parameters quantifying the physics of the electronic structure, \textit{e.g.}\ magnetic exchange, zero-field splitting (ZFS), Zeeman interaction, etc. Optimal values for these parameters are obtained by fitting measured data,\cite{PHI, MAGPACK} such as the temperature variation of $\chi T$, and concluding on a certain set of uniquely defined values from good grounds may require input from first principles calculations.

With simplistic formulations, model Hamiltonians are appealing in experimental contexts and serve as a bridge between experiments and overly complicated theoretical approaches. In the case of model spin Hamiltonians,  only spin operators are at play, with the assumption that the orbital part(s) of the wave function(s) is(are) factored out. In mononuclear complexes, the model Hamiltonian applies to the $\ket{S, M_S}$ components of the ground spin state. In binuclear complexes and beyond, the model space may include more than one spin state. For instance, it may be constituted of the spin components of all the spin states that are triggered by the Heisenberg-Dirac-van Vleck (HDVV) Hamiltonian. \cite{Heisenberg:1926a, Dirac:1929b, VanVleck:1932a} Yet, these types of models reproduce faithfully magnetic data at low temperatures, in the absence of local orbital degeneracy, as soon as the interaction space encodes the effective physics of the studied system.

In polynuclear systems, two main types of models apply. The giant-spin model, \hgs{}, only aims at describing the ZFS of the ground spin state and it is used in a fashion similar to that of mononuclear systems. The multispin model, \hms{}, aims to describe the splitting and mixing of the spin components of the HDVV spectrum. With both model Hamiltonians, the theoretical extraction of relevant parameters is based on an explicit mapping onto an effective Hamiltonian (\heff{}) that is built on top of the \textit{ab initio} calculations.\cite{Maurice:2010a, Maurice:2010b, Maurice:2011a, Autschbach:2018m, Maurice:2023a} Alternatively, one may skip the explicit projection of the wave functions and work within the pseudospin framework.\cite{Chibotaru:2012a} Since \hgs{} targets only the ground spin state it is quite clear that it is not fully relevant in the weak-exchange limit.\cite{Wilson:2006a, Maurice:2010a} \hms{} is better suited in such cases; it is naturally built up in the $\ket{S_i, M_{Si}, \ldots}$ basis (the \textit{uncoupled} basis, the $i$'s denoting the active magnetic centers) and may be further expressed in the $\ket{S, M_S}$ basis (the \textit{coupled} one).\cite{Boca:1999a} 

Calculations based on the complete active space self-consistent field (CASSCF) approach\cite{roos1980b} are appealing in the context of magnetochemistry.\cite{Malrieu:2014a} This approach can describe correctly d$^n$ and f$^n$ near-degenerate configurations and allows for a systematic improvement of the electron correlation in post-CASSCF multireference treatments. Subsequently, the spin-orbit coupling (SOC) can be introduced by diagonalization of an electronic energy plus spin-orbit operator matrix, on the basis of the spin components of the previous spin-free CASSCF states, with the electronic energies potentially replaced by dynamically correlated ones. This is the spirit of (dressed) spin-orbit configuration interaction (SOCI).\cite{malmqvist02cpl} 

Many seminal studies show the use of configuration interaction schemes to calculate magnetic properties.\cite{Calzado:2002a, Calzado:2002b, Broer:2003a, Ganyushin:2006a, Calzado:2009a, Maurice:2009a, Atanasov:2011a, Neese:2011b} It is  worth noting that density functional theory (DFT) is equally involved in rationalizing magnetic data,\cite{Neese:2007a, Schmitt:2011b, Belkhiri:2019a, Cheng:2021a, Rata:2023a, David:2023b, Meskaldji:2023a} as well as single-reference spin-flip approaches,\cite{Mayhall:2014a, Mayhall:2015a, Orms:2018a} and more recently coupled-cluster methods.\cite{Schurkus:2020a, Pokhilko:2020a, Alessio:2021a} In principle, the low-energy \emph{ab initio} spectrum offers sufficient detail for calculating anisotropy parameters and spin-orbit correction to the exchange interaction if a SOCI is performed,\cite{Maurice:2016a, Maurice:2023a} as well as EPR parameters and $\chi T$ profiles provided that the Zeeman interaction is treated.\cite{Bolvin:2006a, vancoillie09jpca, Atanasov:2011a, Ungur:CF-2017a, Dey:2022a} This is the starting point of our work.

Extraction of magnetic parameters from \hms{} may be straightforward in centrosymmetric binuclear complexes. In such cases, all rank-2 tensors involved in \hms{}  (\emph{vide infra}) have the same principal axes,  or principal axis frames (PAFs). Such a frame, which may be derived from \hgs{} if not directly from symmetry arguments,\cite{Maurice:2010a, Maurice:2010b, Maurice:2010c} not only simplifies the model construction but also helps in shortcutting the parameter extraction through the effective Hamiltonian theory. In practice, cases were identified in which matrix elements are nil in the model but non-nil in \heff{}.\cite{Maurice:2010c} Such inequalities were proposed to arise from the lack of a rank-4, biquadratic anisotropy exchange tensor in \hms{}. The case of low-symmetry binuclear complexes is even more complicated. Although the molecular orientation may correspond to a molecular PAF, derived somehow from \hgs{} or from the diagonalization of the EPR $g$-matrix in the case of pseudospin Hamiltonians,\cite{Chibotaru:2012a} this \emph{xyz} frame may not correspond at all to the local or specific PAFs of all the rank-2 tensors of \hms{}. This must lead to nonzero off-diagonal elements of these tensors if expressed in the molecular frame, which is almost always neglected in experimental studies. Hence, the use of \hms{} in current magnetochemistry applications still appears problematic.

\begin{figure}[t!]
\centering
  \includegraphics[width=0.7\textwidth]{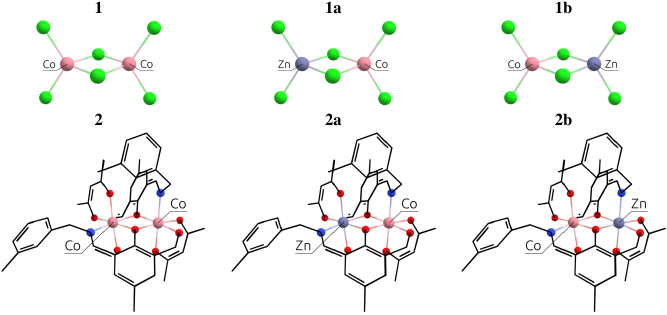}
  \caption{Molecular geometries of the binuclear complexes \textbf{1} and \textbf{2}, and of the respective \textbf{a} and \textbf{b} models obtained by substituting one Co atom by a Zn one. H atoms are omitted for clarity. Red, green and blue correspond to O, Cl and N atoms, respectively.}
  \label{fig:geoms}
\end{figure}

This article aims to revive the utilization of \hms{} for interpreting the low-temperature magnetic properties of binuclear complexes; it forwards a simple, rigorous, and versatile strategy to resolve the model, irrespective of the coordination symmetry or the regime (weak- or strong-exchange). The technique is showcased using dicobalt(II) complexes, a centrosymmetric one, \ce{[Co2Cl6]^2-} (\textbf{1}),\cite{Sun:1999a, Graaf:2006a} and an unsymmetrical one, \ce{[Co2(L)2(acac)2(H2O)]}\footnote{L = (4-methyl-2-formyl-6-(((2-trifluoromethyl)phenyl)methyliminomethyl)phenol).} (\textbf{2}).\cite{Song:2020a} The molecular structures are shown in Figure \ref{fig:geoms}. Dicobalt(II) systems were chosen simply to show that, by using the proposed strategy, full extraction of magnetic parameters can be easily performed even in challenging cases (here, the HDVV matrix being 16$\times$16 and the local magnetic centers intrinsically displaying Kramers degeneracies). The article concludes that achieving a strong agreement between the matrix elements of \hms{} and \heff{}, necessary for the validation and resolution of the former, is possible in any case, and, in practice, easily achievable by

\begin{itemize}

\item[\textbf{i}] building a first effective Hamiltonian in the \textit{coupled} basis,

\item[\textbf{ii}] determining the molecular PAF, which is identified by resolution of the anisotropic model Hamiltonian \hmod{}=$\hat{S}\bar{\bar{D}}_{S=3}\hat{S}$ effectively at play only in the $\ket{S=3, M_S}$ block of the full model space of \hms{},

\item[\textbf{iii}] recomputing the effective Hamiltonian in the molecular PAF and consistently revising the $\pm$ signs of the cross-blocks matrix elements, coupling $M_S$ components belonging to different $S$-blocks in \heff{}, and

\item[\textbf{iv}] expressing all rank-2 tensors of \hms{} in the molecular PAF to finally determine the missing quantities by minimizing the mismatch between \hms{} and \heff{}.

\end{itemize}

\noindent A recent erratum to Reference \citenum{Gould:2022a} also raised the issue of conflicting signs mentioned at point (iii). In order to properly project \heff{} onto \hms{}, it is essential to use correct prefactors. In fact, the revision of such conflicting signs practically eliminates the need to introduce tensors of rank superior to 2 to achieve agreement between each and every matrix element of \hms{} and \heff{}, regardless of the coordination symmetry. Thus, the present article concludes that the resolution of \hms{} for \ce{[Ni2(en)4Cl2]^{2+}}, en = ethylendiamine,  was in fact straightforward, and already properly done in the seminal publication.\cite{Maurice:2010c}

The subsequent sections commence with a brief overview of the standard \hms{} and its validation by effective Hamiltonians. This is followed by an outline of the proposed strategy for utilizing these two concepts in generally deriving magnetic properties for binuclear complexes. The results and discussion extend over two major sections showcasing the resolution of \hms{}, and the subsequent calculation of the $\chi T$ profile, for the centrosymmetric complex \textbf{1} firstly, and for the unsymmetrical complex \textbf{2}, secondly. The article concludes that, when constructed appropriately, the standard multispin Hamiltonian correctly describes the effective magnetic interactions even in the weak-exchange limit, thus justifying its general relevance in the experimental design of SMMs based on d-elements. Finally, a series of take-home messages will be delivered to the attention of the experimental community involved in molecular magnetism, emphasizing once more the strong need for good, independent, computational input to consistently interpret the data.

\section{Theory and computational strategy}
\label{sec:theory}

\subsection{The multispin model Hamiltonian for binuclear complexes}
\label{subsec:hms}

In the absence local orbital momentum, the model multispin Hamiltonian expression reads:\cite{Kahn:1993b, Boca:1999a, Maurice:2023a}
\begin{equation}
\label{eq:hms}
\text{\hms{} = \js{} + \da{} + \db{} + \dab + \dm{}}
\end{equation}

\noindent where $a$ and $b$ label the two magnetic centers, $\hat{S}_a$ and $\hat{S}_b$ are spin operators, $J$ is the Heisenberg exchange magnetic coupling, $\bar{\bar{D}}_a$ and $\bar{\bar{D}}_b$ are rank-2 tensors describing the local anisotropies, $\bar{\bar{D}}_{ab}$ is a rank-2 tensor describing the symmetric anisotropic exchange, and $\bar{d}$ is a pseudovector describing the antisymmetric component of the anisotropic exchange. In centrosymmetric complexes, the latter term of Equation \ref{eq:hms}, also referred to as the Dzyaloshinskii-Moriya interaction (DMI),\cite{Dzyaloshinskii:1964a, Moriya:1960a, Dmitrienko:2014a, Bouammali:2021b, Bouammali:2021a, Bouammali:2022a} vanishes, and otherwise, may be less important than the other terms unless specific situations are encountered (exotic coordination environment and/or orbital near-degeneracy).  In fact, this work does not specifically focus in the DMI, which will be later justified by comparing \heff{} and \hms{}.  

In order to describe interaction with an applied magnetic field, $\vec{B}$, Equation \ref{eq:hms} is completed by the Zeeman Hamiltonian:

\begin{equation}
\label{eq:hzee}
\text{\hzee{} = \za{} + \zb{}}
\end{equation}

\noindent where $\mu_B$ is the Bohr magneton and $\bar{\bar{g}}_a$ and $\bar{\bar{g}}_b$ are the local Zeeman splitting tensors. The rank-2 tensors of Equations \ref{eq:hms} and \ref{eq:hzee} are represented as 3$\times$3 matrices consisting of up to nine non-zero components in arbitrary \emph{xyz} frames. The tensors are of course diagonal in their respective PAFs. Though the axial ($D_a$, $D_b$, $D_{ab}$) and rhombic ($E_a$, $E_b$, $E_{ab}$) local and exchange anisotropic parameters depend only on the diagonal elements of the respective tensors ($\bar{\bar{D}}_a$, $\bar{\bar{D}}_b$, $\bar{\bar{D}}_{ab}$),  following $D=\nicefrac{3}{2}D_{zz}$ and $E=\nicefrac{1}{2}(D_{xx}-D_{yy})$\footnote{The standard conventions are applied here: tensor is traceless, $\left|D\right| > 3E$ and $E > 0$}, it should be noted that the PAFs of $\bar{\bar{D}}_a$, $\bar{\bar{D}}_b$ and $\bar{\bar{D}}_{ab}$ may not coincide with each other in the general case, nor with those of $\bar{\bar{g}}_a$ and $\bar{\bar{g}}_b$. Thus, symmetry is key to understanding how to practically deal with so many tensors. 

If centrosymmetry is present, one may expect that all the tensors of Equations \ref{eq:hms} and \ref{eq:hzee} are diagonal in the same coordinate frame, which effectively corresponds to the molecular PAF. This facilitates the straightforward construction of \hms{} and the extraction of relevant axial and rhombic parameters using the effective Hamiltonian theory. However, if centrosymmetry is not present, two additional challenges arise. Firstly, one must determine the molecular PAF. Secondly, one must express all the rank-2 tensors in this coordinate frame. In this scenario, the rank-2 tensors may no longer be diagonal, although symmetry may still impose some of the off-diagonal elements to be zero in specific situations. 

\hms{} and \hzee{} are both constructed in the uncoupled spin-basis, $\ket{S_a, M_{Sa}; S_b, M_{Sb}}$, or shortly $\ket{M_{Sa}, M_{Sb}}$. Since we deal with d$^7$ Co(II) centers, $S_a = S_b = \nicefrac{3}{2}$ and $M_{Sa}, M_{Sb}\in \lbrace \pm\nicefrac{3}{2}, \pm\nicefrac{1}{2} \rbrace $, resulting in a 16$\times$16 model space. Thus, the matrices of all terms in Equations \ref{eq:hms} and \ref{eq:hzee} must be expressed in this same 16$\times$16 basis. \hms{} is designed here to reproduce the energy levels generated by the coupling of the local spin-quartets, \emph{i.e.}, the septet ($S=3$), quintet ($S=2$), triplet ($S=1$) and singlet ($S=0$) coupled-spin states. Translation of \hms{} from the uncoupled-spin basis, $\ket{M_{Sa}, M_{Sb}}$, to the coupled-spin basis, $\ket{S, M_S}$ ($S=0, 1, 2, 3$, $M_S=\pm S$), is achieved via the transformation matrix U based on Clebsch-Gordan (CG) coefficients:\cite{Boca:1999a}
\begin{equation}
\label{eq:transform}
\text{\hms{}(coupled)= U$^\mathrm{T} \cdot$ \hms{}(uncoupled) $\cdot$ U}
\end{equation}

\noindent where U$^\mathrm{T}$ is the transposed of U. The 16$\times$16 matrix U of CG coefficients used in this work is provided in Table \ref{tab:umatrix-SI} of the Supporting Information (SI) file.

\subsection{Construction of the effective Hamiltonian}
\label{subsec:heff}

\heff{} spans the same model space as \hms{} and it is initially constructed in the coupled-spin basis, in accord with the expressions of the underlying SOCI wavefunctions, by means of the des Cloizeaux formalism:\cite{desCloizeaux:1960a}
\begin{equation}
\label{eq:heff}
\text{\heff{}}= \sum_K E_K \braket{\phi_i | \psi_K^L} \braket{\psi_K^L | \phi_j}
\end{equation}
\noindent where $\phi_i$, $\phi_j$ are model-space $\ket{S, M_S}$ spin functions, $ \psi_K^L$ are L\"{o}wdin-orthonormalized projections of the SOCI wavefunctions onto the model interaction space, and the $E_K$'s are the lowest SOCI energies (4 in the case of a mononuclear cobalt(II) complex and 16 in the case of a binuclear one). Note that Equation \ref{eq:transform} may also be used to transform \heff{} from the coupled-spin basis to the uncoupled-spin one.

\subsection{Proposed step-by-step strategy for extracting the multispin Hamiltonian}
\label{subsec:strategy}

\begin{table}[b!]
\centering
\renewcommand{\arraystretch}{1.5}
\caption{Matrix elements of \hmod{}=$\hat{S}\bar{\bar{D}}_{S=3}\hat{S}$ expressed in an arbitrary axis frame in the basis of the $\ket{S=3, M_S}$ functions}   
\label{tab:hmod}
\begin{adjustbox}{width=\textwidth}
\begin{tabular}{rccccccc}
\hline
\hmod{}  
& $\ket{3, 3}$ & $\ket{3, 2}$ & $\ket{3, 1}$ & $\ket{3, 0}$ & $\ket{3, -1}$ & $\ket{3, -2}$ & $\ket{3, -3}$ \\
\hline

 $\bra{3, 3}$ &
 $ \frac{3}{2}(D_{xx}+D_{yy}) + 9 D_{zz} $ &
 $ \frac{5 \sqrt{6} \left(D_{xz} - i D_{yz}\right)}{2} $ &
 $ \frac{\sqrt{15} \left(D_{xx} - D_{yy} - 2 i D_{xy}\right)}{2} $ &
 0 &
 0 &
 0 &
 0 \\ 

 $\bra{3, 2}$ &
 $ \frac{5 \sqrt{6} \left(D_{xz} + i D_{yz}\right)}{2} $ &
 $ 4(D_{xx}+D_{yy}+D_{zz}) $ &
 $ \frac{3 \sqrt{10} \left(D_{xz} - i D_{yz}\right)}{2} $ &
 $ \frac{\sqrt{30} \left(D_{xx} - D_{yy} - 2 i D_{xy}\right)}{2} $ &
 0 &
 0 &
 0 \\

 $\bra{3, 1}$ &
 $ \frac{\sqrt{15} \left(D_{xx} - D_{yy} + 2 i D_{xy}\right)}{2} $ &
 $ \frac{3 \sqrt{10} \left(D_{xz} + i D_{yz}\right)}{2} $ &
 $ \frac{11}{2}(D_{xx}+D_{yy}) + D_{zz} $ &
 $ \sqrt{3} \left(D_{xz} - i D_{yz}\right) $ &
 $ 3 D_{xx} - 3 D_{yy} - 6 i D_{xy} $ &
 0 &
 0 \\ 

 $\bra{3, 0}$ &
 0 &
 $ \frac{\sqrt{30} \left(D_{xx} - D_{yy} + 2 i D_{xy}\right)}{2} $ &
 $ \sqrt{3} \left(D_{xz} + i D_{yz}\right) $ &
 $ 6(D_{xx} + D_{yy}) $ &
 $ - \sqrt{3} \left(D_{xz} - i D_{yz}\right) $ &
 $ \frac{\sqrt{30} \left(D_{xx} - D_{yy} - 2 i D_{xy}\right)}{2} $ &
 0 \\ 

 $\bra{3, -1}$ &
 0 &
 0 &
 $ 3 D_{xx} - 3 D_{yy} + 6 i D_{xy} $ &
 $ - \sqrt{3} \left(D_{xz} + i D_{yz}\right) $ &
 $ \frac{11}{2}(D_{xx}+D_{yy}) + D_{zz} $ &
 $ -\frac{3 \sqrt{10} \left(D_{xz} - i D_{yz}\right)}{2} $ &
 $ \frac{\sqrt{15} \left(D_{xx} - D_{yy} - 2 i D_{xy}\right)}{2} $ \\ 

 $\bra{3, -2}$ &
 0 &
 0 &
 0 &
 $ \frac{\sqrt{30} \left(D_{xx} - D_{yy} + 2 i D_{xy}\right)}{2} $ &
 $ - \frac{3 \sqrt{10} \left(D_{xz} + i D_{yz}\right)}{2} $ &
 $ 4(D_{xx} + D_{yy} + D_{zz} $ &
 $ - \frac{5 \sqrt{6} \left(D_{xz} - i D_{yz}\right)}{2} $ \\

 $\bra{3, -3}$ &
 0 &
 0 &
 0 &
 0 &
 $ \frac{\sqrt{15} \left(D_{xx} - D_{yy} + 2 i D_{xy}\right)}{2} $ &
 $ - \frac{5 \sqrt{6} \left(D_{xz} + i D_{yz}\right)}{2} $ &
 $ \frac{3}{2}(D_{xx} + D_{yy}) + 9 D_{zz} $ \\
\hline
\end{tabular}
\end{adjustbox}
\end{table}

textbf{Step 1} The first step involves determining the molecular PAF. Note that, unlike the energies, the composition of the SOCI wavefunctions varies with the molecular orientation such that their interpretation may become needlessly complex in the event that the cartesian $z$-axis does not coincide with the molecular PAF. For both the studied complexes (\textbf{1} and \textbf{2}), the molecular PAF was determined from the resolution of \hmod{}=$\hat{S}\bar{\bar{D}}_{S=3}\hat{S}$, describing only the ground $S=3$ state. A similar strategy was employed for determining the anisotropy axes of \ce{[Ni2(en)4Cl2]^{2+}}.\cite{Maurice:2010c} The procedure consists in:

\begin{itemize}
\item[\textbf{i}] perform a first reference SOCI calculation with an arbitrary \emph{xyz} frame, construct the 16$\times$16 matrix of \heff{} according to Section \ref{subsec:heff} and focus only on the $S=3$ block. This procedure is in essence different from simply following the giant spin approach: in the case of spin mixing, it ensures that we properly extract the actual PAF of the $S$ = 3 block.

\item[\textbf{ii}] derive the analytical matrix representation of \hmod{}=$\hat{S}\bar{\bar{D}}_{S=3}\hat{S}$ in the $\ket{S=3, M_S}$ space and extract $\bar{\bar{D}}_{S=3}$ by best equating \heff{} from point (i) to \hmod{}. Since this matrix is uncommon in the transition metal literature (an $S$ = 3 state is impossible to reach within a d$^n$ manifold), we provide it in Table \ref{tab:hmod}.

\item[\textbf{iii}] rotate the molecular coordinates from the arbitrary \emph{xyz} frame to the molecular PAF: $xyz_i^{\mathrm{MPAF}}=V^{-1} \cdot xyz_i^{\mathrm{arb}}$, where $xyz_i^{\mathrm{arb}}$ are the coordinates of all the $i$ atoms in the arbitrary frame and $V$ is the eigenvector matrix of $\bar{\bar{D}}_{S=3}$ (convention may apply for the respective labeling of the $x$, $y$ and $z$ axes).
\end{itemize}  

\noindent With the molecular geometry rotated in the molecular PAF, the SOCI calculations of Step 1 are repeated and the 16$\times$16 representative matrix of \heff{} is re-built. This is the \heff{} that will be essentially used for the validation and resolution of \hms{}. However, note that the exact same conclusions can be reached by extracting it all in an arbitrary axis frame (in a more tedious way!).

\textbf{Step 2} SOCI calculations are conducted with two model, monomeric, structures defined by the substitution of a magnetic center with a diamagnetic one. In the present cases, models \textbf{1a}, \textbf{1b} and \textbf{2a}, \textbf{2b} (Figure \ref{fig:geoms}) are obtained by replacing one Co with one Zn in the dimeric structures \textbf{1} and \textbf{2} respectively. The goal here is to extract independently the local anisotropy tensors, $\bar{\bar{D}}_a$ and $\bar{\bar{D}}_b$, for the magnetic centers, determine their respective PAFs, and calculate the corresponding axial and rhombic parameters $D_a$, $E_a$, $D_b$, $E_b$. For each magnetic center, the eigenvectors of $\bar{\bar{D}}$, leading to $\bar{\bar{D}}^{\mathrm{diag}}$, provide with the $V$ rotation matrix that can be used to express them back in the molecular PAF, through:

\begin{equation}
\label{eq:rot}
\bar{\bar{D}}(\text{molecular PAF}) = V \bar{\bar{D}}^{\mathrm{diag}} V^{-1}
\end{equation}

\textbf{Step 3} The 16$\times$16 model analytical matrix of \hms{} in the uncoupled-spin basis is derived with the leading terms of it, \emph{i.e.}, the Heisenberg and the local anisotropy terms,  by applying simple spin-operator algebra. This matrix is subsequently transformed in the coupled-spin basis using Equation \ref{eq:transform}. 

Afterwards, the quality of the model matrix can be evaluated by expressing it numerically and comparing it to \heff{}. The less deviation, the better the model. To shortcut the extraction, one may directly use the spin-free $J$ value and the local anisotropy tensors calculated independently at Step 2 (of course, expressed in the molecular PAF for consistency). In principle, this approach should already provide a good representation of \heff{} since (i) the effect of the SOC on $J$ is usually quite small\cite{Maurice:2010c, Maurice:2010b} and (ii) the symmetric anisotropy exchange tensor may be of lesser importance than the local ones. Finally, we may also refine the $J$ value and the components of the local anisotropy tensors as well as introduce the symmetric exchange tensor in our model to further improve it. Refinement is carried out through a least-squares fitting of the model parameters, aiming to minimize the Root Mean Square Deviation (RMSD) between \hms{} and \heff{}.

Here, we have deliberately chosen to make use of momoner calculations to (i) validate and provide additional support for this route, which is one of the most commonly employed in current literature when modeling magnetic properties, for instance followed by the POLY\_ANISO program,\cite{Chibotaru:2008a, Ungur:2009a} and (ii) avoid the extraction of too many parameters simultaneously in unsymmetrical cases, such as that of complex \textbf{2}, thus preventing the occurrence of unnoticed human mistakes by allowing a cross-validation of the extraction procedure based on the much simpler extracted parameters of the monomer calculations. Note that we would be now confident in performing a complete extraction based on a dimer calculation on a new system, or alternatively in performing only monomer calculations if the dimer calculations would prove to be too demanding.

\subsection{Computational details}
\label{subsec:details}

Electronic structure calculations were performed with the ORCA package, v5.0.3.\cite{ORCA, ORCA5} Scalar relativistic effects were introduced by using the Douglas-Kroll-Hess (DKH) Hamiltonian.\cite{Douglas:1974a, Hess:1985a, Hess:1986, Wolf:2002a} The Co, N, O, and Cl electrons were treated with the all-electron, triple-zeta DKH-def2-TZVP basis sets whereas the C, F, and H electrons were treated with the smaller, double-zeta DKH-def2-SVP ones. These bases were derived from the original def2 variants by recontraction within the DKH framework.\cite{weigend2005} To speed off the calculation of electron repulsion integrals, the RIJCOSX ``chain-of-spheres'' density-fitting\cite{Neese:2009a} was applied together with large, automatically-generated auxiliary basis sets.\cite{Stoychev:2017a} The geometries of \textbf{1} and \textbf{2} were extracted from published crystal information data.\cite{Sun:1999a, Song:2020a} Concerning \textbf{2}, the H coordinates were optimized within the DFT framework, using the Perdew-Burke-Ernzerhof generalized gradient approximation,\cite{Perdew:1996b} and otherwise the same details as above hold.  The \emph{xyz} coordinates in the initial, arbitrary, coordinate frames are provided in the SI.

The zero-order wavefunctions were converged within the state-averaged (SA) CASSCF framework, with the active space spanned by the Co(II) d$^7$ shell(s). \emph{I.e.}, CASSCF(14, 10) and CASSCF(7, 5) calculations were performed for the dimeric species \textbf{1} and \textbf{2} and for the monomeric species \textbf{1a}, \textbf{1b}, \textbf{2a} and \textbf{2b}, respectively. The production-level SA schemes included 49 states per $S$ = 3, 2, 1 and 0 block for \textbf{1}, seven $S=\nicefrac{3}{2}$ states for \textbf{1a} and \textbf{1b}, 9 states per $S$ = 3, 2, 1 and 0 block for \textbf{2}, three $S=\nicefrac{3}{2}$ states for \textbf{2a} and \textbf{2b}. Regarding complex \textbf{2} and its monomers, additional SA schemes were explored for validation purposes, which will be discussed in the results sections. The 49-root per block SA scheme employed for \textbf{1} was validated in Reference \citenum{Graaf:2006a}, 49 being the total number of roots generated from the product of the two local (crystal-field split) $^4F$ Co(II) terms. Concerning complex \textbf{2}, the 9-root per block SA scheme has been validated in this work, 9 corresponding to the number of roots generated from the product of two local $^4T_1$ Co(II) terms.

State energies including dynamic correlation effects were calculated with the strongly-contracted, n-electron valence state perturbation theory at second order (NEVPT2).\cite{Angeli:2001a, Angeli:2001b} For \textbf{2} in particular, the relative correlated energies for the lowest $S$ = 3, 2, 1 and 0 states were additionally recorrected by difference dedicated CI (DDCI2) and iterative DDCI2 calculations,\cite{Miralles:1993a, Garcia:1995a} with an orbital-energy cutoff between $-$10 and 1000 Hartree and \emph{tsel} and \emph{tpre} thresholds set to 1e-10 and 1e-6 respectively. In this way, a better $J$ value was obtained, or in other words, a better separation of the spin blocks. In the iterative DDCI2 calculation,  a limit of ten iterations was employed.  

Finally, the production SOCI calculations were performed for all the dimeric and monomeric species. In these calculations, the spin-orbit operator matrix was constructed in the basis of the spin components of the CASSCF states. The SOCI matrix is constituted of off-diagonal elements, triggered by the spin-orbit coupling, and also of diagonal ones, the electronic spin-free energies. The diagonal was dressed with dynamically-correlated energies and the resulting matrix was diagonalized to generate the spin-orbit coupled wavefunctions and energies. Hereafter, SOCI calculations based on NEVPT2 and DDCI2 correlated energies will be referred to as SO-NEVPT2 and SO-DDCI2 respectively. We point out additionally that, in the SO-DDCI2 calculations, only the relative energies of the lowest $S$ = 3, 2, 1, 0 CASSCF states were adjusted by the DDCI2 energy spacings, whereas the NEVPT2 relative energies were retained for the remaining 192 CASSCF states. The intention here was to characterize as best as possible the spin-orbit states that may be populated in the temperature range employed in experimental magnetic susceptibility studies. Since other excited states appear at much higher energies, above 2800 \icm{}, they do not contribute much to the observed properties.

Finally, powder-averaged \emph{ab initio} $\chi T$ curves were generated for complexes \textbf{1} and \textbf{2} directly from the outcomes of the SOCI calculations. Note that, in ORCA, $\chi$ is calculated by finite differentiation of the partition function by using the (field-corrected) spin-orbit states.\cite{Atanasov:2011a} Furthermore, by using the same approach, we have modeled the $\chi T$ curves based on the pre-validated \hms{} + \hzee{} models, through the usual approximation,

\begin{equation}
\chi_\alpha \approx \frac{M}{B} = \frac{k_B T}{\mu_B} \frac{\partial \mathrm{ln} Z}{\partial B_\alpha} \frac{1}{B_\alpha}
\label{eq:chi-mb}
\end{equation}

\noindent as well as by computing the full $\chi$ tensor,

\begin{equation}
\chi_{\alpha, \beta} = \frac{N_A k_B T}{10} \frac{\partial^2 \mathrm{ln} Z}{\partial B_\alpha \partial B_\beta}
\label{eq:chi-full}
\end{equation} 

\noindent where $M$ is the magnetization, $B$ = 0.2 T is the applied magnetic field, $\alpha$ and $\beta$ are two field directions, $Z = \sum_i^{16} e^{-\frac{E_i}{k_B T}} $ is the partition function, and $N_A$ and $k_B$ are the Avogadro and Boltzmann constants, respectively. For the present dicobalt(II) cases, Equations \ref{eq:chi-mb} and \ref{eq:chi-full} led to practically identical $\chi T = f(T)$ model curves. 

\section{Results and discussion}

\subsection{\ce{[Co2Cl6]^{2-}},  a centrosymmetric complex}
\label{sec:res-complex2}

\subsubsection{Introduction and spin-free spectrum}

\begin{figure}[t!]
\centering
  \includegraphics[width=.7\textwidth]{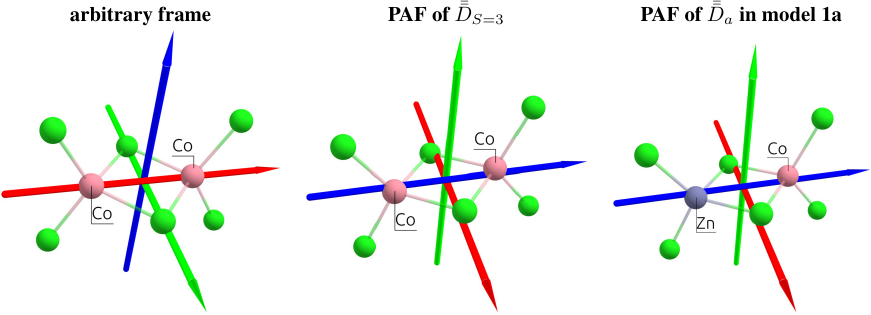}
\caption{Arbitrarily-chosen coordinate frame and calculated PAFs for complex \textbf{1} (from $\bar{\bar{D}}_{S=3}$) and for the model structure \textbf{1a} (from $\bar{\bar{D}}_a$). Axes color code: $z$=red, $x$=blue, $y$=green.}
  \label{fig:c1-frames}
\end{figure}

Magnetic properties in this edge-sharing, bitetrahedral ion were measured by Su et al.\cite{Sun:1999a} Each Co(II) center adopts an orbitally non-degenerate ground state (GS), $^4A_2$,  correlating with the $^4F$ ground term of the isolated ion. The $\chi T$ profile was fitted by an anisotropic model with a relatively large antiferromagnetic Heisenberg exchange coupling, $J$=23.2 \icm{} (\js{} formalism), a large local anisotropy $D$=29 \icm{} parameter, and an isotropic $g$-factor of 2.25. With first-principles calculations, de Graaf and Sousa noted that the anisotropy parameter is well approximated by that of a \ce{[CoCl_4]^{2-}} structural model.\cite{Graaf:2006a} However, the calculated $J$ value,  34.3 \icm{} with second-order CAS prturbation theory (CASPT2),\cite{andersson1992} overreached the fitted $J$. Furthermore, SOCI delivered a low-energy spin-orbit spectrum plagued by a large admixture of septet ($S$=3), quintet ($S$=2), triplet ($S$=1), and singlet ($S$=0) spin components. The authors concluded the impossibility of calculating $J$ and $D$ with \hms{} of Equation \ref{eq:hms} and reiterated the difficulties of describing magnetic properties in the weak-exchange regime.\cite{Boca:2004a} 

Our first-principles calculations support the previous theoretical data. Namely, complex \textbf{1} adopts a spin-singlet GS with a Land\'e ordering for the triplet, quintet, and septet spin-states with antiferromagnetic $J$ of 7.5 (CASSCF) / 11.2 \icm{} (NEVPT2). The $J$ value matches that of Ref. \citenum{Graaf:2006a} within 1 \icm{} at the CASSCF level, but it is three times smaller when the dynamic correlation is included. Although the discrepancy may originate from the choice of the  PT2 flavor, here NEVPT2 vs. CASPT2 in Reference \citenum{Graaf:2006a}, both these methods deliver similar accuracy when evaluated against the fitted $J$ value of 23.2 \icm{}; \textit{i.e.}\ if CASPT2 overestimates it by 11.1 \icm{}, NEVPT2 underestimates it by 12 \icm{}. In this context, DDCI2 calculations led to $J = 21.3$ \icm{}, within $\sim$2 \icm{} of the fitted value. Iterative DDCI2 calculations did not converge well on a particular value and delivered $J$ within the $\sim$18--21.5 \icm{} range in 10 iterations, with the mean $J$ = 20.9 \icm{} still in excellent agreement with the fitted $J$. 

\subsubsection{Spin-orbit spectrum and determination of the molecular PAF}

\begin{table}[b!]
\centering
\caption{Matrix elements, in \icm{}, of the $S = 3$ block within the 16$\times$16 matrix of \heff{}, constructed from SO-NEVPT2 calculations on complex \textbf{1} oriented in the arbitrary coordinate frame}   
\label{tab:c1-heff-s3block}
\renewcommand{\arraystretch}{1.5}
\begin{adjustbox}{width=\textwidth}
\begin{tabular}{rccccccc}
\hline
\hmod{}  
& $\ket{3, 3}$ & $\ket{3, 2}$ & $\ket{3, 1}$ & $\ket{3, 0}$ & $\ket{3, -1}$ & $\ket{3, -2}$ & $\ket{3, -3}$ \\
\hline

$\bra{3, 3}$ &    79.0   &   1.0+2.0$i$   &   $-$15.0$-$6.0$i$   &   0   &   0   &   0   &   0   \\ 
$\bra{3, 2}$ &    1.0$-$2.0$i$   &   87.0   &   1.0+1.0$i$   &   $-$21.0 $-$9.0$i$   &   0   &   0   &   0   \\ 
$\bra{3, 1}$ &   $-$15.0+6.0$i$   &   1.0$-$1.0$i$   &   92.0   &   1.0$i$   &   $-$23.0$-$10.0$i$   &   0   &   0   \\ 
$\bra{3, 0}$ &    0   &   $-$21.0+9.0 $i$   &   $-$1.0$i$   &   93.0   &   $-$1.0$i$   &   $-$21.0$-$9.0$i$   &   0   \\ 
$\bra{3,-1}$ &    0   &   0   &   $-$23.0+10.0$i$   &   1.0$i$   &   92.0   &   $-$1.0$-$1.0$i$   &   $-$15.0$-$6.0$i$   \\  
$\bra{3,-2}$ &    0   &   0   &   0   &   $-$21.0+9.0$i$   &   $-$1.0+1.0$i$   &   87.0   &   $-$1.0$-$2.0$i$   \\  
$\bra{3,-3}$ &    0   &   0   &   0   &   0   &   $-$15.0+6.0$i$   &   $-$1.0+2.0$i$   &   79.0   \\
\hline
\end{tabular}
\end{adjustbox}
\end{table}

SOCI calculations were initially performed with complex \textbf{1} oriented in the arbitrarily-chosen coordinate frame shown in Figure \ref{fig:c1-frames}, \textit{i.e.}\ with \emph{z} and \emph{y} collinear with the Co$_2$ and $\mu$-Cl$_2$ internuclear axis, respectively, and \emph{x} perpendicular to the \ce{Co-($\mu$Cl2)-Co} plane. The SOC splits and intercalates the $\ket{S, M_S}$ components of the spin-free states, resulting in 16 low-lying levels stretching up to 123 \icm{} (see Table \ref{tab:c1-soci-energies-SI}) and a continuum of excited levels above $\sim$2700 \icm{}. The low-energy spectrum originates almost entirely from SOC admixture of the 16 $M_S$ components of the ground $S$ = 3, 2, 1, and 0 spin-free states. The SO-NEVPT2 wavefunctions listed in Table \ref{tab:c1-sopt2-wfns-arb-SI} support this statement and demonstrate that each of these states contains more than 90\% summed contributions from such $\ket{S, M_S}$ components. This indicates that the majority of the physics is captured by the low-energy \emph{ab initio} spectrum, which therefore is adequate for the construction of \heff{} and validation of \hms{}.

Following the strategy presented in Section \ref{sec:theory}, the first step toward the resolution of \hms{} involves the identification of the molecular PAF. Here, the SOCI wavefunctions of Table \ref{tab:c1-sopt2-wfns-arb-SI} already show little spin-mixing due to misalignment of the \emph{z}-axis with the principal magnetic anisotropy axis of the complex, meaning that the arbitrarily-chosen input frame is not very different from the molecular PAF. The numerical matrix elements of the $\ket{S=3, M_S}$ block within \heff{} are shown in Table \ref{tab:c1-heff-s3block}. Comparison with the analytical \hmod{}=$\hat{S}\bar{\bar{D}}_{S=3}\hat{S}$ matrix of Table \ref{tab:hmod} concludes with the perfect one-to-one correspondence of the respective matrix elements showing the validity of the model Hamiltonian. Moreover, it is clear that the matrix elements in \heff{} that should vanish in the molecular PAF are very close to zero, \textit{e.g.}\ $\bra{3, 2} \hat{H}^\mathrm{mod} \ket{3, 3}$, $\bra{3, 0} \hat{H}^\mathrm{mod} \ket{3, 1}$, etc., meaning that the arbitrary axis frame is indeed not far from the molecular PAF. Indeed, extraction and diagonalization of $\bar{\bar{D}}_{S=3}$ led to the molecular PAF shown in Figure \ref{fig:c1-frames}, which essentially differs from the arbitrary frame by axis relabeling, \textit{i.e.}\ $z \rightarrow x$, $x \rightarrow y$ and $y \rightarrow z$. SOCI performed with complex \textbf{1} oriented in the molecular PAF led to the wavefunctions printed in Table \ref{tab:c1-sopt2-wfns-mpaf-SI}; these are slightly cleaner than the previous ones and show large spin-mixings between $S$=0 and $S$=2 spin-components, and between the $S$=1 and $S$=3 ones.

\subsubsection{Extraction and validation of the multispin Hamiltonian}

In order to assemble the model matrix of \hms{} (Equation \ref{eq:hms}), we must determine at first the local anisotropy tensors, $\bar{\bar{D}}_a$ and $\bar{\bar{D}}_b$, associated with the two Co centers, and thereof the axial ($D_a$ and $D_b$) and rhombic ($E_a$ and $E_b$) local ZFS parameters (which are here respectively equal by symmetry). To this end, SO-NEVPT2 calculations were performed on the monomeric model structures \textbf{1a} and \textbf{1b} shown in Figure \ref{fig:geoms}, both oriented in the molecular PAF determined above.  Because of symmetry, one may here skip the computation of \textbf{1b}. However, doing it has two advantages, (i) it allows one to validate spreadsheets and (ii) it is generally applicable also in cases with no symmetry. Resolution of the anisotropic Hamiltonians, $\hat{S_a}\bar{\bar{D}}_a\hat{S_a}$ and $\hat{S_b}\bar{\bar{D}}_b\hat{S_b}$, was achieved following the effective Hamiltonian workflow for mononuclear complexes.\cite{Maurice:2009a} $\bar{\bar{D}}_a$ and $\bar{\bar{D}}_b$ are found identical here, as expected due to the centrosymmetry. Moreover, they are diagonal in the molecular PAF:

\begin{equation}
\bar{\bar{D}}_a = \bar{\bar{D}}_b =  \left[\begin{matrix}-2.1 & 0 & 0\\0 & -9.5 & 0\\0 & 0 & 11.6\end{matrix}\right]
\label{eq:c1-local-frames}
\end{equation}

\noindent which shows that $\bar{\bar{D}}_a$, $\bar{\bar{D}}_b$ and $\bar{\bar{D}}_{S=3}$ share the same PAF. The ZFS parameters derived from matrix \ref{eq:c1-local-frames} are $D_a$ = $D_b$ = 17.4 \icm{} and $E_a$ = $E_b$ = 3.7 \icm{}. 

The 16x16 analytical matrix of \hms{} was initially derived in the uncoupled-spin basis, $\ket{M_{Sa}, M_{Sb}}$. Subsequently, the matrix was translated into the coupled-spin basis, $\ket{S, M_S}$, and it is listed in Tables \ref{tab:hms-arb1-SI}, \ref{tab:hms-arb2-SI}, and \ref{tab:hms-arb3-SI}. In the molecular PAF, the \hms{} matrix simplifies greatly to that shown in Table \ref{tab:hms}, where all elements depend only on the axial and rhombic parameters of the local anisotropies, $D_a$ and $E_a$ (due to centrosymmetry, $D_b$ and $E_b$ can be replaced by $D_a$ and $E_a$ respectively), and of the symmetric exchange anisotropy, $D_{ab}$ and $E_{ab}$. The 16$\times$16 \heff{} matrix derived from the SO-NEVPT2 calculation in the molecular PAF is given in Tables \ref{tab:c1-heff-pt2-unrevsigns-c-SI} and \ref{tab:c1-heff-pt2-unrevsigns-uc-SI} for the coupled- and uncoupled-spin basis respectively.

In a first approximation, one may assume vanishing symmetric anisotropy, \textit{i.e.}\ neglect the contribution of $D_{ab}$ and $E_{ab}$ to Table \ref{tab:hms}, and use the calculated $J$=11.2, $D_a$=17.4 and $E_a$=3.7 \icm{} to evaluate \hms{}. The \hms{} numerical matrix expressed in the coupled- and uncoupled-spin bases is shown in Tables \ref{tab:c1-hms-noDab-pt2-c-SI} and \ref{tab:c1-hms-noDab-pt2-uc-SI}, respectively. Concerning the coupled-spin basis, the correspondence between \hms{} and \heff{} matrices is outstanding at first glance. A closer look reveals, however, that the $\bra{0, 0} \text{\heff{}} \ket{2, M_S} $ and $\bra{1, M_S} \text{\heff{}} \ket{3, M_S}$ elements, as well as their complex-conjugates, have opposite sign compared to counterparts in \hms{}. The sign discrepancy reported and clarified in the erratum to Reference \citenum{Gould:2022a}, occurs since the $\ket{S, M_S}$ spin functions enter with arbitrary phases in the \emph{ab initio} spin-orbit eigenvectors, and may show up with random $\pm$1 prefactors between different \emph{ab initio} runs or runs on different computers. The sign arbitrariness becomes problematic when \heff{} is translated in the uncoupled-spin basis using tabulated CG coefficients that follow a specific sign convention. The ones used  in this work, shown in Table \ref{tab:umatrix-SI}, follow the Condon and Shortley convention.\cite{Condon:1951a, Alder:1971a} Indeed, comparison between the \hms{} and \heff{} numerical matrices in the uncoupled-spin basis, \ref{tab:c1-hms-noDab-pt2-uc-SI} \textit{vs.}\ \ref{tab:c1-heff-pt2-unrevsigns-uc-SI}, is rather poor, generating misleading conclusions. In order to obey sign conventions, one may express the spin-free wavefunctions in a basis of localized orbitals, pick a phase convention, and derive analytically the expressions of the spin-orbit wavefunctions; finally, revise the CG coefficients according to the chosen phase convention. This path has been adopted in Reference \citenum{Bouammali:2021a}. Pursuing such a scheme, although rigorous, is tedious and not appealing in general. Instead, one may follow the path taken in Reference \citenum{Gould:2022a} and revise the conflicting signs in \heff{} such that projection onto \hms{} using CG coefficients in the Condon-Shortley convention gives the smallest deviation. In this work, it is realized from the onset that, by construction, the model \hms{} already yields the correct signs that must be adopted in \heff{} itself. The sign-revised, coupled-spin basis \heff{} is shown in Table \ref{tab:c1-heff-pt2-revsigns-c-SI} and the uncoupled-spin basis matrix derived thereof is shown in Table \ref{tab:c1-heff-pt2-revsigns-uc-SI}. Comparison with \hms{} counterparts (Tables \ref{tab:c1-hms-noDab-pt2-c-SI} and \ref{tab:c1-hms-noDab-pt2-uc-SI}) reveals outstanding agreement regardless of basis, with maximum deviation not larger than 1 and 1.8 \icm{} for the off-diagonal and diagonal elements respectively (see Tables \ref{tab:c1-diff-uc-SI} and \ref{tab:c1-diff-c-SI}). Three highly important conclusions are here drawn, (i) the local anisotropy parameters calculated independently using monomeric structures \textbf{1a} and \textbf{1b} are transferable to calculations on the dimeric complex \textbf{1}, (ii) the local anisotropies are the main contributors to \hms{} and actually bring the model to very close agreement with \heff{}, and (iii) the approximate \hms{}, accounting only for the Heisenberg exchange and local anisotropies, is validated through the effective Hamiltonian theory and may be already used in modeling magnetic properties such as the $\chi T$ curve.

\begin{table}[b!]
\centering
\small
 \setlength{\tabcolsep}{3pt}
\caption{Magnetic parameters, in \icm{}, obtained from the resolution of \hms{} based on \heff{}'s derived from different calculations on complex \textbf{1} in the molecular PAF. The isotropic $g$-factor per Co center, obtained with models \textbf{1a} and \textbf{1b}, is also printed}
\label{tab:c1-parameters}
\renewcommand{\arraystretch}{1.5}
\begin{tabular}{@{}lcccccc@{}}
\toprule
  Method     & $J$  & $D_a$ & $E_a$ & $D_{ab}$ & $E_{ab}$ & $g$ \\ \midrule
SO-NEVPT2 & 10.5 & 17.13  & 3.68   & 0.25$^a$ / $-$0.64$^b$      & $-$0.34$^a$ / 0.045$^b$ & 2.38  \\
SO-DDCI2  & 19.3 & 17.32  & 3.70   & 0.36$^a$ / $-$0.71$^b$     & $-$0.35$^a$ / 0.004$^b$ & n/a  \\
Fit$^c$       & 23.2 & 29    & --    & --       & --  & 2.25     \\ \bottomrule
\end{tabular}
\flushleft{$^a$Obtained from a least-squares fitting procedure minimizing the RMSD between \hms{} and \heff{}; $^b$Recalculated parameters after relabeling of the axes such that the convention $E_{ab}$>0 and $|D_{ab}| > 3 E_{ab}$ is fulfilled: $x \rightarrow z$, $y \rightarrow x$, $z \rightarrow y$ with SO-NEVPT2 and $x \leftrightarrow z$ with SO-DDCI2; $^c$From Reference \citenum{Sun:1999a}.}
\end{table}

Prior to modeling the $\chi T$ curve, it may be noted that the comparison between the analytical matrix of \hms{} (Table \ref{tab:hms}) and the numerical  matrix of \heff{} (Table \ref{tab:c1-heff-pt2-revsigns-c-SI}) provides with enough equations for a full extraction of all the magnetic parameters at once. Thus, a more precise extraction can be performed on \textbf{1} in order to obtain $J$ under the effect of SOC and both the local and the symmetric-exchange anisotropies. Such a full extraction has been performed by least-squares fitting of the \hms{} parameters in order to minimize the RMSD with the \heff{} matrix elements. The extracted parameters, listed in Table \ref{tab:c1-parameters}, lead to \hms{} matrices (Tables \ref{tab:c1-hms-full-pt2-uc-SI} and \ref{tab:c1-hms-full-pt2-c-SI} for the uncoupled- and coupled-spin basis respectively) that agree with \heff{} within 1 \icm{} concerning the off-diagonal and diagonal elements (see Tables \ref{tab:c1-full-diff-uc-SI} and \ref{tab:c1-full-diff-c-SI}). Furthermore, diagonalization of the model \hms{} matrices leads to highly accurate spin-orbit energies, within 1.2 and 1.8 \icm{} of the \emph{ab initio} energies (see Table \ref{tab:c1-soci-energies-SI}), further supporting the validity of \hms{} and of our extraction scheme.

Several aspects may be noted from Table \ref{tab:c1-parameters}: (i) since $J$ only appears on the diagonal of \hms{} in Table \ref{tab:hms}, switching from NEVPT2 to corrected DDCI2 relative energies affects the $J$ values themselves but leaves mostly unchanged the remaining parameters; the SOC reduces $J$ by 0.7 and 2 \icm{} at the SO-NEVPT2 and SO-DDCI2 levels, respectively, (ii) the re-extracted local anisotropy parameters at the SO-NEVPT2 level, $D_a$=17.1 and $E_a$=3.7 \icm{} are in complete agreement with the counterparts extracted independently with the \textbf{1a} and \textbf{1b} monomeric structures, $D_a$=17.4 and $E_a$=3.7 \icm{}, reaffirming the transferability of these parameters from monomer to dimer calculations, and (iii) the symmetric exchange anisotropy is indeed minor compared to local anisotropies and thus neglecting it from the construction of \hms{} is not a bad approximation. It is worth noting for specialists that re-labeling the axes describing the PAF of $\bar{\bar{D}}_{ab}$ is necessary in order to respect the conventions $E_{ab}$>0 and $|D_{ab}| > 3E_{ab}$. In particular, the easy axis of magnetization of the symmetric anisotropy, which is collinear with the \ce{Co2} internuclear axis, falls perpendicular to the molecular easy axis of local anisotropy, which in turn is almost collinear with the $\mu$-\ce{Cl2} internuclear axis.

\begin{figure}[t!]
\centering
\includegraphics[width=.45\textwidth]{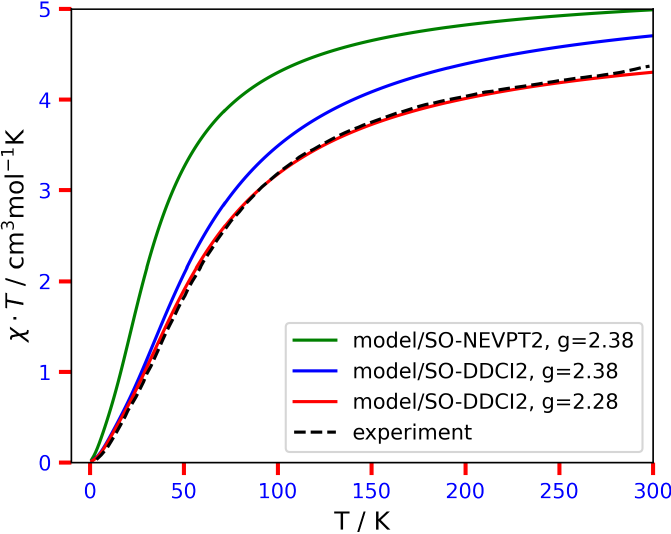}
\caption{$\chi T = f(T)$ experimental curve of complex \textbf{1}, digitized from Reference \citenum{Sun:1999a}, and modeled with the spin Hamiltonian $\hat{H} = \hat{H}^\mathrm{MS} + \hat{H}^\mathrm{zee}$, dressed with parameters listed in Table \ref{tab:c1-parameters}. The best agreement with experiment is obtained with the SO-DDCI2 parameters and a re-adjusted isotropic $g$-factor of 2.28.}
\label{fig:c1-chit}
\end{figure}

Finally, the fully extracted \hms{} is used to model the $\chi T$ curve of complex \textbf{1}. For this task, one needs to express in matrix form the Zeeman terms of the two magnetic centers, $\mu_\mathrm{B} \vec{B} \bar{\bar{g}}_a \hat{S_a}$ and $\mu_\mathrm{B} \vec{B} \bar{\bar{g}}_b \hat{S_b}$, and add them to \hms{}. Working in the uncoupled-spin basis, the 16$\times$16 analytical matrices of the Zeeman terms are derived in Tables \ref{tab:zeeman-a-SI} and \ref{tab:zeeman-b-SI}. These matrices are evaluated with $\bar{\bar{g}}_a$ and $\bar{\bar{g}}_b$ tensors obtained \emph{via} the effective Hamiltonian theory from SO-NEVPT2 calculations performed on structures \textbf{1a} and \textbf{1b} in the molecular PAF:

\begin{equation}
\bar{\bar{g}}_a = \bar{\bar{g}}_b =  \left[\begin{matrix} 2.40 & 0 & 0 \\ 0 & 2.50 & 0 \\ 0 & 0 & 2.23 \end{matrix}\right]
\label{eq:c1-gmat}
\end{equation}

\noindent The diagonal form of Equation \ref{eq:c1-gmat} shows, as expected, that the local PAFs of the $\bar{\bar{g}}$ tensors also coincide with the molecular PAF. The isotropic $g$-factor of 2.38 is very close to the fitted value in Reference \citenum{Sun:1999a}, $g$=2.25.

\begin{landscape}
\begin{table}
\caption{\ Analytical matrix elements of \hms{} = \js{} + \da{} + \db{} + \dab{} for binuclear Co(II) complexes in the  molcular PAF expressed in the coupled spin-basis, $\ket{S, M_S}$}   
\label{tab:hms}
\setlength{\tabcolsep}{0pt}
\renewcommand{\arraystretch}{1.5}
\begin{adjustbox}{width=1.45\textwidth}
\begin{tabular}{lcccccccccccccccc}
\hline
$H^{MS}$  
& $\ket{3, 3}$ & $\ket{3, 2}$ & $\ket{3, 1}$ & $\ket{3, 0}$ & $\ket{3, -1}$ & $\ket{3, -2}$ & $\ket{3, -3}$ \\
\hline

$\bra{3, 3}$ &
$D_{a} + D_{b} + \frac{3 D_{ab}}{2} + \frac{9 J}{4}$ &
0 &
$\frac{\sqrt{15} \left(2 E_{a} + 2 E_{b} + 3 E_{ab}\right)}{10}$ &
0 &
0 &
0 &
0 \\

$\bra{3, 2}$ &
0 &
$\frac{9 J}{4}$ &
0 &
$\frac{\sqrt{30} \left(2 E_{a} + 2 E_{b} + 3 E_{ab}\right)}{10}$ &
0 &
0 &
0 \\

$\bra{3, 1}$ &
$\frac{\sqrt{15} \left(2 E_{a} + 2 E_{b} + 3 E_{ab}\right)}{10}$ &
0 &
$-\frac{3 D_{a}}{5} - \frac{3 D_{b}}{5} - \frac{9 D_{ab}}{10} + \frac{9 J}{4}$ &
0 &
$\frac{6 E_{a}}{5} + \frac{6 E_{b}}{5} + \frac{9 E_{ab}}{5}$ &
0 &
0 \\

$\bra{3, 0}$ &
0 &
$\frac{\sqrt{30} \left(2 E_{a} + 2 E_{b} + 3 E_{ab}\right)}{10}$ &
0 &
$-\frac{4 D_{a}}{5} - \frac{4 D_{b}}{5} - \frac{3 D_{ab}}{5} + \frac{9 E_{ab}}{5} + \frac{9 J}{4}$ &
0 &
$\frac{\sqrt{30} \left(2 E_{a} + 2 E_{b} + 3 E_{ab}\right)}{10}$ &
0 \\

$\bra{3, -1}$ &
0 &
0 &
$\frac{6 E_{a}}{5} + \frac{6 E_{b}}{5} + \frac{9 E_{ab}}{5}$ &
0 &
$- \frac{3 D_{a}}{5} - \frac{3 D_{b}}{5} - \frac{9 D_{ab}}{10} + \frac{9 J}{4}$ &
0 &
$\frac{\sqrt{15} \left(2 E_{a} + 2 E_{b} + 3 E_{ab}\right)}{10}$ \\

$\bra{3, -2}$ &
0 &
0 &
0 &
$\frac{\sqrt{30} \left(2 E_{a} + 2 E_{b} + 3 E_{ab} \right)}{10}$ &
0 &
$\frac{9 J}{4}$ &
0 \\

$\bra{3, -3}$ &
0 &
0 &
0 &
0 &
$\frac{\sqrt{15} \left(2 E_{a} + 2 E_{b} + 3 E_{ab} \right)}{10}$ &
0 &
$D_{a} + D_{b} + \frac{3 D_{ab}}{2} + \frac{9 J}{4}$  \\

$\bra{2, 2}$ &
0 &
$D_{a} - D_{b}$ &
0 &
$\frac{\sqrt{30} \left(E_{a} - E_{b}\right)}{10}$ &
0 &
0 &
0  \\

$\bra{2, 1}$ &
$\frac{\sqrt{6} \left(- E_{a} + E_{b}\right)}{2}$ &
0 &
$\frac{\sqrt{10} \left(D_{a} - D_{b}\right)}{5}$ &
0 &
$\frac{3 \sqrt{10} \left(E_{a} - E_{b}\right)}{10}$ &
0 &
0 \\

$\bra{2, 0}$ &
0 &
$\frac{\sqrt{6} \left(- E_{a} + E_{b}\right)}{2}$ &
0 &
0 &
0 &
$\frac{\sqrt{6} \left(E_{a} - E_{b}\right)}{2}$ &
0 \\

$\bra{2, -1}$ &
0 &
0 &
$\frac{3 \sqrt{10} \left(- E_{a} + E_{b}\right)}{10}$ &
0 &
$\frac{\sqrt{10} \left(- D_{a} + D_{b}\right)}{5}$ &
0 &
$\frac{\sqrt{6} \left(E_{a} - E_{b}\right)}{2}$  \\

$\bra{2, -2}$ &
0 &
0 &
0 &
$\frac{\sqrt{30} \left(- E_{a} + E_{b}\right)}{10}$ &
0 &
$- D_{a} + D_{b}$ &
0 \\

$\bra{1, 1}$ &
$\frac{3 \sqrt{10} \left(E_{a} + E_{b} - E_{ab} \right)}{10}$ &
0 &
$\frac{\sqrt{6} \left(D_{a} + D_{b} - D_{ab} \right)}{5}$ &
0 &
$\frac{\sqrt{6} \left(E_{a} + E_{b} - E_{ab} \right)}{10}$ &
0 &
0 \\

$\bra{1, 0}$ &
0 &
$\frac{\sqrt{30} \left(E_{a} + E_{b} - E_{ab} \right)}{10}$ &
0 &
$\frac{3 D_{a}}{5} + \frac{3 D_{b}}{5} - \frac{4 D_{ab}}{5} - \frac{3 E_{ab}}{5}$ &
0 &
$\frac{\sqrt{30} \left(E_{a} + E_{b} - E_{ab} \right)}{10}$ &
0 \\

$\bra{1, -1}$ &
0 &
0 &
$\frac{\sqrt{6} \left(E_{a} + E_{b} - E_{ab} \right)}{10}$ &
0 &
$\frac{\sqrt{6} \left(D_{a} + D_{b} - D_{ab} \right)}{5}$ &
0 &
$\frac{3 \sqrt{10} \left(E_{a} + E_{b} - E_{ab} \right)}{10}$ \\

$\bra{0, 0}$ &
0 &
0 &
0 &
0 &
0 &
0 &
0 \\ \hline

& $\ket{2, 2}$ & $\ket{2, 1}$ & $\ket{2, 0}$ & $\ket{2, -1}$ & $\ket{2, -2}$ & $\ket{1, 1}$ & $\ket{1, 0}$ & $\ket{1, -1}$ &  $\ket{0, 0}$ \\
\hline

$\bra{3, 3}$ &
0 &
$\frac{\sqrt{6} \left(- E_{a} + E_{b}\right)}{2}$ &
0 &
0 &
0 &
$\frac{3 \sqrt{10} \left(E_{a} + E_{b} - E_{ab}\right)}{10}$ &
0 &
0 &
0 \\

$\bra{3, 2}$ &
$D_{a} - D_{b}$ &
0 &
$\frac{\sqrt{6} \left(- E_{a} + E_{b}\right)}{2}$ &
0 &
0 &
0 &
$\frac{\sqrt{30} \left(E_{a} + E_{b} - E_{ab}\right)}{10}$ &
0 &
0 \\

$\bra{3, 1}$ &
0 &
$\frac{\sqrt{10} \left(D_{a} - D_{b}\right)}{5}$ &
0 &
$\frac{3 \sqrt{10} \left(- E_{a} + E_{b}\right)}{10}$ &
0 &
$\frac{\sqrt{6} \left(D_{a} + D_{b} - D_{ab}\right)}{5}$ &
0 &
$\frac{\sqrt{6} \left(E_{a} + E_{b} - E_{ab}\right)}{10}$ &
0 \\

$\bra{3, 0}$ &
$\frac{\sqrt{30} \left(E_{a} - E_{b}\right)}{10}$ &
0 &
0 &
0 &
$\frac{\sqrt{30} \left(- E_{a} + E_{b}\right)}{10}$ &
0 &
$\frac{3 D_{a}}{5} + \frac{3 D_{b}}{5} - \frac{4 D_{ab}}{5} - \frac{3 E_{ab}}{5}$ &
0 &
0 \\

$\bra{3, -1}$ &
0 &
$\frac{3 \sqrt{10} \left(E_{a} - E_{b}\right)}{10}$ &
0 &
$\frac{\sqrt{10} \left(- D_{a} + D_{b}\right)}{5}$ &
0 &
$\frac{\sqrt{6} \left(E_{a} + E_{b} - E_{ab} \right)}{10}$ &
0 &
$\frac{\sqrt{6} \left(D_{a} + D_{b} - D_{ab} \right)}{5}$ &
0 \\

$\bra{3, -2}$ &
0 &
0 &
$\frac{\sqrt{6} \left(E_{a} - E_{b}\right)}{2}$ &
0 &
$- D_{a} + D_{b}$ &
0 &
$\frac{\sqrt{30} \left(E_{a} + E_{b} - E_{ab} \right)}{10}$ &
0 &
0 \\

$\bra{3, -3}$ &
0 &
0 &
0 &
$\frac{\sqrt{6} \left(E_{a} - E_{b}\right)}{2}$ &
0 &
0 &
0 &
$\frac{3 \sqrt{10} \left(E_{a} + E_{b} - E_{ab} \right)}{10}$ &
0 \\

$\bra{2, 2}$ &
$D_{ab} - \frac{3 J}{4}$ &
0 &
$\frac{\sqrt{6} E_{ab}}{2}$ &
0 &
0 &
0 &
$\frac{\sqrt{30} \left(- E_{a} + E_{b}\right)}{5}$ &
0 &
$\frac{\sqrt{6} \left(E_{a} + E_{b} - E_{ab} \right)}{2}$  \\

$\bra{2, 1}$ &
0 &
$- \frac{D_{ab}}{2} - \frac{3 J}{4}$ &
0 &
$\frac{3 E_{ab}}{2}$ &
0 &
$\frac{\sqrt{15} \left(D_{a} - D_{b}\right)}{5}$ &
0 &
$\frac{\sqrt{15} \left(- E_{a} + E_{b}\right)}{5}$ &
0 \\

$\bra{2, 0}$ &
$\frac{\sqrt{6} E_{ab}}{2}$ &
0 &
$- \frac{4 D_{ab}}{3} - E_{ab} - \frac{3 J}{4}$ &
0 &
$\frac{\sqrt{6} E_{ab}}{2}$ &
0 &
0 &
0 &
$D_{a} + D_{b} - \frac{2 D_{ab}}{3} + E_{ab}$    \\

$\bra{2, -1}$ &
0 &
$\frac{3 E_{ab}}{2}$ &
0 &
$- \frac{D_{ab}}{2} - \frac{3 J}{4}$ &
0 &
$\frac{\sqrt{15} \left(E_{a} - E_{b}\right)}{5}$ &
0 &
$\frac{\sqrt{15} \left(- D_{a} + D_{b}\right)}{5}$ &
0 \\

$\bra{2, -2}$ &
0 &
0 &
$\frac{\sqrt{6} E_{ab}}{2}$ &
0 &
$D_{ab} - \frac{3 J}{4}$ &
0 &
$\frac{\sqrt{30} \left(E_{a} - E_{b}\right)}{5}$ &
0 &
$\frac{\sqrt{6} \left(E_{a} + E_{b} - E_{ab} \right)}{2}$ \\

$\bra{1, 1}$ &
0 &
$\frac{\sqrt{15} \left(D_{a} - D_{b}\right)}{5}$ &
0 &
$\frac{\sqrt{15} \left(E_{a} - E_{b}\right)}{5}$ &
0 &
$- \frac{2 D_{a}}{5} - \frac{2 D_{b}}{5} + \frac{17 D_{ab}}{30} - \frac{11 J}{4}$ &
0 &
$- \frac{6 E_{a}}{5} - \frac{6 E_{b}}{5} + \frac{17 E_{ab}}{10}$ &
0  \\

$\bra{1, 0}$ &
$\frac{\sqrt{30} \left(- E_{a} + E_{b}\right)}{5}$ &
0 &
0 &
0 &
$\frac{\sqrt{30} \left(E_{a} - E_{b}\right)}{5}$ &
0 &
$\frac{4 D_{a}}{5} + \frac{4 D_{b}}{5} - \frac{16 D_{ab}}{15} + \frac{E_{ab}}{5} - \frac{11 J}{4}$ &
0 &
0 \\

$\bra{1, -1}$ &
0 &
$\frac{\sqrt{15} \left(- E_{a} + E_{b}\right)}{5}$ &
0 &
$\frac{\sqrt{15} \left(- D_{a} + D_{b}\right)}{5}$ &
0 &
$- \frac{6 E_{a}}{5} - \frac{6 E_{b}}{5} + \frac{17 E_{ab}}{10}$ &
0 &
$- \frac{2 D_{a}}{5} - \frac{2 D_{b}}{5} + \frac{17 D_{ab}}{30} - \frac{11 J}{4}$ &
0 \\

$\bra{0, 0}$ &
$\frac{\sqrt{6} \left(E_{a} + E_{b} - E_{ab}\right)}{2}$ &
0 &
$D_{a} + D_{b} - \frac{2 D_{ab}}{3} + E_{ab}$ &
0 &
$\frac{\sqrt{6} \left(E_{a} + E_{b} - E_{ab} \right)}{2}$ &
0 &
0 &
0 &
$- \frac{D_{ab}}{3} - E_{ab} - \frac{15 J}{4}$ \\

\hline
\end{tabular}
\end{adjustbox}
\end{table}
\end{landscape}

The modeled $\chi T = f(T)$ curves with magnetic parameters from Table \ref{tab:c1-parameters}, shown in Figure \ref{fig:c1-chit}, highlight primarily the role of the magnetic coupling $J$ in achieving agreement with the experiment. Since the $J$=10.5 \icm{} extracted from the SO-NEVPT2 calculation is too small, the $\chi T$ curve increases steadily and diverges from the reference experimental curve. The revised $J$=19.3 \icm{} extracted out of the SO-DDCI2 calculation leads to $\chi T$ in a much closer agreement, especially in the low-temperature range, $\sim$0--50 K. Furthermore, with a slightly adjusted isotropic $g$-factor of 2.28, which is even closer to the fitted 2.25 value,\cite{Sun:1999a} the modeled $\chi T$ curve with the SO-DDCI2 parameters fits almost perfectly the experimental curve in the whole temperature range. The present article does not aim to delve into higher-level theoretical approaches that could potentially improve the description of $g$ from the onset. Therefore, we conclude that the successful resolution and utilization of the standard multispin model Hamiltonian for calculating magnetic properties in the centrosymmetric complex \textbf{1} have been here neatly demonstrated.

\subsection{\ce{[Co2(L)2(acac)2(H2O)]}, an unsymmetrical complex}
\label{sec:res-complex1}
\subsubsection{Introduction and local, Co(II), electronic structures}

The $\mu$-\ce{O2}(phenoxo)-bridged binuclear complex \textbf{2} displays Co(II) centers coordinated by O and N atoms. The local coordination around both metals is near-$O_h$, with bond angles between 85--96$^\circ$ and Co--O and Co--N mean distances of $\sim$2.07 and 2.08 \AA{} respectively. Recent work recorded a $\chi T$ profile consistent with weak ferromagnetic behavior and an interesting maximum below 10 K.\cite{Song:2020a} Fitting of the $\chi T$ curve resulted in $J = -3.74$ \icm (\js{} convention). The study concluded that overall, the spin alignment is favored by the spin-orbit coupling (SOC) and the local coordination geometry around the metals. It is important to note that the local orbital degeneracy of the 3d $t_{2g}$ orbitals in $O_h$ leads to non-zero orbital  momentum in the Co(II) d$^7$ configuration,  thus theoretically invalidating the use of the spin Hamiltonian approach. However, we will show here that the local distortions at play in complex \textbf{2} are large enough to make the spin Hamiltonian approach relevant. Note that it is possible to generalize our approach to account for orbital momentum, which will be the subject of a dedicated work. Therefore, we begin detailing the local many-electron states at the Co(II) centers as provided by NEVPT2 calculations conducted with the \textbf{2a} and \textbf{2b} monomeric structures.

\begin{figure}[t!]
\centering
  \includegraphics[width=.5\textwidth]{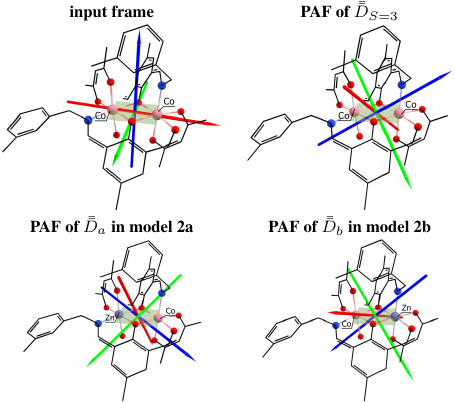}
  \caption{Arbitrarily-chosen coordinate frame and calculated PAFs for complex \textbf{2} (from $\bar{\bar{D}}_{S=3}$) and for the model structures \textbf{2a} (from $\bar{\bar{D}}_a$) and \textbf{2b} (from $\bar{\bar{D}}_b$). Axes color code: $x$=blue, $y$=green, $z$=red.}
  \label{fig:c2-frames}
\end{figure}

\begin{table}[b!]
\caption{Spin-free (SF) and spin-orbit (SO) energies and local magnetic parameters for the Co(II) centers of complex \textbf{2}, obtained with NEVPT2 calculations and different SA/SI schemes$^a$}
\label{tab:c2-monomers-elstr}
\small
\renewcommand{\arraystretch}{1.5}
\begin{tabular}{@{}llllllll@{}}
\toprule
        & \multicolumn{3}{c}{\textbf{2a}} & & \multicolumn{3}{c}{\textbf{2b}} \\ \cmidrule{2-4} \cmidrule{6-8}
SF      & 10Q+40D   & 7Q   & 3Q  & & 10Q+40D   & 7Q   & 3Q  \\ \midrule

$^4T_1$ & 0    & 0    & 0    & & 0    & 0    & 0    \\
        & 622  & 600  & 610  & & 460  & 446  & 468  \\
        & 1079 & 1044 & 1059 & & 1175 & 1143 & 1154 \\

$^4T_2$ & 8952  & 8673  & n/a & & 9189  & 8895 & n/a \\
        & 9341  & 9050  & n/a & & 9352  & 9059 & n/a \\
        & 10555 & 10211 & n/a & & 9583  & 9280 & n/a \\
        
$^4A_2$ & 19969 & 19344 & n/a & & 19190 & 18584 & n/a \\ \midrule
SO$^b$ \\
KD$_1$        & 0   & 0   & 0   & & 0    & 0   & 0 \\
KD$_2$        & 147 & 157 & 178 & & 190  & 194 & 205 \\
KD$_3$        & 721 & 708 & 733 & & 664  & 659 & 688 \\   \midrule

\multicolumn{8}{c}{Local ZFS parameters and isotropic $g$-factors}  \\

& $D_a$ & $E_a$ & $g_a$  & & $D_b$ & $E_b$ & $g_b$ \\
& 86.31 & 11.91 & 2.30 & & 93.14 & 25.02 & 2.31 \\

\bottomrule
\end{tabular}
\\ $^a$Relative energies and ZFS parameters in cm$^{-1}$; letters D and Q are used to denote spin-doublet and spin-quartet states; structures \textbf{2a} and \textbf{2b} are shown in Figure \ref{fig:geoms}. $^b$KD in the labeling of the SO states stands for Kramers Doublet.
\end{table}

\begin{table}[t!]
\centering
\caption{Matrix elements, in cm$^{-1}$, of the $S=3$ block within the 16$\times$16 matrix of \heff{}, constructed from SO-NEVPT2 calculations on complex \textbf{2} in the arbitrary coordinate frame}   
\label{tab:c2-heff-s3block}
\renewcommand{\arraystretch}{1.5}
\begin{adjustbox}{width=\textwidth}
\begin{tabular}{rccccccc}
\hline
\hmod{}  
& $\ket{3, 3}$ & $\ket{3, 2}$ & $\ket{3, 1}$ & $\ket{3, 0}$ & $\ket{3, -1}$ & $\ket{3, -2}$ & $\ket{3, -3}$ \\
\hline

$\bra{3, 3}$ & $ 134.0 $ & $ 18.0 + 44.0 i $ & $ 5.0 + 62.0 i $ & $ 0 $ & $ 0 $ & $ 0 $ & $ 0 i$ \\
$\bra{3, 2}$ & $ 18.0 - 44.0 i $ & $ 188.0 $ & $ 14.0 + 34.0 i $ & $ 8.0 + 88.0 i $ & $ 0 $ & $ 0 $ & $ 0 $ \\
$\bra{3, 1}$ & $ 5.0 - 62.0 i $ & $ 14.0 - 34.0 i $ & $ 221.0 $ & $ 5.0 + 13.0 i $ & $ 8.0 + 97.0 i $ & $ 0 $ & $ 0 $ \\
$\bra{3, 0}$ & $ 0 $ & $ 8.0 - 88.0 i $ & $ 5.0 - 13.0 i $ & $ 232.0 $ & $ -5.0 - 13.0 i $ & $ 8.0 + 88.0 i $ & $ 0 $ \\
$\bra{3, -1}$ & $ 0 $ & $ 0 $ & $ 8.0 - 97.0 i $ & $ -5.0 + 13.0 i $ & $ 221.0 $ & $ -14.0 - 34.0 i $ & $ 5.0 + 62.0 i $ \\
$\bra{3, -2}$ & $ 0 $ & $ 0 $ & $ 0 $ & $ 8.0 - 88.0 i $ & $ -14.0 + 34.0 i $ & $ 188.0 $ & $ -18.0 - 44.0 i $ \\ 
$\bra{3, -3}$ & $ 0 $ & $ 0 $ & $ 0 $ & $ 0 $ & $ 5.0 - 62.0 i $ & $ -18.0 + 44.0 i $ & $ 134.0 $ \\
\hline
\end{tabular}
\end{adjustbox}
\end{table}

The ground term of the free-ion Co(II), denoted as $^4F$, splits in (weak) $O_h$ ligand fields, resulting in a ground $^4T_1$ state and excited $^4T_2$ and $^4A_2$  states. The threefold degeneracy of the orbital-triplet states is then lifted by distortions of the local $O_h$ coordination. According to Table \ref{tab:c2-monomers-elstr}, both Co(II) centers exhibit spin-free energy levels with a $^4F$ parentage that are well isolated within specific ranges: approximately 0-1100 \icm{} for $^4T_1$, 8600-10500 \icm{} for $^4T_2$, and above 18500 \icm{} for $^4A_2$. The energy variation of these individual levels on the employed SA scheme is minimal. If the overall degeneracy lift of $^4T_1$ is similar for both Co centers (around 1100 \icm{}), the splitting of $^4T_2$ is three times larger in \textbf{1a} (around 1500 \icm{}) than in \textbf{1b} (around 400 \icm{}). Taken together with the fact that the spectral widths of the states listed in Table \ref{tab:c2-monomers-elstr} is about 750 \icm{} higher in \textbf{2a} compared to \textbf{2b}, these data reflect the slightly more distorted octahedron around the magnetic center of the former structure. Overall, based on energy gaps alone, one may conclude that the three levels of $^4T_1$ parentage are the primary contributors shaping the local spin-orbit electronic structures. Consequently, performing SOC calculations with a SA/SI scheme involving these three quartets should be sufficient to address the local ZFS parameters. 

Indeed, we found that the lowest-energy Kramers doublets (KDs) only marginally change when increasing the SA/SI scheme beyond the first three quartet roots. As listed in Table \ref{tab:c2-monomers-elstr}, each Co(II) center exhibits one KD below approximately 200 \icm{} and the others above 700 \icm{}. Evidently, all these Kramers doublets (KDs) fully arise as admixtures of $\ket{S=\nicefrac{3}{2}, M_S}$ components of the spin-free states correlating with $^4T_1$. Furthermore, the spin-free quartet GS predominantly contributes to the wavefunctions of KD$_1$ ($\sim$75\%) and KD$_2$ ($\sim$ 90\%) in both \textbf{2a} and \textbf{2b}. The next significant contribution to KD$_1$, approximately 20\%, comes from the first excited spin-quartet root.  Anyway, one can confidently extract local ZFS parameters from the usual anisotropic spin Hamiltonians $\hat{S_a}\bar{\bar{D}}_a\hat{S_a}$ and $\hat{S_b}\bar{\bar{D}}_b\hat{S_b}$, expressed in the $\ket{S=\nicefrac{3}{2}, M_S}$ basis of the ground quartet state (the model space thus consisting of the leading contributors to both KD$_1$ and KD$_2$).

The extracted axial, $D_a$ and $D_b$, and rhombic, $E_a$ and $E_b$, ZFS parameters are listed at the bottom of Table \ref{tab:c2-monomers-elstr}. The PAFs of the  local rank-2 tensors, $\bar{\bar{D}}_a$ and $\bar{\bar{D}}_b$, shown in Figure \ref{fig:c2-frames}, are very distinct this time, with orientations that may not have been guessed without performing those calculations. In comparison with complex \textbf{1}, the axial parameters $D_{a/b} \simeq 90$ \icm{} are at least fivefold larger. Furthermore, $E_b$ = 25 \icm{} is twice larger than $E_a$ = 12 \icm, and both are at least three times larger than the local rhombic parameter in complex \textbf{1}, 3.7 \icm{}. On another hand, the Co centers share similar isotropic $g$-factor, $g_a \simeq g_b \simeq 2.30$, with that of the Co centers in complex \textbf{1}, $g_a = g_b = 2.35$. Here, however, the spans of $\bar{\bar{g}}_a$ and $\bar{\bar{g}}_b$, 1.86--2.66 and 1.81--2.84 respectively, are much larger than the $\bar{\bar{g}}$-tensor span in complex \textbf{2}, 2.23--2.40, also highlighting the much larger anisotropy in complex \textbf{2} in terms of the local $\bar{\bar{g}}$'s. 

\subsubsection{Spin-free and spin-orbit spectra and determination of the molecular PAF}

\begin{table}[b!]
\small
\caption{Low-energy spin-orbit spectrum of complex \textbf{2}$^a$}
\label{tab:c2-soci}
\renewcommand{\arraystretch}{1.5}
\begin{tabular}{@{}clcclc@{}}
\toprule
SO state & $\Delta E$ & $\sum \ket{S, M_S}^b$ & SO state  & $\Delta E$ & $\sum \ket{S, M_S}^b$ \\ \midrule
$\Psi_1$   & 0          & 0.52    & $\Psi_9$    & 206.4      & 0.66    \\
$\Psi_2$   & 1.9        & 0.51    & $\Psi_{10}$ & 206.9      & 0.67    \\
$\Psi_3$   & 3.6        & 0.50    & $\Psi_{11}$ & 209.6      & 0.65    \\
$\Psi_4$   & 7.7        & 0.50    & $\Psi_{12}$ & 212.2      & 0.66    \\
$\Psi_5$   & 177.0      & 0.62    & $\Psi_{13}$ & 379.7      & 0.80    \\
$\Psi_6$   & 177.7      & 0.64    & $\Psi_{14}$ & 380.0      & 0.80    \\
$\Psi_7$   & 180.7      & 0.64    & $\Psi_{15}$ & 388.4      & 0.69    \\
$\Psi_8$   & 182.2      & 0.61    & $\Psi_{16}$ & 389.0      & 0.81    \\ \bottomrule
\end{tabular}
\\$^a$SO-NEVPT2 calculations, relative energies in cm$^{-1}$. Other excited energy levels start at 691 cm$^{-1}$; $^b$Total contribution from the spin-components of the lowest-energy $S$ = 3, 2, 1 and 0 spin-free states.
\end{table}
 
Calculations for complex \textbf{2} were initially performed in an arbitrary coordinate system (see Figure \ref{fig:c2-frames},  the \ce{Co2} and $\mu$-\ce{O2} internuclear axis correspond to the $z$ and $y$ directions, respectively, and the $x$ axis is perpendicular to the Co-[$\mu$\ce{O2}]-Co plane). Spin-free NEVPT2 calculations were conducted, revealing a Land\'e type spectrum with an $S$ = 3 GS followed by a first set of $S$ = 2, 1 and 0 excited states, mapped with a ferromagnetic $J$ ($-2.85$ \icm{}). This value is within 1 \icm{} of the $J=-3.74$ \icm{} obtained from fitting the experimental $\chi T$ curve.\cite{Song:2020a} When considering the SOC, there are 16 lowest-lying energy levels within 389 \icm (see Tables \ref{tab:c2-soci} and \ref{tab:c2-soci-energies-SI}). In comparison to complex \textbf{1}, not only is this energy range three times larger, but the continuum of excited levels begins at a much lower energy, 691 \icm{}. Importantly, the total contribution of the lowest-energy $S$=3, 2, 1, 0 spin-free states to the wavefunctions of those 16 spin-orbit levels gradually increases, from about 50\% in $\Psi_1$ to 81\% in $\Psi_{16}$. Although these weights are lower than what was observed in \textbf{1}, it should be noted that our SOCI wave functions of reference are still dominated by the standard model space (the projection on the model space being more than 50\%) and that \textit{in fine} will we show that our approach allows us to quite accurately reproduce the experimental $\chi T$ curve.

A quick look at Table \ref{tab:c2-sopt2-wfns-arb-SI} reveals that the wavefunctions of the 16 lowest-lying spin-orbit levels are plagued by serious spin-mixing, which may be less important if the molecular PAF is used. We proceeded, therefore, as before to derive the molecular PAF. The matrix elements of the $S=3$ block within the 16$\times$16 \heff{}, shown in Table \ref{tab:c2-heff-s3block}, perfectly match the \hmod{} analytical matrix elements listed in Table \ref{tab:hmod}. Unlike in the previous case of complex \textbf{1}, matrix elements such as $\bra{3, 2} \hat{H}^\mathrm{eff} \ket{3, 3}$ or $\bra{3, 1} \hat{H}^\mathrm{eff} \ket{3, 2}$ deviate significantly from zero, whereas they should be zero if the input \emph{xyz} frame is the molecular PAF. Extraction and diagonalization of $\bar{\bar{D}}_{S=3}$ led to the molecular PAF depicted in Figure \ref{fig:c2-frames}. This frame is not only distinct from the initial input frame but also distinct from the local PAFs (of $\bar{\bar{D}}_a$ and $\bar{\bar{D}}_b$). The SO-NEVPT2 calculation was repeated in the molecular PAF, leading to much cleaner wavefunctions for the 16 lowest-lying energy levels (see Table \ref{tab:c2-sopt2-wfns-mpaf-SI}).


\subsubsection{Extraction of the multispin Hamiltonian}

Figure \ref{fig:c2-frames} clearly highlights that the local anisotropy tensors, $\bar{\bar{D}}_a$ and $\bar{\bar{D}}_b$, cannot be diagonal in the molecular PAF. Since our procedure does not require any assumption on the local PAFs, we have re-computed (rotated) them in the molecular PAF prior to assembling \hms{} according to Equation \ref{eq:hms}:

\begin{equation}
\small
\bar{\bar{D}}_a = \left[\begin{matrix}-33.1 & 7.2 & -15.9\\7.2 & -19.4 & 14.3\\-15.9 & 14.3 & 52.6\end{matrix}\right];
\bar{\bar{D}}_b = \left[\begin{matrix}-1.8 & -6.2 & 16.3\\-6.2 & -53.8 & -14.6\\16.3 & -14.6 & 55.7\end{matrix}\right]
\label{eq:c2-localaniso}
\end{equation}

\noindent It should be stressed that no specific relationship appears between the elements of these tensors, in accord with the C$_1$ symmetry point group of complex \textbf{2}. 

We are now in the position to best equate the 16$\times$16 matrix of \hms{} with the representative matrix of \heff{}. We first consider the expressions reported in Tables \ref{tab:hms-arb1-SI} and \ref{tab:hms-arb2-SI}, meaning that we neglect the symmetric exchange tensor. By using the $-$2.85 \icm{} spin-free $J$ value and the full $\bar{\bar{D}}_{a}$ and $\bar{\bar{D}}_{a}$ tensors (Equation \ref{eq:c2-localaniso}), the estimated numerical matrix of \hms{} is given in Tables \ref{tab:c2-hms-noDab-pt2-uc-SI} and \ref{tab:c2-hms-noDab-pt2-c-SI} in the uncoupled $\ket{M_{Sa}, M_{Sb}}$ and coupled $\ket{S, M_S}$ spin basis, respectively. These are brought in correspondence with the effective Hamiltonians shown in Tables \ref{tab:c2-heff-pt2-c-SI} (coupled-spin basis) and \ref{tab:c2-heff-pt2-uc-SI} (uncoupled-spin basis).  As was the case with complex \textbf{1}, outstanding agreement is obtained from the comparison regardless of the spin basis. Indeed, the difference between \heff{} and \hms{} is already smaller than 1 and 2.7 \icm{} concerning the off-diagonal and diagonal matrix elements, respectively. It is worth emphasizing once again that the independently calculated local anisotropy tensors, $\bar{\bar{D}}_a$ and $\bar{\bar{D}}_b$, within the framework of the  \textbf{2a} and \textbf{2b} monomeric structures, can be safely transferred to the dimeric structure \textbf{2}.

As with complex \textbf{1}, we proceeded to extract $J$ under the influence of SOC, as well as the symmetric anisotropy tensor $\bar{\bar{D}}_{ab}$, without revising the local anisotropies. In other words, we constructed \hms{} as follows:
\begin{equation}
\text{\hms{}} = \text{\js{}} + \left[ \text{\da{}} + \text{\db{}} \right]_\mathrm{mono} + \text{\dab{}}
\label{eq:c2-hms-fullextraction}
\end{equation}

\begin{table}[t!]
\centering
\small
\caption{Magnetic parameters, in \icm{}, obtained from the complete resolution of \hms{} for complex \textbf{2} oriented in the molecular PAF }
\label{tab:c2-parameters}
\renewcommand{\arraystretch}{1.5}
\begin{tabular}{@{}cccccccccccccccccc@{}}
\toprule

$J$ & $D_{ab}$ & $E_{ab}$  & $D_a$ & $E_a$ & $D_b$ & $E_b$  \\ \midrule
$-$2.28 & $-$0.07$^a$ / 0.22$^b$ & $-$0.13$^a$ / 0.03$^b$ & 86.31 & 11.91 & 93.14 & 25.02 \\

\bottomrule
\end{tabular}
\flushleft{$^a$Generated from a least-squares fitting of the model parameters in order to minimize the RMSD between \hms{} and \heff{}; $^b$Recalculated parameters after relabeling of the MPAF axes such that the convention $E_{ab}$>0 and $|D_{ab}| > 3 E_{ab}$ is fulfilled: $y \rightarrow z$, $z \rightarrow x$, $x \rightarrow y$.}
\end{table}

\noindent where ``mono'' refers to the monomer calculations. The contribution of \dab{} to \hms{}, expressed in an arbitrary axis frame and in the coupled-spin basis, is displayed in Table \ref{tab:hms-arb3-SI}. The contribution of \dab{} to \hms{} in its own PAF is given in Table \ref{tab:deab-mpaf-SI}.  A careful inspection of all the relevant tables revealed that in fact the previous deviations between the estimated \hms{} and \heff{} can essentially be explained by Table \ref{tab:deab-mpaf-SI}. In other words, the PAF of  \dab{} practically matches the molecular PAF, and thus we only need to fit the $J$, $D_{ab}$, and $E_{ab}$ parameters to refine \hms{}. 

Table \ref{tab:c2-parameters} summarizes all the key magnetic parameters resulting from the resolution of \hms{} for the case of complex \textbf{2} (we recall that the $\bar{\bar{D}}_a$ and $\bar{\bar{D}}_b$ tensors are not diagonal in the molecular PAF). In the coupled-spin basis, the numerical matrix of \hms{} is shown in Table \ref{tab:c2-hms-full-pt2-c-SI}. This model is of course in even better agreement with the effective Hamiltonian (Table \ref{tab:c2-heff-pt2-c-SI}). The difference matrix between \hms{} and \heff{}, displayed in Table \ref{tab:c2-full-diff-c-SI}, now shows that all the elements are generally smaller than 1 \icm{}. Furthermore, diagonalization of our model \hms{} yields highly accurate spin-orbit energies, with an average deviation of only 1.2 \icm{} and a maximum deviation of 2.6 \icm{}. Consequently, we have successfully applied our new procedure to properly extract all the rank-2 tensors of \hms{} and, ultimately, validated the standard multispin Hamiltonian for any binuclear complex since the key step of  unsymmetrical dicobalt(II) complexes is now solved.

\begin{figure}[t!]
\centering
  \includegraphics[width=.5\textwidth]{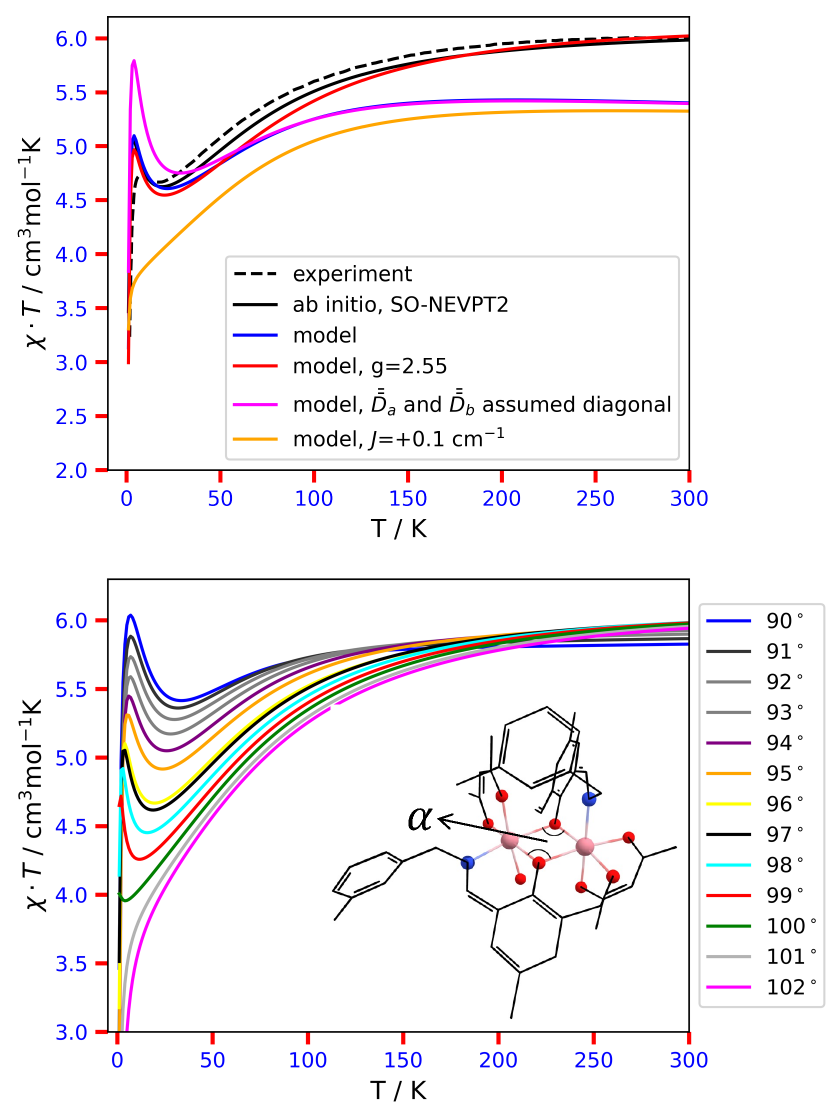}
  \caption{Top: Experimental $\chi T = f(T)$ curve of complex \textbf{2}, digitized from Reference \citenum{Song:2020a}, and generated from \emph{ab initio} calculations and multispin Hamiltonian models. Bottom: $\chi T = f(T)$ curves obtained with SO-NEVPT2 as a function of the angle $\alpha^\circ = \angle$Co--O--Co. In the crystal structure, $\alpha = 97^\circ$, and the black curve represents the best approximation of the experimental $\chi T$.}
  \label{fig:c2-chit}
\end{figure}

Finally, we take a step forward and attempt to model the powder-averaged $\chi T$ profile of complex \textbf{2} using the validated \hms{}. For this purpose, we use again the analytical expressions of the Zeeman Hamiltonians, derived in Tables \ref{tab:zeeman-a-SI} and \ref{tab:zeeman-b-SI}, using the $\bar{\bar{g}}_a$ and $\bar{\bar{g}}_b$ tensors obtained with the \textbf{2a} and \textbf{2b} structural models,  based on calculations performed in the molecular PAF:

\begin{equation}
\small
\bar{\bar{g}}_a = \left[\begin{matrix}2.59 &	-0.10 &	0.13 \\ -0.09 &	2.43 & -0.11 \\ 0.12 &	-0.08 &	1.89
\end{matrix}\right];
\bar{\bar{g}}_b = \left[\begin{matrix}2.23 &	0.06 & -0.11 \\ 0.06 & 2.83 & 0.13 \\-0.11&0.12&1.86\end{matrix}\right]
\label{eq:c2-gtensors}
\end{equation}

\noindent Those matrices are not diagonal in the molecular PAF and in fact, the corresponding local PAFs also do not strictly match one another (as was observed for $\bar{\bar{D}}_a$ and $\bar{\bar{D}}_b$). Since the off-diagonal elements are smaller than the diagonal ones in Equations \ref{eq:c2-gtensors},  one can still consider that $\bar{\bar{g}}_a$ and $\bar{\bar{g}}_b$ are close to being diagonal in the molecular PAF, even if a detailed analysis may reveal an interchange between the respective hard and intermediate axes of magnetization between $\bar{\bar{g}}_a$ and $\bar{\bar{g}}_b$.

Figure \ref{fig:c2-chit}, top panel, demonstrates the excellent agreement between the reference, experimental $\chi T$ curve\cite{Song:2020a} and the one obtained directly out of the SO-NEVPT2 calculation. Therefore, we can be confident in the quality of our \textit{ab initio} calculations and now aim at producing good quality model curves, in view of further supporting the validity of both \hms{} and of our procedure to extract the magnetic parameters. The model curve, obtained from \hms{} + \hzee{} dressed with quantities from Table \ref{tab:c2-parameters} and Equation \ref{eq:c2-gtensors}, agrees with the reference data in the low-temperature region. As the temperature increases, it deviates more and more, eventually reaching a plateau around 150 K which is quite lower than the experimental curve. The behavior may be caused by (i) the lack of excited states in our model that may be populated at some point and in particular at 300 K and (ii) an underestimation of the isotropic $g$ value for each Co center. Concerning point (i), the sum of the Boltzmann populations of the first 16 energy levels is 94\% at 300 K.  We hypothesize that the remaining 6\% in terms of population are not mandatory to explain the discrepancy between this first model curve and the \textit{ab initio} one. Furthermore, it appears that the model curve is too low for a much larger temperature range, in particular from 30 to 300 K. At 30 K, it is clear that only the first 16 energy levels are populated, hence the problem may fully lie on the isotropic $g$ values, as was observed in complex \textbf{1} (in this case, the isotropic $g$ value had to be tempered to a lower value). By following this hypothesis, we here need to increase the isotropic $g$ value, which leads to a revised model curve that is more satisfactory for the whole 30-300 K problematic range, without compromising the already good 0--30 K range. The fitted $g=2.55$ value is not completely random as it may be justified based on the measured room-temperature $\chi T$ value of $6.03$ $\mathrm{cm^3mol^{-1}K}$, or rather $3.015$ $\mathrm{cm^3mol^{-1}K}$ per Co center, according to:

\begin{equation}
g^2 = \frac{\chi_M \cdot T \cdot 3 k_B}{N_A \cdot S(S+1) \cdot \mu_B^2}
\end{equation}

\noindent This expression leads to $g = 2.54$ with the Boltzmann constant, $k_B = 0.695$  $\mathrm{cm^{-1}K^{-1}}$, $\mu_B = 0.467$, $\mathrm{cm^{-1}T^{-1}}$, and $N_A \cdot \mu_B = 0.558$ $\mathrm{cm^3mol^{-1}T}$. Note that since our main point here is to understand the zero-field behavior, we do not aim to delve deeper into explaining the reasons behind this revision of the SO-NEVPT2 $g$ value.

Two additional models are displayed in Figure \ref{fig:c2-chit}, top panel, evidencing the shape of $\chi T$ if one assumes that the local anisotropy tensors are diagonal in the molecular frame and if the axial and rhombic parameters displayed in Table \ref{tab:c2-parameters} are used, or if $J$ is artificially set to a weakly antiferromagnetic value.  The first scenario artificially pushes up the model curve in the low-temperature range,  meaning that the maximum is too high. While it is quite intuitive that the mismatch of the actual PAFs of the local anisotropy tensors should lower the model curve, we should stress that if one assumes that the tensors are diagonal in the molecular PAF and if one fits the local ZFS parameter values, as could be done to fit the experimental data, this should lead to too weak ZFS parameters. In other words, it is crucial to account for the mismatch of the local PAFs in the model, otherwise, meaningless parameters would be obtained. This is exactly where experimental extractions of \hms{} parameters should not be performed independently from computational chemistry data. 

We now aim to better explain the occurrence of a local maximum of $\chi T$ at low temperatures. From the previous paragraph, we have learned that the mismatch between the local PAFs works towards enhancing this $\chi T$ maximum. Such maximum does not occur if the antiferromagnetic coupling scenario is retained (see Figure \ref{fig:c2-chit}). Thus, one could naively think that $J$ must be maximized to favor the occurrence of such a maximum. In practice, to get a local maximum of $\chi T$, we believe that it requires population of a state that is much less magnetic than the ones before and after it. A close inspection of Table \ref{tab:c2-sopt2-wfns-mpaf-SI} reveals that the fourth spin-orbit level is essentially grounded of the $M_S$ = 0 components of the spin-quintet and spin-singlet states so that this must be the states that are looked for. For this state to pop up before all the components of the $S$ = 3 state, one needs to be in the weak-exchange limit (note that this also strengthens the spin-mixing with the singlet state, which also pushes down this state in the model).  Additionally, this state must be well separated in energy from the ones that are higher, otherwise, a continuous enhancement of $\chi T$ would be observed. According to Table \ref{tab:c2-soci}, a gap of about 170 \icm{} occurs in complex \textbf{2}, which corroborates our interpretation.

To strengthen the conclusion drawn earlier, the bottom panel of Figure \ref{fig:c2-chit} illustrates the variation of $\chi T$, obtained directly from SO-NEVPT2 calculations, as a function of the Co--O--Co angles. At the crystal-structure value of 97$^\circ$, the calculated curve best approximates the reference data. Angles below 97$^\circ$ promote stronger ferromagnetic coupling (without getting rid of the weak exchange regime). Consequently, as the ``less magnetic'' state shifts toward higher energies, the spike in $\chi T$ is both enhanced and slightly shifted to higher temperatures. On the other hand, angles above 100$^\circ$ promote antiferromagnetism. As a result, the $\chi T$ curves no longer exhibit local maxima. The overall structural change, induced when the Co--O--Co angle is changed from 97 to, for instance, 101$^\circ$, is characterized by an RMSD of 0.08 \AA{}, it is thus negligible. Therefore, one can regard the unfolding of $\chi T$ in complex \textbf{2} as an intermediate between the ferromagnetic and antiferromagnetic regimes. This is due to the fact that even a tiny displacement in the molecular structure promptly alters $\chi T$, transitioning between these two regimes.

\subsection{Reinterpretation of the \ce{[Ni2(en)4Cl2]^{2+}} case,  a centrosymmetric dinickel(II) complex in the weak-exchange regime}
\label{sec:results-nickel}

\begin{table}[t!]
\centering
\small
\caption{Numerical matrix elements, in cm$^{-1}$, of \hms{} for the \ce{[Ni2(en)4Cl2]^{2+}}  complex in the coupled-spin basis$^a$ }   
\label{tab:nickel-hms-c}
\renewcommand{\arraystretch}{1.5}
\begin{tabular}{rrrrrrrrrr}
\hline
\hms{}  
& $\ket{2, 2}$ & $\ket{2, 1}$ & $\ket{2, 0}$ & $\ket{2, -1}$ & $\ket{2, -2}$ & $\ket{1, 1}$ & $\ket{1, 0}$ & $\ket{1, -1}$ & $\ket{0, 0}$  \\
\hline

$ \bra{2, 2} $ & 1.67 &   0.00 &   1.62 &   0.00 &   0.00 &   0.00 &   0.00 &   0.00 &   2.39 \\
$ \bra{2, 1} $ & 0.00 &  10.74 &   0.00 &   1.99 &   0.00 &   0.00 &   0.00 &   0.00 &   0.00 \\ 
$ \bra{2, 0} $ & 1.62 &   0.00 &  13.76 &   0.00 &   1.62 &   0.00 &   0.00 &   0.00 &$-$9.07 \\
$ \bra{2,-1} $ & 0.00 &   1.99 &   0.00 &  10.74 &   0.00 &   0.00 &   0.00 &   0.00 &   0.00 \\ 
$ \bra{2,-2} $ & 0.00 &   0.00 &   1.62 &   0.00 &   1.67 &   0.00 &   0.00 &   0.00 &   2.39 \\
$ \bra{1, 1} $ & 0.00 &   0.00 &   0.00 &   0.00 &   0.00 &  21.81 &   0.00 &$-$2.09 &   0.00 \\
$ \bra{1, 0} $ & 0.00 &   0.00 &   0.00 &   0.00 &   0.00 &   0.00 &  12.01 &   0.00 &   0.00 \\ 
$ \bra{1,-1} $ & 0.00 &   0.00 &   0.00 &   0.00 &   0.00 &$-$2.09 &   0.00 &  21.81 &   0.00 \\
$ \bra{0, 0} $ & 2.39 &   0.00 &$-$9.07 &   0.00 &   2.39 &   0.00 &   0.00 &   0.00 &  23.96 \\
\hline
\end{tabular}
\\$^a$Obtained by evaluating the analytical expressions reported by Reference \citenum{Maurice:2010c}.
\end{table}

\begin{table}[b!]
\centering
\caption{Numerical matrix elements of \heff{} expressed in the coupled-spin basis for the \ce{[Ni2(en)4Cl2]^{2+}} complex in the molecular PAF$^a$}   
\label{tab:nickel-heff-c}
\renewcommand{\arraystretch}{1.5}
\begin{adjustbox}{width=\textwidth}
\begin{tabular}{rrrrrrrrrrr}
\hline
\hms{}  
& $\ket{2, 2}$ & $\ket{2, 1}$ & $\ket{2, 0}$ & $\ket{2, -1}$ & $\ket{2, -2}$ & $\ket{1, 1}$ & $\ket{1, 0}$ & $\ket{1, -1}$ & $\ket{0, 0}$  \\
\hline

$ \bra{2, 2} $ & 1.67 &   0.00   &   1.62   &     0.00   &   0.00   &   0.00   &   0.00         &   0.00   &   $-$2.37+0.02$i$ \\
$ \bra{2, 1} $ & 0.00 &  10.74   &   0.00   &     1.99   &   0.00   &   0.00   &   0.00         &   0.00   &   $-$0.02$-$0.09$i$ \\
$ \bra{2, 0} $ & 1.62 &   0.00   &  13.79   &     0.00   &   1.62   &   0.00   &   0.00         &   0.00   &   9.06 \\
$ \bra{2,-1} $ & 0.00 &   1.99   &   0.00   &    10.74   &   0.00   &   0.00   &   0.00         &   0.00   &   0.02$-$0.09$i$  \\
$ \bra{2,-2} $ & 0.00 &   0.00   &   1.62   &    0.00    &   1.67   &   0.00   &   0.00         &   0.00   &   $-$2.37$-$0.02$i$ \\
$ \bra{1, 1} $ & 0.00 &   0.00   &   0.00   &    0.00    &   0.00   &  21.81   &   0.02+0.07$i$   &  $-$2.10 +0.03$i$ &   0.00  \\
$ \bra{1, 0} $ & 0.00 &   0.00   &   0.00   &    0.00    &   0.00   &   0.02$-$0.07$i$ &  12.02   &  $-$0.02$-$0.07$i$ &   0.00  \\
$ \bra{1,-1} $ & 0.00 &   0.00   &   0.00   &    0.00    &   0.00   &  $-$2.10$-$0.03$i$ &  $-$0.02+0.07$i$ &  21.81   &   0.00  \\   
$ \bra{0, 0} $ & $-$2.37$-$0.02$i$ &  $-$0.02+0.09$i$ &  9.06   &   0.02+0.09$i$ &  $-$2.37+0.02$i$ &   0.00   &   0.00   &   0.00   &  24.12 \\
\hline
\end{tabular}
\end{adjustbox}
\flushleft{$^a$Reproduced from Ref. \citenum{Maurice:2010c}.}
\end{table}

An accurate determination of magnetic parameters in \ce{[Ni(en)4Cl2]Cl2} has been reported more than a decade ago in the context of high-field EPR experiments.\cite{Herchel:2007a} The complex is weakly ferromagnetic, with $J = -9.66$, $D_a = -4.78$, and $D_{ab} = -0.64$ cm$^{-1}$. Maurice et al. performed an extraction of these parameters from the anisotropic multispin Hamiltonian through the effective Hamiltonian theory.\cite{Maurice:2010c} The generated parameters, $J = -5.415$, $D_a = -9.437$, $E_a = 2.042$, $D_{ab} = 0.367$ and $E_{ab} = -0.052$ \icm{}, which are in reasonable agreement with the experimental counterparts, allow for the derivation of the \hms{} numerical matrix listed in Table \ref{tab:nickel-hms-c}. Comparison with the numerical \heff{} matrix, reproduced from Reference \citenum{Maurice:2010c} in Table \ref{tab:nickel-heff-c}, shows outstanding agreement down to only $\pm$ prefactors for the elements coupling the $S = 2$ and $S = 0$ blocks in \heff{}. Performing a basis-change to the uncoupled basis using tabulated CG coefficients following the Condon-Shortley convention, these conflicting signs lead to unexpected matrix elements, such as $\bra{1, -1}$\heff{}$\ket{-1, 1}$ = 8.6 \icm{}, that should be null according to the standard \hms{}.\cite{Maurice:2010c} As discussed in this article, in \emph{ab initio} calculations, the conflicting signs arise from arbitrary phases and must be adjusted based on the model matrix. This sign adjustment results in a nearly identical model and effective matrices in both the coupled-spin and uncoupled-spin bases, thereby fully validating the standard multispin Hamiltonian for \ce{[Ni(en)4Cl2]Cl2}. Consequently, in this context, the introduction of rank-4, biquadratic exchange tensors in \hms{} is no longer required, unless high accuracy is sought after. Therefore, it is important to stress that the weak-exchange limit was correctly solved in the previous experimental and theoretical works.\cite{Herchel:2007a, Maurice:2010c} However, the experimental extraction was based on the assumption that rhombicity is negligible, which is not supported by the calculations. Therefore, the experimental $J = -9.66$, $D_a = -4.78$, and $D_{ab} = -0.64$ cm$^{-1}$ values may be the subject of small uncertainties due to the neglect of rhombicity, though the important features (weak-exchange limit, local easy-axis anisotropies, $|D_{ab}|<|D_a|$) are now no doubt.

\section{Conclusion}
\label{sec:conclusion}

By studying dicobalt(II) complexes, we have confirmed the validity of the standard \hms{} independent from the weak- or strong-exchange regime. Using \textit{ab initio} calculations, we have demonstrated that it is possible to extract the full tensors that make up the model, without making any assumptions about their principal axis frames (PAFs). Furthermore, our analysis, based on model $\chi T$ curves, has revealed that assuming the local anisotropy tensors are diagonal in the molecular PAF should lead to erroneous local anisotropy parameters in the fitting process. Therefore, concerning unsymmetrical or low-symmetry binuclear complexes, a rigorous interpretation of low-temperature magnetic data should retain key inputs from quantum mechanical calculations similar to those used in this work, \textit{i.e.} multiconfigurational and relativistic wave functions methods. We hope that the present article will trigger new joint theory/experiment studies based on this renewed perspective of explicitly mapping \emph{ab initio} data onto \hms{}.

\begin{suppinfo}
Supporting Information (SI) available: Complimentary tables reporting numerical and/or analytical matrix elements of model and effective Hamiltionians, wavefunctions and energies, and XYZ coordinates for the two dicobalt(II) complexes under study.
\end{suppinfo}

\section{Author Contributions}
RM conceived the study. DCS, RM and BLG secured funding. DCS performed all the theoretical calculations and implemented all codes utilized in generating the data reported in this manuscript. All authors were equally involved in the data analysis. DCS wrote the first draft of the manuscript. RM and BLG shaped the final form of the manuscript.


\section{Acknowledgements}
DCS acknowledges support from the European Union’s Horizon 2020 Research and Innovation Program under the Marie Sklodowska-Curie Grant Agreement No. 899546, through the  BIENVEN\"{U}E COFUND program. DCS also acknowledges the infrastructure support provided through the RECENT AIR grant agreement MySMIS no. 127324.

\newcommand{\Aa}[0]{Aa}
\providecommand*{\mcitethebibliography}{\thebibliography}
\csname @ifundefined\endcsname{endmcitethebibliography}
{\let\endmcitethebibliography\endthebibliography}{}

\clearpage



\section*{Supplementary material}

Supporting Information for:

# The resolution of the weak-exchange limit made rigorous, simple and general in binuclear complexes

Dumitru-Claudiu Sergentu,[a,b] Boris Le Guennic,[a] and Rémi Maurice[*a]

[a] *Univ Rennes, CNRS ISCR (Institut des Sciences Chimiques de Rennes) – UMR 6226, Rennes, France.*
[b] *Universitatea Alexandru Ioan Cuza din Iași, Laboratorul RA-03 (RECENT AIR), Iași, Romania.*

## Contents

# S1 Supporting tables

Table S1 Matrix $U$ used to translate $\hat{H}^{\text{eff}}$ and $\hat{H}^{\text{MS}}$ between the coupled-spin and uncoupled-spin bases.

| $U$ | $|3,3\rangle$ | $|3,2\rangle$ | $|3,1\rangle$ | $|3,0\rangle$ | $|3,-1\rangle$ | $|3,-2\rangle$ | $|3,-3\rangle$ | $|2,2\rangle$ | $|2,1\rangle$ | $|2,0\rangle$ | $|2,-1\rangle$ | $|2,-2\rangle$ | $|1,1\rangle$ | $|1,0\rangle$ | $|1,-1\rangle$ | $|0,0\rangle$ |
|---|---|---|---|---|---|---|---|---|---|---|---|---|---|---|---|---|
| $\langle 3/2, 3/2 |$ | 1 | 0 | 0 | 0 | 0 | 0 | 0 | 0 | 0 | 0 | 0 | 0 | 0 | 0 | 0 | 0 |
| $\langle 3/2, 1/2 |$ | 0 | $\frac{\sqrt{2}}{2}$ | 0 | 0 | 0 | 0 | 0 | $\frac{\sqrt{2}}{2}$ | 0 | 0 | 0 | 0 | 0 | 0 | 0 | 0 |
| $\langle 1/2, 3/2 |$ | 0 | $\frac{\sqrt{2}}{2}$ | 0 | 0 | 0 | 0 | 0 | $-\frac{\sqrt{2}}{2}$ | 0 | 0 | 0 | 0 | 0 | 0 | 0 | 0 |
| $\langle 3/2, -1/2 |$ | 0 | 0 | $\frac{\sqrt{5}}{5}$ | 0 | 0 | 0 | 0 | 0 | $\frac{\sqrt{2}}{2}$ | 0 | 0 | 0 | $\frac{\sqrt{30}}{10}$ | 0 | 0 | 0 |
| $\langle 1/2, 1/2 |$ | 0 | 0 | $\frac{\sqrt{15}}{5}$ | 0 | 0 | 0 | 0 | 0 | 0 | 0 | 0 | 0 | $-\frac{\sqrt{10}}{10}$ | 0 | 0 | 0 |
| $\langle -1/2, 3/2 |$ | 0 | 0 | $\frac{\sqrt{5}}{5}$ | 0 | 0 | 0 | 0 | 0 | $-\frac{\sqrt{2}}{2}$ | 0 | 0 | 0 | $\frac{\sqrt{30}}{10}$ | 0 | 0 | 0 |
| $\langle 3/2, -3/2 |$ | 0 | 0 | 0 | $\frac{\sqrt{5}}{10}$ | 0 | 0 | 0 | 0 | 0 | $\frac{1}{2}$ | 0 | 0 | 0 | $\frac{3\sqrt{5}}{10}$ | 0 | $\frac{1}{2}$ |
| $\langle 1/2, -1/2 |$ | 0 | 0 | 0 | $\frac{3\sqrt{5}}{10}$ | 0 | 0 | 0 | 0 | 0 | $\frac{1}{2}$ | 0 | 0 | 0 | $-\frac{\sqrt{5}}{10}$ | 0 | $-\frac{1}{2}$ |
| $\langle -1/2, 1/2 |$ | 0 | 0 | 0 | $\frac{3\sqrt{5}}{10}$ | 0 | 0 | 0 | 0 | 0 | $-\frac{1}{2}$ | 0 | 0 | 0 | $-\frac{\sqrt{5}}{10}$ | 0 | $\frac{1}{2}$ |
| $\langle -3/2, 3/2 |$ | 0 | 0 | 0 | $\frac{\sqrt{5}}{10}$ | 0 | 0 | 0 | 0 | 0 | $-\frac{1}{2}$ | 0 | 0 | 0 | $\frac{3\sqrt{5}}{10}$ | 0 | $-\frac{1}{2}$ |
| $\langle 1/2, -3/2 |$ | 0 | 0 | 0 | 0 | $\frac{\sqrt{5}}{5}$ | 0 | 0 | 0 | 0 | 0 | $\frac{\sqrt{2}}{2}$ | 0 | 0 | 0 | $\frac{\sqrt{30}}{10}$ | 0 |
| $\langle -1/2, -1/2 |$ | 0 | 0 | 0 | 0 | $\frac{\sqrt{15}}{5}$ | 0 | 0 | 0 | 0 | 0 | 0 | 0 | 0 | 0 | $-\frac{\sqrt{10}}{5}$ | 0 |
| $\langle -3/2, 1/2 |$ | 0 | 0 | 0 | 0 | $\frac{\sqrt{5}}{5}$ | 0 | 0 | 0 | 0 | 0 | $-\frac{\sqrt{2}}{2}$ | 0 | 0 | 0 | $\frac{\sqrt{30}}{10}$ | 0 |
| $\langle -1/2, -3/2 |$ | 0 | 0 | 0 | 0 | 0 | $\frac{\sqrt{2}}{2}$ | 0 | 0 | 0 | 0 | 0 | $\frac{\sqrt{2}}{2}$ | 0 | 0 | 0 | 0 |
| $\langle -3/2, -1/2 |$ | 0 | 0 | 0 | 0 | 0 | $\frac{\sqrt{2}}{2}$ | 0 | 0 | 0 | 0 | 0 | $-\frac{\sqrt{2}}{2}$ | 0 | 0 | 0 | 0 |
| $\langle -3/2, -3/2 |$ | 0 | 0 | 0 | 0 | 0 | 0 | 1 | 0 | 0 | 0 | 0 | 0 | 0 | 0 | 0 | 0 |

Table S2 SOCI energies, in cm$^{-1}$, corresponding to the 16 low-lying spin-orbit states of complex 1, calculated with various approaches.

| SOCI state | SO-CASSCF | SO-NEVPT2 | model//SO-NEVPT2 | SO-DDCI2 | model//SO-DDCI2 |
|---|---|---|---|---|---|
| $\psi_1$ | 0.0 | 0.0 | 0.0 | 0.0 | 0.0 |
| $\psi_2$ | 4.8 | 6.6 | 5.9 | 11.2 | 10.3 |
| $\psi_3$ | 21.8 | 24.1 | 23.4 | 30.1 | 29.5 |
| $\psi_4$ | 29.0 | 39.9 | 39.0 | 56.1 | 55.4 |
| $\psi_5$ | 56.2 | 55.9 | 54.6 | 76.7 | 74.7 |
| $\psi_6$ | 56.8 | 56.3 | 55.4 | 77.2 | 75.8 |
| $\psi_7$ | 56.9 | 56.8 | 55.6 | 77.7 | 75.8 |
| $\psi_8$ | 57.4 | 57.0 | 55.8 | 77.9 | 76.1 |
| $\psi_9$ | 57.6 | 63.7 | 63.0 | 94.2 | 92.8 |
| $\psi_{10}$ | 70.6 | 72.2 | 71.7 | 106.9 | 106.1 |
| $\psi_{11}$ | 71.4 | 78.2 | 78.0 | 108.5 | 107.6 |
| $\psi_{12}$ | 78.5 | 79.6 | 78.9 | 122.9 | 122.6 |
| $\psi_{13}$ | 92.7 | 87.1 | 86.6 | 133.8 | 133.2 |
| $\psi_{14}$ | 100.0 | 96.1 | 95.7 | 139.4 | 138.9 |
| $\psi_{15}$ | 123.2 | 123.2 | 122.5 | 170.2 | 169.3 |
| $\psi_{16}$ | 123.4 | 123.3 | 122.7 | 170.4 | 169.5 |

Table S3 Contribution of the $|S, M_S\rangle$ components of the low-lying $S = 0, 1, 2, 3$ states to the wavefunctions of the 16 low-energy spin-orbit states, obtained from a SO-NEVPT2 calculation on complex 1 in the arbitrary *xyz* frame.[a]

| SOCI state | $|3,3\rangle$ | $|3,2\rangle$ | $|3,1\rangle$ | $|3,0\rangle$ | $|3,-1\rangle$ | $|3,-2\rangle$ | $|3,-3\rangle$ | $|2,2\rangle$ | $|2,1\rangle$ | $|2,0\rangle$ | $|2,-1\rangle$ | $|2,-2\rangle$ | $|1,1\rangle$ | $|1,0\rangle$ | $|1,-1\rangle$ | $|0,0\rangle$ |
|---|---|---|---|---|---|---|---|---|---|---|---|---|---|---|---|---|
| $\psi_1$ | 0 | 0 | 0 | 0 | 0 | 0 | 0 | 0.13 | 0 | 0 | 0 | 0.13 | 0 | 0 | 0 | 0.65 |
| $\psi_2$ | 0.03 | 0 | 0 | 0 | 0 | 0 | 0.03 | 0 | 0 | 0 | 0 | 0 | 0.42 | 0.00 | 0.42 | 0 |
| $\psi_3$ | 0 | 0.05 | 0 | 0.03 | 0 | 0.05 | 0 | 0 | 0 | 0 | 0 | 0 | 0 | 0.78 | 0 | 0 |
| $\psi_4$ | 0.13 | 0 | 0.11 | 0 | 0.11 | 0 | 0.13 | 0 | 0 | 0 | 0 | 0 | 0.21 | 0 | 0.21 | 0 |
| $\psi_5$ | 0 | 0 | 0 | 0 | 0 | 0 | 0 | 0.45 | 0.01 | 0.01 | 0.01 | 0.45 | 0 | 0 | 0 | 0 |
| $\psi_6$ | 0 | 0 | 0 | 0 | 0 | 0 | 0 | 0 | 0.46 | 0 | 0.46 | 0 | 0 | 0 | 0 | 0 |
| $\psi_7$ | 0 | 0 | 0 | 0 | 0 | 0 | 0 | 0.02 | 0.02 | 0.87 | 0.02 | 0.02 | 0 | 0 | 0 | 0 |
| $\psi_8$ | 0 | 0 | 0 | 0 | 0 | 0 | 0 | 0.01 | 0.45 | 0.014 | 0.45 | 0.01 | 0 | 0 | 0 | 0 |
| $\psi_9$ | 0 | 0.19 | 0.01 | 0.39 | 0.01 | 0.19 | 0 | 0 | 0 | 0 | 0 | 0 | 0 | 0.13 | 0 | 0 |
| $\psi_{10}$ | 0 | 0.01 | 0.27 | 0.01 | 0.27 | 0.01 | 0 | 0 | 0 | 0 | 0 | 0 | 0.18 | 0 | 0.18 | 0 |
| $\psi_{11}$ | 0.36 | 0 | 0.07 | 0.01 | 0.07 | 0 | 0.36 | 0 | 0 | 0 | 0 | 0 | 0.03 | 0 | 0.03 | 0 |
| $\psi_{12}$ | 0 | 0 | 0 | 0 | 0 | 0 | 0 | 0.32 | 0 | 0.02 | 0 | 0.32 | 0 | 0 | 0 | 0.28 |
| $\psi_{13}$ | 0 | 0.45 | 0.01 | 0 | 0.01 | 0.45 | 0 | 0 | 0 | 0 | 0 | 0 | 0 | 0 | 0 | 0 |
| $\psi_{14}$ | 0.32 | 0.01 | 0.06 | 0 | 0.06 | 0.01 | 0.32 | 0 | 0 | 0 | 0 | 0 | 0.08 | 0 | 0.08 | 0 |
| $\psi_{15}$ | 0 | 0.2 | 0.03 | 0.45 | 0.03 | 0.2 | 0 | 0 | 0 | 0 | 0 | 0 | 0 | 0.01 | 0 | 0 |
| $\psi_{16}$ | 0.06 | 0.02 | 0.36 | 0.04 | 0.36 | 0.02 | 0.06 | 0 | 0 | 0 | 0 | 0 | 0 | 0 | 0 | 0 |

[a] The corresponding state energies are printed in Table S2.

Table S4 Contribution of the $|S, M_S\rangle$ components of the low-lying $S = 0, 1, 2, 3$ states to the wavefunctions of the 16 low-lying spin-orbit states, obtained from a SO-NEVPT2 calculation on complex 2 in the molecular PAF.[a]

| SOCI state | $|3,3\rangle$ | $|3,2\rangle$ | $|3,1\rangle$ | $|3,0\rangle$ | $|3,-1\rangle$ | $|3,-2\rangle$ | $|3,-3\rangle$ | $|2,2\rangle$ | $|2,1\rangle$ | $|2,0\rangle$ | $|2,-1\rangle$ | $|2,-2\rangle$ | $|1,1\rangle$ | $|1,0\rangle$ | $|1,-1\rangle$ | $|0,0\rangle$ |
|---|---|---|---|---|---|---|---|---|---|---|---|---|---|---|---|---|
| $\psi_1$ | 0 | 0 | 0 | 0 | 0 | 0 | 0 | 0.02 | 0 | 0.23 | 0 | 0.02 | 0 | 0 | 0 | 0.65 |
| $\psi_2$ | 0 | 0 | 0.03 | 0 | 0.03 | 0 | 0 | 0 | 0 | 0 | 0 | 0 | 0.43 | 0 | 0.43 | 0 |
| $\psi_3$ | 0 | 0 | 0.07 | 0 | 0.07 | 0 | 0 | 0 | 0 | 0 | 0 | 0 | 0.39 | 0 | 0.39 | 0 |
| $\psi_4$ | 0 | 0 | 0 | 0.5 | 0 | 0 | 0 | 0 | 0 | 0 | 0 | 0 | 0 | 0.41 | 0 | 0 |
| $\psi_5$ | 0 | 0 | 0 | 0 | 0 | 0 | 0 | 0 | 0.46 | 0 | 0.46 | 0 | 0 | 0 | 0 | 0 |
| $\psi_6$ | 0 | 0 | 0 | 0 | 0 | 0 | 0 | 0.46 | 0 | 0 | 0 | 0.46 | 0 | 0 | 0 | 0 |
| $\psi_7$ | 0 | 0 | 0 | 0 | 0 | 0 | 0 | 0.4 | 0 | 0.11 | 0 | 0.4 | 0 | 0 | 0 | 0 |
| $\psi_8$ | 0 | 0 | 0 | 0 | 0 | 0 | 0 | 0 | 0.46 | 0 | 0.46 | 0 | 0 | 0 | 0 | 0 |
| $\psi_9$ | 0.01 | 0 | 0.39 | 0 | 0.39 | 0 | 0.01 | 0 | 0 | 0 | 0 | 0 | 0.07 | 0 | 0.07 | 0 |
| $\psi_{10}$ | 0 | 0.17 | 0 | 0.23 | 0 | 0.17 | 0 | 0 | 0 | 0 | 0 | 0 | 0 | 0.36 | 0 | 0 |
| $\psi_{11}$ | 0.01 | 0 | 0.42 | 0 | 0.42 | 0 | 0.01 | 0 | 0 | 0 | 0 | 0 | 0.03 | 0 | 0.03 | 0 |
| $\psi_{12}$ | 0 | 0 | 0 | 0 | 0 | 0 | 0 | 0.04 | 0 | 0.57 | 0 | 0.04 | 0 | 0 | 0 | 0.28 |
| $\psi_{13}$ | 0 | 0.46 | 0 | 0 | 0 | 0.46 | 0 | 0 | 0 | 0 | 0 | 0 | 0 | 0 | 0 | 0 |
| $\psi_{14}$ | 0 | 0.29 | 0 | 0.19 | 0 | 0.29 | 0 | 0 | 0 | 0 | 0 | 0 | 0 | 0.16 | 0 | 0 |
| $\psi_{15}$ | 0.46 | 0 | 0 | 0 | 0 | 0 | 0.46 | 0 | 0 | 0 | 0 | 0 | 0 | 0 | 0 | 0 |
| $\psi_{16}$ | 0.45 | 0 | 0.01 | 0 | 0.01 | 0 | 0.45 | 0 | 0 | 0 | 0 | 0 | 0 | 0 | 0 | 0 |

[a] The corresponding state energies are printed in Table S2.

Table S5 Matrix elements of $\hat{H}^{MS} = J\hat{S}_a\hat{S}_b + \hat{S}_a\bar{\bar{D}}_a\hat{S}_a + \hat{S}_b\bar{\bar{D}}_b\hat{S}_b$ for dicobalt(II) complexes in an arbitrary axis frame expressed in the $|S, M_S\rangle$ coupled spin-basis spanned by the $S = 0, 1, 2, 3$ spin functions. Notations such as $D_{xx}^{a+b}$ and $-D_{xx}^{a-b}$ should be read as $D_{xx}^a + D_{xx}^b$ and $-D_{xx}^a + D_{xx}^b$ respectively. This Table is continued by Table S6. The symmetric anisotropy exchange part, $\hat{S}_a\bar{\bar{D}}_{ab}\hat{S}_b$, is provided separately in Table S7.

Table S6 Table S5 continued.

| $H^{MS}$ | $|1,1\rangle$ | $|1,0\rangle$ | $|1,-1\rangle$ | $|0,0\rangle$ |
|---|---|---|---|---|
| $\langle 3,3|$ | $\dfrac{3\sqrt{10}\left(D_{xx}^{a+b}-2iD_{xy}^{a+b}-D_{yy}^{a+b}\right)}{20}$ | 0 | 0 | 0 |
| $\langle 3,2|$ | $\dfrac{\sqrt{15}\left(-D_{xz}^{a+b}+iD_{yz}^{a+b}\right)}{5}$ | $\dfrac{\sqrt{30}\left(D_{xx}^{a+b}-2iD_{xy}^{a+b}-D_{yy}^{a+b}\right)}{20}$ | 0 | 0 |
| $\langle 3,1|$ | $\dfrac{\sqrt{6}\left(-D_{xx}^{a+b}-D_{yy}^{a+b}+2D_{zz}^{a+b}\right)}{20}$ | $\dfrac{2\sqrt{3}\left(-D_{xz}^{a+b}+iD_{yz}^{a+b}\right)}{5}$ | $\dfrac{\sqrt{6}\left(D_{xx}^{a+b}-2iD_{xy}^{a+b}-D_{yy}^{a+b}\right)}{20}$ | 0 |
| $\langle 3,0|$ | $\dfrac{3\sqrt{2}\left(D_{xz}^{a+b}+iD_{yz}^{a+b}\right)}{10}$ | $-\dfrac{3D_{xx}^{a+b}}{10}-\dfrac{3D_{yy}^{a+b}}{10}+\dfrac{3D_{zz}^{a+b}}{5}$ | $\dfrac{3\sqrt{2}\left(-D_{xz}^{a+b}+iD_{yz}^{a+b}\right)}{10}$ | 0 |
| $\langle 3,-1|$ | $\dfrac{\sqrt{6}\left(D_{xx}^{a+b}+2iD_{xy}^{a+b}-D_{yy}^{a+b}\right)}{20}$ | $-\dfrac{2\sqrt{3}\left(D_{xz}^{a+b}+iD_{yz}^{a+b}\right)}{5}$ | $\dfrac{\sqrt{6}\left(-D_{xx}^{a+b}-D_{yy}^{a+b}+2D_{zz}^{a+b}\right)}{10}$ | 0 |
| $\langle 3,-2|$ | 0 | $\dfrac{\sqrt{30}\left(D_{xx}^{a+b}+2iD_{xy}^{a+b}-D_{yy}^{a+b}\right)}{20}$ | $\dfrac{\sqrt{15}\left(D_{xz}^{a+b}+iD_{yz}^{a+b}\right)}{5}$ | 0 |
| $\langle 3,-3|$ | 0 | 0 | $\dfrac{3\sqrt{10}\left(D_{xx}^{a+b}+2iD_{xy}^{a+b}-D_{yy}^{a+b}\right)}{20}$ | 0 |
| $\langle 2,2|$ | $\dfrac{\sqrt{15}\left(-D_{xz}^{a-b}+iD_{yz}^{a-b}\right)}{5}$ | $\dfrac{\sqrt{30}\left(-D_{xx}^{a-b}+2iD_{xy}^{a-b}+D_{yy}^{a-b}\right)}{10}$ | 0 | $\dfrac{\sqrt{6}\left(D_{xx}^{a+b}-2iD_{xy}^{a+b}-D_{yy}^{a+b}\right)}{4}$ |
| $\langle 2,1|$ | $\dfrac{\sqrt{15}\left(-D_{xx}^{a-b}-D_{yy}^{a-b}+2D_{zz}^{a-b}\right)}{10}$ | $\dfrac{\sqrt{30}\left(D_{xz}^{a-b}-iD_{yz}^{a-b}\right)}{10}$ | 0 | $\sqrt{6}\left(-D_{xz}^{a+b}+iD_{yz}^{a+b}\right)$ |
| $\langle 2,0|$ | $\dfrac{3\sqrt{10}\left(D_{xz}^{a-b}+iD_{yz}^{a-b}\right)}{10}$ | $\dfrac{\sqrt{30}\left(D_{xx}^{a-b}-D_{yy}^{a-b}\right)}{10}$ | $\dfrac{3\sqrt{10}\left(D_{xz}^{a-b}-iD_{yz}^{a-b}\right)}{10}$ | $-\dfrac{D_{xx}^{a+b}}{2}-\dfrac{D_{yy}^{a+b}}{2}+D_{zz}^{a+b}$ |
| $\langle 2,-1|$ | $\dfrac{\sqrt{15}\left(D_{xx}^{a-b}+D_{yy}^{a-b}-2D_{zz}^{a-b}\right)}{10}$ | $\dfrac{\sqrt{15}\left(-D_{xz}^{a-b}-iD_{yz}^{a-b}\right)}{5}$ | $\dfrac{\sqrt{15}\left(-D_{xx}^{a-b}-D_{yy}^{a-b}+2D_{zz}^{a-b}\right)}{10}$ | $\dfrac{\sqrt{6}\left(D_{xx}^{a+b}+2iD_{xy}^{a+b}-D_{yy}^{a+b}\right)}{4}$ |
| $\langle 2,-2|$ | 0 | $\dfrac{3\sqrt{2}\left(-D_{xx}^{a-b}+D_{yy}^{a-b}\right)}{10}$ | $-\dfrac{3D_{xx}^{a-b}}{5}+\dfrac{6iD_{xy}^{a-b}}{5}+\dfrac{3D_{yy}^{a-b}}{5}$ | 0 |
| $\langle 1,1|$ | $-\dfrac{11J}{4}+\dfrac{29D_{xx}^{a+b}}{20}+\dfrac{29D_{yy}^{a+b}}{20}+\dfrac{17D_{zz}^{a+b}}{20}$ | $-\dfrac{11J}{4}+\dfrac{17D_{xx}^{a+b}}{20}+\dfrac{17D_{yy}^{a+b}}{20}+\dfrac{41D_{zz}^{a+b}}{20}$ | $-\dfrac{3D_{xx}^{a+b}}{5}+\dfrac{6iD_{xy}^{a+b}}{5}+\dfrac{3D_{yy}^{a+b}}{5}$ | 0 |
| $\langle 1,0|$ | $\dfrac{3\sqrt{2}\left(-D_{xz}^{a+b}-iD_{yz}^{a+b}\right)}{5}$ | $\dfrac{3\sqrt{2}\left(D_{xz}^{a+b}+iD_{yz}^{a+b}\right)}{5}$ | $\dfrac{3\sqrt{2}\left(D_{xz}^{a+b}+iD_{yz}^{a+b}\right)}{5}$ | 0 |
| $\langle 1,-1|$ | $\dfrac{3D_{xx}^{a+b}}{5}-\dfrac{6iD_{xy}^{a+b}}{5}+\dfrac{3D_{yy}^{a+b}}{5}$ | 0 | $-\dfrac{11J}{4}+\dfrac{29D_{xx}^{a+b}}{20}+\dfrac{29D_{yy}^{a+b}}{20}+\dfrac{17D_{zz}^{a+b}}{20}$ | 0 |
| $\langle 0,0|$ | 0 | 0 | 0 | $-\dfrac{15J}{4}+\dfrac{5D_{xx}^{a+b}}{4}+\dfrac{5D_{yy}^{a+b}}{4}+\dfrac{5D_{zz}^{a+b}}{4}$ |

Table S7 Matrix elements of $\hat{H}^{MS}$ for dicobalt(II) complexes in an arbitrary axis frame expressed in the $|S, M_S\rangle$ coupled spin-basis defined by $S = 0, 1, 2, 3$ spin functions. This table provides the symmetric anisotropy exchange part, $\hat{S}_a \bar{\bar{D}}_{ab} \hat{S}_b$.

Table S8 Complex 1: Matrix elements, in cm$^{-1}$, of $\hat{H}^{\text{eff}}$ expressed in the $|S, M_S\rangle$ coupled-spin basis, obtained from a SO-NEVPT2 calculation on complex 1 in the molecular PAF.

| $\hat{H}^{\text{MS}}$ | $|3,3\rangle$ | $|3,2\rangle$ | $|3,1\rangle$ | $|3,0\rangle$ | $|3,-1\rangle$ | $|3,-2\rangle$ | $|3,-3\rangle$ | $|2,2\rangle$ | $|2,1\rangle$ | $|2,0\rangle$ | $|2,-1\rangle$ | $|2,-2\rangle$ | $|1,1\rangle$ | $|1,0\rangle$ | $|1,-1\rangle$ | $|0,0\rangle$ |
|---|---|---|---|---|---|---|---|---|---|---|---|---|---|---|---|---|
| $\langle 3,3|$ | 58.2 | 0 | 5.1 | 0 | 0 | 0 | 0 | 0 | 0 | 0 | 0 | 0 | $-7.8+0.6i$ | 0 | 0 | 0 |
| $\langle 3,2|$ | 0 | 23.3 | 0 | 7.2 | 0 | 0 | 0 | 0 | 0 | 0 | 0 | 0 | 0 | $-4.5+0.3i$ | 0 | 0 |
| $\langle 3,1|$ | 5.1 | 0 | 2.5 | 0 | 7.9 | 0 | 0 | 0 | 0 | 0 | 0 | 0 | $-16.7$ | 0 | $-2.0+0.1i$ | 0 |
| $\langle 3,0|$ | 0 | 7.2 | 0 | $-4.5$ | 0 | 7.2 | 0 | 0 | 0 | 0 | 0 | 0 | 0 | $-20.4$ | 0 | 0 |
| $\langle 3,-1|$ | 0 | 0 | 7.9 | 0 | 2.5 | 0 | 5.1 | 0 | 0 | 0 | 0 | 0 | $-2.0-0.1i$ | 0 | $-16.7$ | 0 |
| $\langle 3,-2|$ | 0 | 0 | 0 | 7.2 | 0 | 23.3 | 0 | 0 | 0 | 0 | 0 | 0 | 0 | $-4.5-0.3i$ | 0 | 0 |
| $\langle 3,-3|$ | 0 | 0 | 0 | 0 | 5.1 | 0 | 58.2 | 0 | 0 | 0 | 0 | 0 | 0 | 0 | $-7.8-0.6i$ | 0 |
| $\langle 2,2|$ | 0 | 0 | 0 | 0 | 0 | 0 | 0 | $-7.4$ | $0.1i$ | $-0.4+0.2i$ | 0 | 0 | $-42.4$ | 0 | 0 | $-9.6+0.5i$ |
| $\langle 2,1|$ | 0 | 0 | 0 | 0 | 0 | 0 | 0 | $-0.1i$ | $-7.3$ | 0 | $-0.5+0.3i$ | 0 | 0.1 | $-1.8$ | $-0.1$ | $-0.1+0.1i$ |
| $\langle 2,0|$ | 0 | 0 | 0 | 0 | 0 | 0 | 0 | $-0.4-0.2i$ | 0 | $-7.3$ | 0 | $-0.4+0.2i$ | 0 | 0 | 0 | $-33.7$ |
| $\langle 2,-1|$ | 0 | 0 | 0 | 0 | 0 | 0 | 0 | 0 | $-0.5-0.3i$ | 0 | $-7.3$ | $-0.1i$ | 0.1 | $-0.1$ | $-42.4$ | $0.1+0.1i$ |
| $\langle 2,-2|$ | 0 | 0 | 0 | 0 | 0 | 0 | 0 | 0 | 0 | $-0.4-0.2i$ | $0.1i$ | $-7.4$ | 0 | 0 | 0 | $-9.6-0.5i$ |
| $\langle 1,1|$ | $-7.8-0.6i$ | 0 | $-16.7$ | 0 | $-2.0+0.1i$ | 0 | 0 | $-42.4$ | 0.1 | 0 | 0.1 | 0 | 0.1 | $-9.3+0.4i$ | 0 | 0 |
| $\langle 1,0|$ | 0 | $-4.5-0.3i$ | 0 | $-20.4$ | 0 | $-4.5+0.3i$ | 0 | 0 | $-1.8$ | 0 | $-0.1$ | 0 | $-9.3-0.4i$ | 0 | $-9.3+0.4i$ | 0 |
| $\langle 1,-1|$ | 0 | 0 | $-2.0-0.1i$ | 0 | $-16.7$ | 0 | $-7.8+0.6i$ | 0 | $-0.1$ | 0 | $-42.4$ | 0 | 0 | $-9.3-0.4i$ | $-42.4$ | 0 |
| $\langle 0,0|$ | 0 | 0 | 0 | 0 | 0 | 0 | 0 | $-9.6-0.5i$ | $-0.1-0.1i$ | $-33.7$ | $0.1-0.1i$ | $-9.6+0.5i$ | 0 | 0 | 0 | $-40.2$ |

Table S9 Complex 1: Matrix elements, in cm$^{-1}$, of $\hat{H}^{\text{eff}}$ (Table S8) expressed in the uncoupled-spin basis, $|M_{Sa}, M_{Sb}\rangle$.

| $\hat{H}^{\text{MS}}$ | $|\tfrac{3}{2},\tfrac{3}{2}\rangle$ | $|\tfrac{3}{2},\tfrac{1}{2}\rangle$ | $|\tfrac{1}{2},\tfrac{3}{2}\rangle$ | $|\tfrac{3}{2},-\tfrac{1}{2}\rangle$ | $|\tfrac{1}{2},\tfrac{1}{2}\rangle$ | $|-\tfrac{1}{2},\tfrac{3}{2}\rangle$ | $|\tfrac{3}{2},-\tfrac{3}{2}\rangle$ | $|\tfrac{1}{2},-\tfrac{1}{2}\rangle$ | $|-\tfrac{1}{2},\tfrac{1}{2}\rangle$ | $|-\tfrac{3}{2},\tfrac{3}{2}\rangle$ | $|\tfrac{1}{2},-\tfrac{3}{2}\rangle$ | $|-\tfrac{1}{2},-\tfrac{1}{2}\rangle$ | $|-\tfrac{3}{2},\tfrac{1}{2}\rangle$ | $|-\tfrac{1}{2},-\tfrac{3}{2}\rangle$ | $|-\tfrac{3}{2},-\tfrac{1}{2}\rangle$ | $|-\tfrac{3}{2},-\tfrac{3}{2}\rangle$ |
|---|---|---|---|---|---|---|---|---|---|---|---|---|---|---|---|---|
| $\langle \tfrac{3}{2},\tfrac{3}{2}|$ | 58.2 | 0 | 0 | $-2.0+0.3i$ | $8.8-0.3i$ | $-2.0+0.3i$ | 0 | $-4.5+0.4i$ | $7.3-0.1i$ | $0.9+0.1i$ | $2.6-0.1i$ | 0 | 0 | 0 | 0 | 0 |
| $\langle \tfrac{3}{2},\tfrac{1}{2}|$ | 0 | 8.0 | 15.4 | 0 | 0 | 0 | $-4.5+0.4i$ | $7.3-0.1i$ | $0.9+0.1i$ | $7.3-0.1i$ | $-4.5-0.1i$ | $2.6-0.1i$ | 0 | 0 | 0 | 0 |
| $\langle \tfrac{1}{2},\tfrac{3}{2}|$ | 0 | 15.4 | 8.0 | 0 | 0 | 0 | $2.6-0.1i$ | $0.9+0.1i$ | $7.3-0.1i$ | $-4.5-0.1i$ | 0 | 0 | 0 | 0 | 0 | 0 |
| $\langle \tfrac{3}{2},-\tfrac{1}{2}|$ | $-2.0-0.3i$ | 0 | 0 | 0 | $-24.0$ | $-16.7$ | 0 | $-2.5+0.3i$ | $5.7-0.1i$ | $-1.9+0.1i$ | $-0.1i$ | 0 | $0.1i$ | 0 | $-16.7$ | 0 |
| $\langle \tfrac{1}{2},\tfrac{1}{2}|$ | $8.8+0.3i$ | 0 | 0 | $-24.0$ | 13.2 | 0.9 | 0 | $5.7-0.1i$ | $3.0-0.1i$ | $5.7-0.1i$ | 0 | $-35.9$ | 0 | $-16.7$ | 13.2 | 0 |
| $\langle -\tfrac{1}{2},\tfrac{3}{2}|$ | $-2.0-0.3i$ | 0 | 0 | $-16.7$ | 13.2 | $-24.0$ | $-0.1i$ | $-1.9+0.1i$ | $5.7-0.1i$ | $-2.5+0.3i$ | 0 | $-16.8$ | $-0.4$ | 13.2 | $-24.0$ | 0 |
| $\langle \tfrac{3}{2},-\tfrac{3}{2}|$ | 0 | $-4.5+0.4i$ | $2.6-0.1i$ | 0 | 0 | $0.1i$ | $-35.9$ | 0 | $-0.4$ | $21.6$ | $-16.8$ | 9.0 | $-0.4$ | 0 | 0 | $-2.0+0.3i$ |
| $\langle \tfrac{1}{2},-\tfrac{1}{2}|$ | 0 | $7.3+0.1i$ | $0.9-0.1i$ | $-4.5+0.4i$ | $5.7-0.1i$ | $-1.9+0.1i$ | $-0.4$ | $-16.8$ | $-1.0$ | $9.0$ | $-1.0$ | 0 | 0 | 0 | 0 | $8.8-0.3i$ |
| $\langle -\tfrac{1}{2},\tfrac{1}{2}|$ | 0 | $0.9-0.1i$ | $7.3+0.1i$ | $7.3-0.1i$ | $3.0+0.1i$ | $5.7-0.1i$ | $-16.8$ | $-1.0$ | $9.0$ | $-1.0$ | $-16.8$ | $0.9+0.1i$ | $7.3-0.1i$ | 0 | 0 | $-2.0+0.3i$ |
| $\langle -\tfrac{3}{2},\tfrac{3}{2}|$ | 0 | $2.6+0.1i$ | $-4.5-0.4i$ | $-1.9-0.1i$ | $5.7-0.1i$ | $-2.5-0.3i$ | $21.6$ | $9.0$ | $-0.4$ | $-35.9$ | 0 | 0 | 0 | 0 | 0 | 0 |
| $\langle \tfrac{1}{2},-\tfrac{3}{2}|$ | 0 | 0 | 0 | $-2.5-0.3i$ | $5.7+0.1i$ | $-1.9-0.1i$ | 0 | $-24.0$ | 13.2 | 0.9 | 0 | $-16.8$ | $-1.0$ | $-16.7$ | 13.2 | 0 |
| $\langle -\tfrac{1}{2},-\tfrac{1}{2}|$ | 0 | 0 | 0 | $5.7+0.1i$ | $3.0-0.1i$ | $5.7+0.1i$ | 13.2 | 0.9 | 13.2 | 0 | $-16.8$ | 0 | $9.0$ | 13.2 | 0.9 | 0 |
| $\langle -\tfrac{3}{2},\tfrac{1}{2}|$ | 0 | 0 | 0 | $-1.9-0.1i$ | $5.7+0.1i$ | $-2.5-0.3i$ | $-16.7$ | 13.2 | $-24.0$ | $-0.4$ | $-1.0$ | $9.0$ | 0 | $-16.7$ | 13.2 | 0 |
| $\langle -\tfrac{1}{2},-\tfrac{3}{2}|$ | 0 | 0 | 0 | 0 | 0 | 0 | 0 | $-16.7$ | 13.2 | $-24.0$ | $-0.4$ | $-16.8$ | 0 | 0 | 15.4 | 8.0 |
| $\langle -\tfrac{3}{2},-\tfrac{1}{2}|$ | 0 | 0 | 0 | 0 | 0 | 0 | 0 | 0 | $-16.7$ | $-24.0$ | 13.2 | 0.9 | 13.2 | 15.4 | 8.0 | 0 |
| $\langle -\tfrac{3}{2},-\tfrac{3}{2}|$ | 0 | 0 | 0 | 0 | 0 | 0 | $-2.0-0.3i$ | $8.8+0.3i$ | $-2.0-0.3i$ | 0 | 0 | $-4.5+0.1i$ | $7.3+0.1i$ | 0 | 0 | 58.2 |

Table S10 Complex 1: Matrix elements, in cm$^{-1}$, of $\hat{H}^{MS}=J\hat{S}_a\hat{S}_b+\hat{S}_a\bar{\bar{D}}_a\hat{S}_a+\hat{S}_b\bar{\bar{D}}_b\hat{S}_b$ expressed in the $|S,M_S\rangle$ coupled-spin basis, obtained with the spin-free NEVPT2 $J=11.2$ cm$^{-1}$ and the SO-NEVPT2 ZFS parameters $D_a=D_b=17.4$ cm$^{-1}$ and $E_a=E_b=3.7$ cm$^{-1}$.

| $\hat{H}^{MS}$ | $\|3,3\rangle$ | $\|3,2\rangle$ | $\|3,1\rangle$ | $\|3,0\rangle$ | $\|3,-1\rangle$ | $\|3,-2\rangle$ | $\|3,-3\rangle$ | $\|2,2\rangle$ | $\|2,1\rangle$ | $\|2,0\rangle$ | $\|2,-1\rangle$ | $\|2,-2\rangle$ | $\|1,1\rangle$ | $\|1,0\rangle$ | $\|1,-1\rangle$ | $\|0,0\rangle$ |
|---|---|---|---|---|---|---|---|---|---|---|---|---|---|---|---|---|
| $\langle 3,3\|$ | 59.8 | −0.1 | 5.7 | 0 | 0 | 0 | 0 | 0 | 0 | 0 | 0 | 0 | 7.0 | 0 | 0 | 0 |
| $\langle 3,2\|$ | −0.1 | 25.1 | −0.1 | 8.1 | 0 | 0 | 0 | 0 | 0 | 0 | 0 | 0 | 0.1 | 4.0 | 0 | 0 |
| $\langle 3,1\|$ | 5.7 | −0.1 | 4.3 | 0 | 8.9 | 0 | 0 | 0 | 0 | 0 | 0 | 0 | 17.0 | 0 | 1.8 | 0 |
| $\langle 3,0\|$ | 0 | 8.1 | 0 | −2.7 | 0 | 8.1 | 0 | 0 | 0 | 0 | 0 | 0 | 0 | 20.8 | 0 | 0 |
| $\langle 3,-1\|$ | 0 | 0 | 8.9 | 0 | 4.3 | 0.1 | 5.7 | 0 | 0 | 0 | 0 | 0 | 1.8 | 0 | 17.0 | 0 |
| $\langle 3,-2\|$ | 0 | 0 | 0 | 8.1 | 0.1 | 25.1 | 0.1 | 0 | 0 | 0 | 0 | 0 | 0 | 4.0 | −0.1 | 0 |
| $\langle 3,-3\|$ | 0 | 0 | 0 | 0 | 5.7 | 0.1 | 59.8 | 0 | 0 | 0 | 0 | 0 | 0 | 0 | 7.0 | 0 |
| $\langle 2,2\|$ | 0 | 0 | 0 | 0 | 0 | 0 | 0 | −8.4 | 0 | 0 | 0 | 0 | 0 | 0 | 0 | 9.1 |
| $\langle 2,1\|$ | 0 | 0 | 0 | 0 | 0 | 0 | 0 | 0 | −8.4 | 0 | 0 | 0 | 0 | 0 | 0 | 0.1 |
| $\langle 2,0\|$ | 0 | 0 | 0 | 0 | 0 | 0 | 0 | 0 | 0 | −8.4 | 0 | 0 | 0 | 0 | 0 | 34.7 |
| $\langle 2,-1\|$ | 0 | 0 | 0 | 0 | 0 | 0 | 0 | 0 | 0 | 0 | −8.4 | 0 | 0 | 0 | 0 | −0.1 |
| $\langle 2,-2\|$ | 0 | 0 | 0 | 0 | 0 | 0 | 0 | 0 | 0 | 0 | 0 | −8.4 | 0 | 0 | 0 | 9.1 |
| $\langle 1,1\|$ | 7.0 | 0.1 | 17.0 | 0 | 1.8 | 0 | 0 | 0 | 0 | 0 | 0 | 0 | −44.6 | 0.1 | −8.9 | 0 |
| $\langle 1,0\|$ | 0 | 4.0 | 0 | 20.8 | 0 | 4.0 | 0 | 0 | 0 | 0 | 0 | 0 | 0.1 | −2.9 | −0.1 | 0 |
| $\langle 1,-1\|$ | 0 | 0 | 1.8 | 0 | 17.0 | −0.1 | 7.0 | 0 | 0 | 0 | 0 | 0 | −8.9 | −0.1 | −44.6 | 0 |
| $\langle 0,0\|$ | 0 | 0 | 0 | 0 | 0 | 0 | 0 | 9.1 | 0.1 | 34.7 | −0.1 | 9.1 | 0 | 0 | 0 | −41.9 |

Table S11 Complex 1: Numerical matrix elements (cm$^{-1}$) of $\hat{H}^{MS}=J\hat{S}_a\hat{S}_b+\hat{S}_a\bar{\bar{D}}_a\hat{S}_a+\hat{S}_b\bar{\bar{D}}_b\hat{S}_b$ expressed in the $|M_{Sa},M_{Sb}\rangle$ uncoupled-spin basis, obtained with the spin-free NEVPT2 $J=11.2$ cm$^{-1}$ and the SO-NEVPT2 ZFS parameters $D_a=D_b=17.4$ cm$^{-1}$ and $E_a=E_b=3.7$ cm$^{-1}$.

| $\hat{H}^{MS}$ | $\|\tfrac{3}{2},\tfrac{3}{2}\rangle$ | $\|\tfrac{3}{2},\tfrac{1}{2}\rangle$ | $\|\tfrac{1}{2},\tfrac{3}{2}\rangle$ | $\|\tfrac{3}{2},-\tfrac{1}{2}\rangle$ | $\|\tfrac{1}{2},\tfrac{1}{2}\rangle$ | $\|-\tfrac{1}{2},\tfrac{3}{2}\rangle$ | $\|\tfrac{3}{2},-\tfrac{3}{2}\rangle$ | $\|\tfrac{1}{2},-\tfrac{1}{2}\rangle$ | $\|-\tfrac{1}{2},\tfrac{1}{2}\rangle$ | $\|-\tfrac{3}{2},\tfrac{3}{2}\rangle$ | $\|\tfrac{1}{2},-\tfrac{3}{2}\rangle$ | $\|-\tfrac{1}{2},-\tfrac{1}{2}\rangle$ | $\|-\tfrac{3}{2},\tfrac{1}{2}\rangle$ | $\|-\tfrac{1}{2},-\tfrac{3}{2}\rangle$ | $\|-\tfrac{3}{2},-\tfrac{1}{2}\rangle$ | $\|-\tfrac{3}{2},-\tfrac{3}{2}\rangle$ |
|---|---|---|---|---|---|---|---|---|---|---|---|---|---|---|---|---|
| $\langle\tfrac{3}{2},\tfrac{3}{2}\|$ | 59.8 | −0.1 | −0.1 | −0.1 | 6.4 | 0 | 0 | 0 | 0 | 0 | 0 | 0 | 0 | 0 | 0 | 0 |
| $\langle\tfrac{3}{2},\tfrac{1}{2}\|$ | −0.1 | 8.4 | 16.7 | 0 | 0 | 6.4 | 0 | 0.1 | 0 | 0 | 0 | 0 | 0 | 0 | 0 | 0 |
| $\langle\tfrac{1}{2},\tfrac{3}{2}\|$ | −0.1 | 16.7 | 8.4 | 0 | 0 | −0.1 | 0 | 0 | 6.4 | 0 | 0 | 0 | 0 | 0 | 0 | 0 |
| $\langle\tfrac{3}{2},-\tfrac{1}{2}\|$ | −0.1 | 0 | 0 | −8.4 | 19.3 | 0 | 0.1 | −0.1 | 0 | 0 | 0 | 0 | 6.4 | 0 | 0 | 0 |
| $\langle\tfrac{1}{2},\tfrac{1}{2}\|$ | 6.4 | 0 | 0 | 19.3 | −31.9 | 19.3 | 0 | 0 | 0 | 0 | 6.4 | 0 | 0 | 6.4 | 0 | 0 |
| $\langle-\tfrac{1}{2},\tfrac{3}{2}\|$ | 0 | 6.4 | −0.1 | 0 | 19.3 | −8.4 | 0 | 0 | −0.1 | 0.1 | 0 | 0 | 0 | 0 | 6.4 | 0 |
| $\langle\tfrac{3}{2},-\tfrac{3}{2}\|$ | 0 | 0 | 0 | 0.1 | 0 | 0 | 9.6 | 16.7 | 0 | 0 | −0.1 | 0 | 0 | 0 | 0 | 0 |
| $\langle\tfrac{1}{2},-\tfrac{1}{2}\|$ | 0 | 0.1 | 0 | −0.1 | 0 | 0 | 16.7 | −37.5 | 22.3 | 0 | 0 | 0.1 | 0 | 0 | 0 | 0.1 |
| $\langle-\tfrac{1}{2},\tfrac{1}{2}\|$ | 0 | 0 | 6.4 | 0 | 0 | −0.1 | 0 | 22.3 | −37.5 | 16.7 | 0 | 0 | −0.1 | 0.1 | 0 | 0 |
| $\langle-\tfrac{3}{2},\tfrac{3}{2}\|$ | 0 | 0 | 0 | 0 | 0 | 0.1 | 0 | 0 | 16.7 | 9.6 | 0 | 0 | 0 | 0 | −0.1 | 0 |
| $\langle\tfrac{1}{2},-\tfrac{3}{2}\|$ | 0 | 0 | 0 | 0 | 6.4 | 0 | −0.1 | 0 | 0 | 0 | −8.4 | 19.3 | 0 | 0 | 0 | 6.4 |
| $\langle-\tfrac{1}{2},-\tfrac{1}{2}\|$ | 0 | 0 | 0 | 0 | 0 | 6.4 | 0 | 0.1 | 0 | 0 | 19.3 | −31.9 | 19.3 | 0 | 0.1 | 0 |
| $\langle-\tfrac{3}{2},\tfrac{1}{2}\|$ | 0 | 0 | 0 | 6.4 | 0 | 0 | 0 | 0 | −0.1 | 0 | 0 | 19.3 | −8.4 | 0 | 0 | 6.4 |
| $\langle-\tfrac{1}{2},-\tfrac{3}{2}\|$ | 0 | 0 | 0 | 0 | 6.4 | 0 | 0 | 0 | 0.1 | 0 | 0 | 0 | 0 | 8.4 | 16.7 | 0.1 |
| $\langle-\tfrac{3}{2},-\tfrac{1}{2}\|$ | 0 | 0 | 0 | 0 | 0 | 6.4 | 0 | 0 | 0 | −0.1 | 0 | 0.1 | 0 | 16.7 | 8.4 | 0.1 |
| $\langle-\tfrac{3}{2},-\tfrac{3}{2}\|$ | 0 | 0 | 0 | 0 | 0 | 0 | 0 | 0.1 | 0 | 0 | 6.4 | 0 | 6.4 | 0.1 | 0.1 | 59.8 |

Table S12 Complex 1: Revised matrix elements, in cm$^{-1}$, of $\hat{H}^{\text{eff}}$ expressed in the $|S, M_S\rangle$ coupled-spin basis, obtained from a SO-NEVPT2 calculation on complex 1 in the molecular PAF. It was generated from the original $\hat{H}^{\text{eff}}$ matrix of Table S8 by multiplying by a factor of $-1$ the elements of the $\langle 0, 0|$ $\hat{H}^{\text{eff}}$ $|2, M_S\rangle$ and $\langle 1, M_S|$ $\hat{H}^{\text{eff}}$ $|3, M_S\rangle$ blocks.

| $\hat{H}^{\text{MS}}$ | $\|3,3\rangle$ | $\|3,2\rangle$ | $\|3,1\rangle$ | $\|3,0\rangle$ | $\|3,-1\rangle$ | $\|3,-2\rangle$ | $\|3,-3\rangle$ | $\|2,2\rangle$ | $\|2,1\rangle$ | $\|2,0\rangle$ | $\|2,-1\rangle$ | $\|2,-2\rangle$ | $\|1,1\rangle$ | $\|1,0\rangle$ | $\|1,-1\rangle$ | $\|0,0\rangle$ |
|---|---|---|---|---|---|---|---|---|---|---|---|---|---|---|---|---|
| $\langle 3,3\|$ | 58.2 | 0 | 5.1 | 0 | 0 | 0 | 0 | 0 | 0 | 0 | 0 | 0 | 7.8−0.6i | 0 | 0 | 0 |
| $\langle 3,2\|$ | 0 | 23.3 | 0 | 7.2 | 0 | 0 | 0 | 0 | 0 | 0 | 0 | 0 | 0 | 4.5−0.3i | 0 | 0 |
| $\langle 3,1\|$ | 5.1 | 0 | 2.5 | 0 | 7.9 | 0 | 0 | 0 | 0 | 0 | 0 | 0 | 16.7 | 0 | 2.0−0.1i | 0 |
| $\langle 3,0\|$ | 0 | 7.2 | 0 | −4.5 | 0 | 7.2 | 0 | 0 | 0 | 0 | 0 | 0 | 0 | 20.4 | 0 | 0 |
| $\langle 3,-1\|$ | 0 | 0 | 7.9 | 0 | 2.5 | 0 | 5.1 | 0 | 0 | 0 | 0 | 0 | 2.0+0.1i | 0 | 16.7 | 0 |
| $\langle 3,-2\|$ | 0 | 0 | 0 | 7.2 | 0 | 23.3 | 0 | 0 | 0 | 0 | 0 | 0 | 0 | 4.5+0.3i | 0 | 0 |
| $\langle 3,-3\|$ | 0 | 0 | 0 | 0 | 5.1 | 0 | 58.2 | 0 | 0 | 0 | 0 | 0 | 0 | 0 | 7.8+0.6i | 0 |
| $\langle 2,2\|$ | 0 | 0 | 0 | 0 | 0 | 0 | 0 | −7.4 | −0.1i | 0.4−0.2i | 0 | 0 | 0 | 0 | 0 | 9.6−0.5i |
| $\langle 2,1\|$ | 0 | 0 | 0 | 0 | 0 | 0 | 0 | 0.1i | −7.3 | 0 | 0.5−0.3i | 0 | 0 | 0 | 0 | 0.1−0.1i |
| $\langle 2,0\|$ | 0 | 0 | 0 | 0 | 0 | 0 | 0 | 0.4+0.2i | 0 | −7.3 | 0 | 0.4−0.2i | 0 | 0 | 0 | 33.7 |
| $\langle 2,-1\|$ | 0 | 0 | 0 | 0 | 0 | 0 | 0 | 0 | 0.5+0.3i | 0 | −7.3 | 0.1i | 0 | 0 | 0 | −0.1−0.1i |
| $\langle 2,-2\|$ | 0 | 0 | 0 | 0 | 0 | 0 | 0 | 0 | 0 | −0.4−0.2i | −0.1i | −7.4 | 0 | 0 | 0 | 9.6+0.5i |
| $\langle 1,1\|$ | 7.8+0.6i | 0 | 16.7 | 0 | 2.0−0.1i | 0 | 0 | 0 | 0 | 0 | 0 | 0 | −42.4 | 0.1 | −9.3+0.4i | 0 |
| $\langle 1,0\|$ | 0 | 4.5+0.3i | 0 | 20.4 | 0 | 4.5−0.3i | 0 | 0 | 0 | 0 | 0 | 0 | 0.1 | −1.8 | −0.1 | 0 |
| $\langle 1,-1\|$ | 0 | 0 | 2.0+0.1i | 0 | 16.7 | 0 | 7.8−0.6i | 0 | 0 | 0 | 0 | 0 | −9.3−0.4i | −0.1 | −42.4 | 0 |
| $\langle 0,0\|$ | 0 | 0 | 0 | 0 | 0 | 0 | 0 | 9.6+0.5i | 0.1+0.1i | 33.7 | −0.1+0.1i | 9.6−0.5i | 0 | 0 | 0 | −40.2 |

Table S13 Complex 1: Matrix elements, in cm$^{-1}$, of $\hat{H}^{\text{eff}}$ expressed in the $|M_{S_a}, M_{S_b}\rangle$ uncoupled-spin basis, obtained by basis transformation of the sign-revised, coupled-spin basis $\hat{H}^{\text{eff}}$ of Table S12.

| $\hat{H}^{\text{MS}}$ | $\|\tfrac{3}{2},\tfrac{3}{2}\rangle$ | $\|\tfrac{3}{2},\tfrac{1}{2}\rangle$ | $\|\tfrac{1}{2},\tfrac{3}{2}\rangle$ | $\|\tfrac{3}{2},-\tfrac{1}{2}\rangle$ | $\|\tfrac{1}{2},\tfrac{1}{2}\rangle$ | $\|-\tfrac{1}{2},\tfrac{3}{2}\rangle$ | $\|\tfrac{3}{2},-\tfrac{3}{2}\rangle$ | $\|\tfrac{1}{2},-\tfrac{1}{2}\rangle$ | $\|-\tfrac{1}{2},\tfrac{1}{2}\rangle$ | $\|-\tfrac{3}{2},\tfrac{3}{2}\rangle$ | $\|\tfrac{1}{2},-\tfrac{3}{2}\rangle$ | $\|-\tfrac{1}{2},-\tfrac{1}{2}\rangle$ | $\|-\tfrac{3}{2},\tfrac{1}{2}\rangle$ | $\|-\tfrac{1}{2},-\tfrac{3}{2}\rangle$ | $\|-\tfrac{3}{2},-\tfrac{1}{2}\rangle$ | $\|-\tfrac{3}{2},-\tfrac{3}{2}\rangle$ |
|---|---|---|---|---|---|---|---|---|---|---|---|---|---|---|---|---|
| $\langle \tfrac{3}{2},\tfrac{3}{2}\|$ | 58.2 | 0 | 0 | 0 | 0 | 0 | 0 | 0 | 0 | 0 | 0 | 0 | 0 | 0 | 0 | 0 |
| $\langle \tfrac{3}{2},\tfrac{1}{2}\|$ | 0 | 8.0 | 15.4 | 0 | 0 | 0 | 0 | 0 | 0 | 0 | 0 | 0 | 0 | 0 | 0 | 0 |
| $\langle \tfrac{1}{2},\tfrac{3}{2}\|$ | 0 | 15.4 | 8.0 | 0 | 0 | 0 | 0 | 0 | 0 | 0 | 0 | 0 | 0 | 0 | 0 | 0 |
| $\langle \tfrac{3}{2},-\tfrac{1}{2}\|$ | 6.5+0.3i | 0 | 0 | −7.7 | 17.9 | −0.4 | 0 | 0 | 0 | 0 | 0.1i | 6.2+0.2i | −0.5−0.2i | 0 | 0 | 0 |
| $\langle \tfrac{1}{2},\tfrac{1}{2}\|$ | −1.0−0.4i | 0 | 0 | 17.9 | −31.8 | 17.9 | 0 | 0 | 0 | 0 | 6.2+0.2i | −0.9−0.3i | 6.2+0.2i | 0 | 0 | 0 |
| $\langle -\tfrac{1}{2},\tfrac{3}{2}\|$ | 6.5+0.3i | 0 | 0 | −0.4 | 17.9 | −7.7 | 0 | 0 | 0 | 0 | −0.5−0.2i | 6.2+0.2i | 0.1i | 0 | 0 | 0 |
| $\langle \tfrac{3}{2},-\tfrac{3}{2}\|$ | 0 | 6.8+0.4i | −0.3−0.1i | 0 | 0 | 0 | 0 | 10.1 | 16.0 | −0.4 | 0 | 0 | 0 | 6.8−0.4i | −0.3+0.1i | 0 |
| $\langle \tfrac{1}{2},-\tfrac{1}{2}\|$ | 0 | −0.5−0.2i | 5.9 | 0 | 0 | 0 | 0 | 16.0 | −37.0 | 20.5 | 0 | 0 | 0 | −0.5+0.2i | 5.9 | 0 |
| $\langle -\tfrac{1}{2},\tfrac{1}{2}\|$ | 0 | 5.9 | −0.5−0.2i | 0 | 0 | 0 | 0 | 20.5 | −37.0 | 16.0 | 0 | 0 | 0 | 5.9 | −0.5+0.2i | 0 |
| $\langle -\tfrac{3}{2},\tfrac{3}{2}\|$ | 0 | −0.3−0.1i | 6.8+0.4i | 0 | 0 | 0 | 0 | −0.4 | 16.0 | 10.1 | 0 | 0 | 0 | −0.3+0.1i | 6.8−0.4i | 0 |
| $\langle \tfrac{1}{2},-\tfrac{3}{2}\|$ | 0 | 0 | 0 | 0.1i | 6.2+0.2i | −0.5−0.2i | 0 | 0 | 0 | 0 | −7.7 | 17.9 | −0.4 | 0 | 0 | 6.5−0.3i |
| $\langle -\tfrac{1}{2},-\tfrac{1}{2}\|$ | 0 | 0 | 0 | 6.2+0.2i | −0.9−0.3i | 6.2+0.2i | 0 | 0 | 0 | 0 | 17.9 | −31.8 | 17.9 | 0 | 0 | −1.0+0.4i |
| $\langle -\tfrac{3}{2},\tfrac{1}{2}\|$ | 0 | 0 | 0 | −0.5−0.2i | 6.2+0.2i | −0.1i | 0 | 0 | 0 | 0 | −0.4 | 17.9 | −7.7 | 0 | 0 | 6.5−0.3i |
| $\langle -\tfrac{1}{2},-\tfrac{3}{2}\|$ | 0 | 0 | 0 | 0 | 0 | 0 | 6.8+0.4i | −0.5+0.2i | 5.9 | −0.3+0.1i | 0 | 0 | 0 | 8.0 | 15.4 | 0 |
| $\langle -\tfrac{3}{2},-\tfrac{1}{2}\|$ | 0 | 0 | 0 | 0 | 0 | 0 | −0.3+0.1i | 5.9 | −0.5+0.2i | 6.8−0.4i | 0 | 0 | 0 | 15.4 | 8.0 | 0 |
| $\langle -\tfrac{3}{2},-\tfrac{3}{2}\|$ | 0 | 0 | 0 | 0 | 0 | 0 | 0 | 0 | 0 | 0 | 6.5+0.3i | −1.0−0.4i | 6.5+0.3i | 0 | 0 | 58.2 |

Table S14 Complex 1: Matrix elements of $\hat{H}^{\text{MS}} - \hat{H}^{\text{eff}}$ (cm$^{-1}$) expressed in the $|M_{Sa}, M_{Sb}\rangle$ uncoupled-spin basis. $\hat{H}^{\text{MS}}$ and the sign-revised $\hat{H}^{\text{eff}}$ are taken from Tables S11 and S13 respectively.

| Diff. | $\|\frac{3}{2},\frac{3}{2}\rangle$ | $\|\frac{3}{2},\frac{1}{2}\rangle$ | $\|\frac{1}{2},\frac{3}{2}\rangle$ | $\|\frac{3}{2},-\frac{1}{2}\rangle$ | $\|\frac{1}{2},\frac{1}{2}\rangle$ | $\|-\frac{1}{2},\frac{3}{2}\rangle$ | $\|\frac{3}{2},-\frac{3}{2}\rangle$ | $\|\frac{1}{2},-\frac{1}{2}\rangle$ | $\|-\frac{1}{2},\frac{1}{2}\rangle$ | $\|-\frac{3}{2},\frac{3}{2}\rangle$ | $\|\frac{1}{2},-\frac{3}{2}\rangle$ | $\|-\frac{1}{2},-\frac{1}{2}\rangle$ | $\|-\frac{3}{2},\frac{1}{2}\rangle$ | $\|-\frac{1}{2},-\frac{3}{2}\rangle$ | $\|-\frac{3}{2},-\frac{1}{2}\rangle$ | $\|-\frac{3}{2},-\frac{3}{2}\rangle$ |
|---|---|---|---|---|---|---|---|---|---|---|---|---|---|---|---|---|
| $\langle\frac{3}{2},\frac{3}{2}\|$ | 1.7 | −0.1 | −0.1 | −0.1+0.3$i$ | 1.0−0.4$i$ | −0.1+0.3$i$ | 0 | 0 | −0.4+0.4$i$ | 0.5−0.2$i$ | 0 | 0 | 0 | 0 | 0 | 0 |
| $\langle\frac{3}{2},\frac{1}{2}\|$ | −0.1 | 0.4 | 1.4 | 0 | 1.4 | 0.4 | −0.4+0.4$i$ | 0.5−0.2$i$ | 0 | 0 | 0.1$i$ | 0.2+0.1$i$ | 0.5−0.2$i$ | 0 | 0 | 0 |
| $\langle\frac{1}{2},\frac{3}{2}\|$ | −0.1 | 1.4 | 0.4 | 0 | −0.1 | −0.7 | 0.3−0.1$i$ | −0.4+0.4$i$ | 0 | 0 | 0.2+0.1$i$ | 0.9−0.3$i$ | 0.2+0.1$i$ | 0 | 0 | 0 |
| $\langle\frac{3}{2},-\frac{1}{2}\|$ | −0.1−0.3$i$ | 0 | 0 | −0.7 | 1.4 | 0.4 | 0 | 0 | 0.2+0.1$i$ | 0.9−0.3$i$ | 0.5−0.2$i$ | 0.2+0.1$i$ | 0.1$i$ | 0 | 0 | 0 |
| $\langle\frac{1}{2},\frac{1}{2}\|$ | 1.0+0.4$i$ | 0 | 0 | 1.4 | −0.1 | 1.4 | 0 | −0.1$i$ | 0.8 | −0.5 | 0 | 0 | 0 | 0 | 0 | 0 |
| $\langle-\frac{1}{2},\frac{3}{2}\|$ | −0.1−0.3$i$ | 0 | 0 | 0.4 | 1.4 | −0.7 | −0.1$i$ | 0.4 | 1.8 | 0.8 | 0 | 0 | 0 | 0 | 0 | 0 |
| $\langle\frac{3}{2},-\frac{3}{2}\|$ | 0 | −0.4−0.4$i$ | 0.3+0.1$i$ | 0 | 0 | −0.1$i$ | −0.5 | 0.8 | −0.5 | 0.4 | −0.1 | 0.4 | 0 | −0.4+0.4$i$ | 0.3−0.1$i$ | 0 |
| $\langle\frac{1}{2},-\frac{1}{2}\|$ | 0 | 0.5+0.2$i$ | 0.5 | 0 | 0.1$i$ | 0.4 | 0.8 | −0.5 | 1.8 | 0.8 | 0.4 | 1.8 | 0 | 0.5−0.2$i$ | 0.5 | 0 |
| $\langle-\frac{1}{2},\frac{1}{2}\|$ | 0 | 0.5 | 0.5+0.2$i$ | 0 | 0.8 | 1.8 | 0.4 | 1.8 | −0.5 | 0.4 | 0.8 | −0.5 | 0 | 0.5 | 0.5−0.2$i$ | 0 |
| $\langle-\frac{3}{2},\frac{3}{2}\|$ | 0 | 0.3+0.1$i$ | −0.4−0.4$i$ | 0 | −0.5 | 0.8 | −0.1 | 0.4 | 1.8 | −0.7 | 0.4 | 1.4 | 0 | 0.3−0.1$i$ | −0.4+0.4$i$ | 0 |
| $\langle\frac{1}{2},-\frac{3}{2}\|$ | 0 | 0 | 0 | −0.1$i$ | 0.2−0.1$i$ | 0.5+0.2$i$ | −0.7 | 0.4 | 1.4 | 0.4 | −0.1 | 0.4 | −0.1 | 0 | 0 | −0.1+0.3$i$ |
| $\langle-\frac{1}{2},-\frac{1}{2}\|$ | 0 | 0 | 0 | 0.2−0.1$i$ | 0.9+0.3$i$ | 0.2−0.1$i$ | 1.4 | 1.8 | −0.5 | 1.4 | 1.4 | −0.1 | 0.4 | 0 | 0 | 1.0−0.4$i$ |
| $\langle-\frac{3}{2},\frac{1}{2}\|$ | 0 | 0 | 0 | 0.5+0.2$i$ | 0.2−0.1$i$ | 0.1$i$ | −0.7 | 0 | 0 | 0.1$i$ | −0.1 | 1.4 | −0.7 | 0 | 0 | −0.1+0.3$i$ |
| $\langle-\frac{1}{2},-\frac{3}{2}\|$ | 0 | 0 | 0 | 0 | 0 | 0 | 0 | −0.1−0.2$i$ | 0.8+0.3$i$ | 0.2−0.2$i$ | 0 | 0 | 0 | −0.1 | 0 | 1.4 |
| $\langle-\frac{3}{2},-\frac{1}{2}\|$ | 0 | 0 | 0 | 0 | 0 | 0 | 0 | 0.2−0.2$i$ | 0.8+0.3$i$ | −0.1−0.2$i$ | 0 | 0 | 0 | 0 | 0 | 0.1 |
| $\langle-\frac{3}{2},-\frac{3}{2}\|$ | 0 | 0 | 0 | 0 | 0 | 0 | 0 | 0 | 0 | 0 | −0.1−0.3$i$ | 1.0+0.4$i$ | −0.1−0.3$i$ | 0.1 | 0.1 | 1.7 |

Table S15 Complex 1: Matrix elements of $\hat{H}^{\text{MS}} - \hat{H}^{\text{eff}}$ (cm$^{-1}$) expressed in the $|S, M_S\rangle$ coupled-spin basis. The $\hat{H}^{\text{MS}}$ and $\hat{H}^{\text{eff}}$ matrices are taken from Tables S10 and S12 respectively.

| Diff. | $\|3,3\rangle$ | $\|3,2\rangle$ | $\|3,1\rangle$ | $\|3,0\rangle$ | $\|3,-1\rangle$ | $\|3,-2\rangle$ | $\|3,-3\rangle$ | $\|2,2\rangle$ | $\|2,1\rangle$ | $\|2,0\rangle$ | $\|2,-1\rangle$ | $\|2,-2\rangle$ | $\|1,1\rangle$ | $\|1,0\rangle$ | $\|1,-1\rangle$ | $\|0,0\rangle$ |
|---|---|---|---|---|---|---|---|---|---|---|---|---|---|---|---|---|
| $\langle 3,3\|$ | 1.7 | −0.1 | 0.7 | 0 | 0 | 0 | 0 | 0 | 0 | 0 | 0 | 0 | −0.8+0.5$i$ | 0 | 0 | 0 |
| $\langle 3,2\|$ | −0.1 | 1.8 | −0.1 | 0.9 | 0 | 0 | 0 | 0 | 0 | 0 | 0 | 0 | 0 | −0.4+0.3$i$ | 0 | 0 |
| $\langle 3,1\|$ | 0.7 | −0.1 | 1.8 | 0 | 1.0 | 0 | 0 | 0 | 0 | 0 | 0 | 0 | 0.3 | 0 | −0.2+0.1$i$ | 0 |
| $\langle 3,0\|$ | 0 | 0.9 | 0 | 1.8 | 0 | 0.9 | 0 | 0 | 0 | 0 | 0 | 0 | 0 | 0.4 | 0 | 0 |
| $\langle 3,-1\|$ | 0 | 0 | 1.0 | 0 | 1.8 | 0.1 | 0.7 | 0 | 0 | 0 | 0 | 0 | −0.2−0.1$i$ | 0 | 0.3 | 0 |
| $\langle 3,-2\|$ | 0 | 0 | 0 | 0.9 | 0.1 | 1.8 | 0.1 | 0 | 0 | 0 | 0 | 0 | 0 | −0.4−0.3$i$ | 0 | 0 |
| $\langle 3,-3\|$ | 0 | 0 | 0 | 0 | 0.7 | 0.1 | 1.7 | 0 | 0 | 0 | 0 | 0 | 0 | 0 | −0.8−0.5$i$ | 0 |
| $\langle 2,2\|$ | 0 | 0 | 0 | 0 | 0 | 0 | 0 | −1.0 | 0.1$i$−0.4+0.2$i$ | 0 | 0 | 0 | 0 | 0 | −0.5+0.5$i$ | 0 |
| $\langle 2,1\|$ | 0 | 0 | 0 | 0 | 0 | 0 | 0 | −0.1$i$ | −1.1 | −0.5+0.3$i$ | −1.1 | 0 | −0.1$i$ | 0 | 0 | 0.1$i$ |
| $\langle 2,0\|$ | 0 | 0 | 0 | 0 | 0 | 0 | 0 | −0.4−0.2$i$ | 0 | −1.1 | 0 | −0.4+0.2$i$ | 0 | 0 | 0 | 1.0 |
| $\langle 2,-1\|$ | 0 | 0 | 0 | 0 | 0 | 0 | 0 | 0 | −0.5−0.3$i$ | 0 | −1.1 | 0.1$i$ | 0 | 0 | 0 | 0.1$i$ |
| $\langle 2,-2\|$ | 0 | 0 | 0 | 0 | 0 | 0 | 0 | 0 | 0 | 0.4+0.2$i$ | −0.1$i$ | −1.0 | 0 | 0 | 0 | −0.5−0.5$i$ |
| $\langle 1,1\|$ | −0.8−0.5$i$ | 0 | 0.3 | 0 | −0.2+0.1$i$ | 0 | 0 | 0 | 0.1$i$ | 0 | 0 | 0 | −2.2 | 0.1$i$ | 0.4−0.4$i$ | 0 |
| $\langle 1,0\|$ | 0 | −0.4−0.3$i$ | 0 | 0.4 | 0 | −0.4+0.3$i$ | 0 | 0 | 0 | 0 | 0 | 0 | −0.1$i$ | −1.1 | −0.1$i$ | 0 |
| $\langle 1,-1\|$ | 0 | 0 | −0.2−0.1$i$ | 0 | 0.3 | 0 | −0.8+0.5$i$ | 0 | 0 | 0 | 0 | 0.4+0.4$i$ | −2.2 | 0.1$i$ | −2.2 | 0 |
| $\langle 0,0\|$ | 0 | 0 | 0 | 0 | 0 | 0 | 0 | 0 | 0 | 1.0 | −0.1$i$ | 0 | 0 | 0 | 0 | −1.7 |

S10 | S1–S25

Table S16 Complex 1: Matrix elements of $\hat{H}^{MS} = J\hat{S}_a\hat{S}_b + \hat{S}_a\bar{\bar{D}}_a\hat{S}_a + \hat{S}_b\bar{\bar{D}}_b\hat{S}_b + \hat{S}_a\bar{\bar{D}}_{ab}\hat{S}_b$ expressed in the $|M_{Sa}, M_{Sb}\rangle$ uncoupled-spin basis.

| Diff. | $\|\tfrac{3}{2},\tfrac{3}{2}\rangle$ | $\|\tfrac{3}{2},\tfrac{1}{2}\rangle$ | $\|\tfrac{1}{2},\tfrac{3}{2}\rangle$ | $\|\tfrac{3}{2},-\tfrac{1}{2}\rangle$ | $\|\tfrac{1}{2},\tfrac{1}{2}\rangle$ | $\|-\tfrac{1}{2},\tfrac{3}{2}\rangle$ | $\|\tfrac{3}{2},-\tfrac{3}{2}\rangle$ | $\|\tfrac{1}{2},-\tfrac{1}{2}\rangle$ | $\|-\tfrac{1}{2},\tfrac{1}{2}\rangle$ | $\|-\tfrac{3}{2},\tfrac{3}{2}\rangle$ | $\|\tfrac{1}{2},-\tfrac{3}{2}\rangle$ | $\|-\tfrac{1}{2},-\tfrac{1}{2}\rangle$ | $\|-\tfrac{3}{2},\tfrac{1}{2}\rangle$ | $\|-\tfrac{1}{2},-\tfrac{3}{2}\rangle$ | $\|-\tfrac{3}{2},-\tfrac{1}{2}\rangle$ | $\|-\tfrac{3}{2},-\tfrac{3}{2}\rangle$ |
|---|---|---|---|---|---|---|---|---|---|---|---|---|---|---|---|---|
| $\langle\tfrac{3}{2},\tfrac{3}{2}\|$ | 58.2 | 0 | 0 | 0 | 6.4 | −0.5 | 6.4 | 0 | 0 | 0 | 0 | 0 | 0 | 0 | 0 | 0 |
| $\langle\tfrac{3}{2},\tfrac{1}{2}\|$ | 0 | 8.0 | 15.6 | 0 | 0 | 0 | 0 | 6.4 | −0.6 | 6.4 | 0 | 0 | 0 | 0 | 0 | 0 |
| $\langle\tfrac{1}{2},\tfrac{3}{2}\|$ | 0 | 15.6 | 8.0 | 0 | 0 | 0 | 0 | −0.6 | 6.4 | −0.6 | 0 | 0 | 0 | 0 | 0 | 0 |
| $\langle\tfrac{3}{2},-\tfrac{1}{2}\|$ | 0 | 0 | 0 | −8.0 | 18.0 | 0 | 6.4 | −0.7 | 6.4 | 0 | −0.5 | 6.4 | 0 | 0 | 0 | 0 |
| $\langle\tfrac{1}{2},\tfrac{1}{2}\|$ | 6.4 | 0 | 0 | 18.0 | −31.6 | 18.0 | 0 | 6.4 | −36.9 | 20.3 | 6.4 | −0.7 | 6.4 | 0 | 0 | 0 |
| $\langle-\tfrac{1}{2},\tfrac{3}{2}\|$ | −0.5 | 0 | 0 | 0 | 18.0 | −8.0 | 0 | 20.3 | −36.9 | 15.6 | 0 | 6.4 | −0.5 | 0 | 0 | 0 |
| $\langle\tfrac{3}{2},-\tfrac{3}{2}\|$ | 6.4 | 0 | 0 | 0 | 0 | 0 | 10.3 | 15.6 | 0 | 0 | 0 | 0 | 0 | 0 | 0 | 0 |
| $\langle\tfrac{1}{2},-\tfrac{1}{2}\|$ | 0 | 6.4 | −0.6 | 0 | 0 | 0 | 15.6 | −36.9 | 20.3 | 0 | 0 | 0 | 0 | 6.4 | 0 | 0 |
| $\langle-\tfrac{1}{2},\tfrac{1}{2}\|$ | 0 | −0.6 | 6.4 | 0 | 0 | 0 | 0 | 20.3 | −36.9 | 15.6 | 0 | 0 | 0 | −0.6 | 6.4 | 0 |
| $\langle-\tfrac{3}{2},\tfrac{3}{2}\|$ | 0 | 6.4 | −0.6 | 0 | 0 | 0 | 0 | 0 | 15.6 | 10.3 | 0 | 0 | 0 | 6.4 | −0.6 | 0 |
| $\langle\tfrac{1}{2},-\tfrac{3}{2}\|$ | 0 | 0 | 0 | −0.5 | 6.4 | 0 | 0 | 0 | 0 | 0 | −8.0 | 18.0 | 0 | 0 | 6.4 | 0 |
| $\langle-\tfrac{1}{2},-\tfrac{1}{2}\|$ | 0 | 0 | 0 | 6.4 | −0.7 | 6.4 | 0 | 0 | 0 | 0 | 18.0 | −31.6 | 18.0 | 0 | 0 | −0.5 |
| $\langle-\tfrac{3}{2},\tfrac{1}{2}\|$ | 0 | 0 | 0 | 0 | 6.4 | −0.5 | 0 | 0 | 0 | 0 | 0 | 18.0 | −8.0 | 0 | 0 | 6.4 |
| $\langle-\tfrac{1}{2},-\tfrac{3}{2}\|$ | 0 | 0 | 0 | 0 | 0 | 0 | 0 | 6.4 | −0.6 | 6.4 | 0 | 0 | 0 | 8.0 | 15.6 | 0 |
| $\langle-\tfrac{3}{2},-\tfrac{1}{2}\|$ | 0 | 0 | 0 | 0 | 0 | 0 | 0 | 0 | 6.4 | −0.6 | 0 | 0 | 0 | 15.6 | 8.0 | 0 |
| $\langle-\tfrac{3}{2},-\tfrac{3}{2}\|$ | 0 | 0 | 0 | 0 | 0 | 0 | 0 | 0 | 0 | 0 | 6.4 | −0.5 | 6.4 | 0 | 0 | 58.2 |

Table S17 Complex 1: Matrix elements of $\hat{H}^{MS} = J\hat{S}_a\hat{S}_b + \hat{S}_a\bar{\bar{D}}_a\hat{S}_a + \hat{S}_b\bar{\bar{D}}_b\hat{S}_b + \hat{S}_a\bar{\bar{D}}_{ab}\hat{S}_b$ expressed in the $|S, M_S\rangle$ coupled-spin basis.

| Diff. | $\|3,3\rangle$ | $\|3,2\rangle$ | $\|3,1\rangle$ | $\|3,0\rangle$ | $\|3,-1\rangle$ | $\|3,-2\rangle$ | $\|3,-3\rangle$ | $\|2,2\rangle$ | $\|2,1\rangle$ | $\|2,0\rangle$ | $\|2,-1\rangle$ | $\|2,-2\rangle$ | $\|1,1\rangle$ | $\|1,0\rangle$ | $\|1,-1\rangle$ | $\|0,0\rangle$ |
|---|---|---|---|---|---|---|---|---|---|---|---|---|---|---|---|---|
| $\langle 3,3\|$ | 58.2 | 0 | 5.3 | 0 | 0 | 0 | 0 | 0 | 0 | 0 | 0 | 0 | 7.3 | 0 | 0 | 0 |
| $\langle 3,2\|$ | 0 | 23.6 | 0 | 7.5 | 0 | 0 | 0 | 0 | 0 | 0 | 0 | 0 | 0 | 4.2 | 0 | 0 |
| $\langle 3,1\|$ | 5.3 | 0 | 2.8 | 0 | 8.2 | 0 | 0 | 0 | 0 | 0 | 0 | 0 | 16.7 | 0 | 1.9 | 0 |
| $\langle 3,0\|$ | 0 | 7.5 | 0 | −4.6 | 0 | 7.5 | 0 | 0 | 0 | 0 | 0 | 0 | 0 | 20.6 | 0 | 0 |
| $\langle 3,-1\|$ | 0 | 0 | 8.2 | 0 | 2.8 | 0 | 5.3 | 0 | 0 | 0 | 0 | 0 | 1.9 | 0 | 16.7 | 0 |
| $\langle 3,-2\|$ | 0 | 0 | 0 | 7.5 | 0 | 23.6 | 0 | 0 | 0 | 0 | 0 | 0 | 0 | 4.2 | 0 | 0 |
| $\langle 3,-3\|$ | 0 | 0 | 0 | 0 | 5.3 | 0 | 58.2 | 0 | 0 | 0 | 0 | 0 | 0 | 0 | 7.3 | 0 |
| $\langle 2,2\|$ | 0 | 0 | 0 | 0 | 0 | 0 | 0 | −7.6 | 0 | −0.4 | 0 | 0 | 0 | 0 | 0 | 9.4 |
| $\langle 2,1\|$ | 0 | 0 | 0 | 0 | 0 | 0 | 0 | 0 | −8.0 | 0 | −0.5 | 0 | 0 | 0 | 0 | 0 |
| $\langle 2,0\|$ | 0 | 0 | 0 | 0 | 0 | 0 | 0 | −0.4 | 0 | −7.9 | 0 | −0.4 | 0 | 0 | 0 | 33.8 |
| $\langle 2,-1\|$ | 0 | 0 | 0 | 0 | 0 | 0 | 0 | 0 | −0.5 | 0 | −8.0 | 0 | 0 | 0 | 0 | 0 |
| $\langle 2,-2\|$ | 0 | 0 | 0 | 0 | 0 | 0 | 0 | 0 | 0 | −0.4 | 0 | −7.6 | 0 | 0 | 0 | 9.4 |
| $\langle 1,1\|$ | 7.3 | 0 | 16.7 | 0 | 1.9 | 0 | 0 | 0 | 0 | 0 | 0 | 0 | −42.4 | 0 | −9.4 | 0 |
| $\langle 1,0\|$ | 0 | 4.2 | 0 | 20.6 | 0 | 4.2 | 0 | 0 | 0 | 0 | 0 | 0 | 0 | −1.7 | 0 | 0 |
| $\langle 1,-1\|$ | 0 | 0 | 1.9 | 0 | 16.7 | 0 | 7.3 | 0 | 0 | 0 | 0 | 0 | −9.4 | 0 | −42.4 | 0 |
| $\langle 0,0\|$ | 0 | 0 | 0 | 0 | 0 | 0 | 0 | 9.4 | 0 | 33.8 | 0 | 9.4 | 0 | 0 | 0 | −39.0 |

Table S18 Complex 1: Difference between the numerical matrix elements of $\hat{H}^{\text{MS}} = J\hat{S}_a\hat{S}_b + \hat{S}_a\bar{\bar{D}}_a\hat{S}_a + \hat{S}_b\bar{\bar{D}}_b\hat{S}_b + \hat{S}_a\bar{\bar{D}}_{ab}\hat{S}_b$ (Table S16) and $\hat{H}^{\text{eff}}$ (Table S13) expressed in the $|M_{S_a}, M_{S_b}\rangle$ uncoupled-spin basis.

| Diff. | $|\tfrac{3}{2},\tfrac{3}{2}\rangle$ | $|\tfrac{3}{2},\tfrac{1}{2}\rangle$ | $|\tfrac{1}{2},\tfrac{3}{2}\rangle$ | $|\tfrac{3}{2},-\tfrac{1}{2}\rangle$ | $|\tfrac{1}{2},\tfrac{1}{2}\rangle$ | $|-\tfrac{1}{2},\tfrac{3}{2}\rangle$ | $|\tfrac{3}{2},-\tfrac{3}{2}\rangle$ | $|\tfrac{1}{2},-\tfrac{1}{2}\rangle$ | $|-\tfrac{1}{2},\tfrac{1}{2}\rangle$ | $|-\tfrac{3}{2},\tfrac{3}{2}\rangle$ | $|\tfrac{1}{2},-\tfrac{3}{2}\rangle$ | $|-\tfrac{1}{2},-\tfrac{1}{2}\rangle$ | $|-\tfrac{3}{2},\tfrac{1}{2}\rangle$ | $|-\tfrac{1}{2},-\tfrac{3}{2}\rangle$ | $|-\tfrac{3}{2},-\tfrac{1}{2}\rangle$ | $|-\tfrac{3}{2},-\tfrac{3}{2}\rangle$ |
|---|---|---|---|---|---|---|---|---|---|---|---|---|---|---|---|---|
| $\langle\tfrac{3}{2},\tfrac{3}{2}|$ | 0 | 0 | 0 | 0 | $-0.1+0.3i$ | $0.5-0.4i$ | $-0.1+0.3i$ | 0 | 0 | 0 | 0 | 0 | 0 | 0 | 0 | 0 |
| $\langle\tfrac{3}{2},\tfrac{1}{2}|$ | 0 | 0 | 0.2 | 0 | 0 | 0 | 0 | $-0.4+0.4i$ | $0.3-0.1i$ | 0 | 0 | 0 | 0 | 0 | 0 | 0 |
| $\langle\tfrac{1}{2},\tfrac{3}{2}|$ | 0 | 0.2 | 0 | 0 | 0 | 0 | 0 | $0.3-0.1i$ | | | | | | | | |
| $\langle\tfrac{3}{2},-\tfrac{1}{2}|$ | $-0.1-0.3i$ | 0 | 0 | 0 | $-0.3$ | 0.4 | 0 | 0 | $0.4$ | $-0.1-0.2i$ | 0 | 0 | 0 | 0 | 0 | 0 |
| $\langle\tfrac{1}{2},\tfrac{1}{2}|$ | $0.5+0.4i$ | 0 | 0 | $-0.3$ | 0.1 | 0.2 | 0 | 0 | 0 | 0 | | | | | | |
| $\langle-\tfrac{1}{2},\tfrac{3}{2}|$ | $-0.1-0.3i$ | 0 | 0 | 0.4 | 0.1 | $-0.3$ | 0 | 0 | 0 | 0 | | | | | | |
| $\langle\tfrac{3}{2},-\tfrac{3}{2}|$ | 0 | $-0.4-0.4i$ | $0.3+0.1i$ | 0 | 0 | 0 | 0.2 | $-0.4$ | 0.4 | $-0.1$ | 0 | 0 | 0 | 0 | 0 | 0 |
| $\langle\tfrac{1}{2},-\tfrac{1}{2}|$ | 0 | $-0.1+0.2i$ | 0.4 | 0 | 0 | 0 | $-0.4$ | 0 | $-0.2$ | 0.4 | | | | | | |
| $\langle-\tfrac{1}{2},\tfrac{1}{2}|$ | 0 | 0.4 | $-0.1+0.2i$ | 0 | 0 | 0 | 0.4 | $-0.2$ | 0 | $-0.4$ | | | | | | |
| $\langle-\tfrac{3}{2},\tfrac{3}{2}|$ | 0 | $0.3+0.1i$ | $-0.4-0.4i$ | 0 | 0 | 0 | $-0.1$ | 0.4 | $-0.4$ | 0.2 | | | | | | |
| $\langle\tfrac{1}{2},-\tfrac{3}{2}|$ | 0 | 0 | 0 | $-0.5-0.1i$ | $0.1-0.2i$ | $0.5+0.2i$ | 0 | 0 | 0 | 0 | 0 | $-0.3$ | 0.4 | $-0.4+0.4i$ | $0.3-0.1i$ | 0 |
| $\langle-\tfrac{1}{2},-\tfrac{1}{2}|$ | 0 | 0 | 0 | $0.1-0.2i$ | $0.3+0.3i$ | $0.1-0.2i$ | 0 | 0 | 0 | 0 | $-0.3$ | 0.1 | 0.1 | $-0.1-0.2i$ | 0.4 | 0 |
| $\langle-\tfrac{3}{2},\tfrac{1}{2}|$ | 0 | 0 | 0 | $0.5+0.2i$ | $0.1-0.2i$ | $-0.5-0.1i$ | 0 | 0 | 0 | 0 | 0.4 | 0.1 | $-0.3$ | $0.4-0.1-0.2i$ | $-0.4+0.4i$ | 0 |
| $\langle-\tfrac{1}{2},-\tfrac{3}{2}|$ | 0 | 0 | 0 | 0 | $0.2+0.3i$ | $0.1-0.2i$ | $-0.1-0.3i$ | 0 | 0 | 0 | 0 | 0 | 0 | $0.3-0.1i$ | $-0.4+0.4i$ | $-0.1+0.3i$ |
| $\langle-\tfrac{3}{2},-\tfrac{1}{2}|$ | 0 | 0 | 0 | 0 | $0.1-0.2i$ | $0.2+0.3i$ | $-0.1-0.3i$ | 0 | 0 | 0 | 0 | 0 | 0.2 | 0 | 0.4 | $0.5-0.4i$ |
| $\langle-\tfrac{3}{2},-\tfrac{3}{2}|$ | 0 | 0 | 0 | 0 | 0 | 0 | 0 | $-0.1-0.3i$ | $0.5+0.4i$ | $-0.1-0.3i$ | 0 | 0 | 0 | 0 | 0 | $-0.1+0.3i$ |

Table S19 Complex 1: Difference between the numerical matrix elements of $\hat{H}^{\text{MS}} = J\hat{S}_a\hat{S}_b + \hat{S}_a\bar{\bar{D}}_a\hat{S}_a + \hat{S}_b\bar{\bar{D}}_b\hat{S}_b + \hat{S}_a\bar{\bar{D}}_{ab}\hat{S}_b$ (Table S17) and $\hat{H}^{\text{eff}}$ (Table S12) expressed in the $|S, M_S\rangle$ coupled-spin basis.

| Diff. | $|3,3\rangle$ | $|3,2\rangle$ | $|3,1\rangle$ | $|3,0\rangle$ | $|3,-1\rangle$ | $|3,-2\rangle$ | $|3,-3\rangle$ | $|2,2\rangle$ | $|2,1\rangle$ | $|2,0\rangle$ | $|2,-1\rangle$ | $|2,-2\rangle$ | $|1,1\rangle$ | $|1,0\rangle$ | $|1,-1\rangle$ | $|0,0\rangle$ |
|---|---|---|---|---|---|---|---|---|---|---|---|---|---|---|---|---|
| $\langle 3,3|$ | 0 | 0 | 0.2 | 0 | 0 | 0 | 0 | 0 | 0 | 0 | 0 | 0 | $-0.5+0.6i$ | 0 | 0 | 0 |
| $\langle 3,2|$ | 0 | 0.3 | 0 | 0.3 | 0 | 0 | 0 | 0.3 | $0.1i$ | 0 | 0 | 0 | 0 | $-0.2+0.3i$ | 0 | 0 |
| $\langle 3,1|$ | 0.2 | 0 | 0.3 | 0 | 0.3 | 0 | 0 | 0 | $-0.7$ | 0 | 0 | 0 | 0 | 0 | $-0.1+0.1i$ | 0 |
| $\langle 3,0|$ | 0 | 0.3 | 0 | $-0.1$ | 0 | 0.3 | 0 | 0 | $-1.0-0.3i$ | $-0.9+0.2i$ | $-0.7$ | 0 | 0 | 0.1 | 0 | 0 |
| $\langle 3,-1|$ | 0 | 0 | 0.3 | 0 | 0.3 | 0 | 0.2 | 0 | 0 | $-1.0+0.3i$ | 0 | $-0.1i$ | 0 | 0 | $-0.1-0.1i$ | 0 |
| $\langle 3,-2|$ | 0 | 0 | 0 | 0.3 | 0 | 0.3 | 0 | 0 | 0 | 0 | $-0.9+0.2i$ | $-0.2$ | 0 | $-0.2-0.3i$ | 0 | 0 |
| $\langle 3,-3|$ | 0 | 0 | 0 | 0 | 0.2 | 0 | 0 | 0 | 0 | 0 | 0 | 0 | 0 | 0 | $-0.5-0.6i$ | 0 |
| $\langle 2,2|$ | 0 | 0 | 0 | 0 | 0 | 0 | 0 | $-0.2$ | $0.1i$ | $-0.6$ | 0 | 0 | 0 | 0 | 0 | $-0.2+0.5i$ |
| $\langle 2,1|$ | 0 | 0 | 0 | 0 | 0 | 0 | 0 | $-0.1i$ | $-0.7$ | 0 | $-1.0+0.3i$ | 0 | $-0.1$ | 0 | 0 | $-0.1+0.1i$ |
| $\langle 2,0|$ | 0 | 0 | 0 | 0 | 0 | 0 | 0 | $-0.9-0.2i$ | 0 | 0 | 0 | $-0.9-0.2i$ | 0 | 0 | 0 | 0 |
| $\langle 2,-1|$ | 0 | 0 | 0 | 0 | 0 | 0 | 0 | 0 | $-1.0-0.3i$ | 0 | $-0.7$ | $-0.1i$ | 0 | 0 | $0.1+0.1i$ | 0 |
| $\langle 2,-2|$ | 0 | 0 | 0 | 0 | 0 | 0 | 0 | 0 | 0 | $0.2i$ | $0.1i$ | $-0.2$ | 0 | 0 | 0 | $-0.2-0.5i$ |
| $\langle 1,1|$ | $-0.5-0.6i$ | 0 | 0 | 0 | 0 | 0 | 0 | 0 | 0 | $-0.6$ | 0 | 0 | 0 | $-0.1$ | $-0.1-0.4i$ | 0 |
| $\langle 1,0|$ | 0 | $-0.2-0.3i$ | 0 | 0.1 | 0 | $-0.2+0.3i$ | 0 | 0 | 0 | $0.2i$ | 0 | 0 | $-0.1$ | 0 | 0.1 | 0 |
| $\langle 1,-1|$ | 0 | 0 | $-0.1-0.1i$ | 0 | 0 | 0 | $-0.5+0.6i$ | 0 | 0 | 0 | 0 | $-0.1+0.4i$ | $-0.1+0.4i$ | 0.1 | 0 | 0 |
| $\langle 0,0|$ | 0 | 0 | 0 | 0 | 0 | 0 | 0 | $-0.2-0.5i$ | $-0.1-0.1i$ | 0 | $0.1-0.1i$ | $-0.2+0.5i$ | 0 | 0 | 0 | 1.1 |

S12 | S1–S25

Table S20 Analytical matrix elements of $\hat{H}^{zee} = \mu_B \vec{B} \hat{g}_a \hat{S}_a$ expressed in the $|M_{Sa}, M_{Sb}\rangle$ uncoupled-spin basis.[a]

| $\mu_B \vec{B} \hat{g}_a \hat{S}_a$ | $\|\tfrac{3}{2},\tfrac{3}{2}\rangle$ | $\|\tfrac{3}{2},\tfrac{1}{2}\rangle$ | $\|\tfrac{1}{2},\tfrac{3}{2}\rangle$ | $\|\tfrac{3}{2},-\tfrac{1}{2}\rangle$ | $\|\tfrac{1}{2},\tfrac{1}{2}\rangle$ | $\|-\tfrac{1}{2},\tfrac{3}{2}\rangle$ | $\|\tfrac{3}{2},-\tfrac{3}{2}\rangle$ | $\|\tfrac{1}{2},-\tfrac{1}{2}\rangle$ | $\|-\tfrac{1}{2},\tfrac{1}{2}\rangle$ | $\|-\tfrac{3}{2},\tfrac{3}{2}\rangle$ | $\|\tfrac{1}{2},-\tfrac{3}{2}\rangle$ | $\|-\tfrac{1}{2},-\tfrac{1}{2}\rangle$ | $\|-\tfrac{3}{2},\tfrac{1}{2}\rangle$ | $\|-\tfrac{1}{2},-\tfrac{3}{2}\rangle$ | $\|-\tfrac{3}{2},-\tfrac{1}{2}\rangle$ | $\|-\tfrac{3}{2},-\tfrac{3}{2}\rangle$ |
|---|---|---|---|---|---|---|---|---|---|---|---|---|---|---|---|---|
| $\langle\tfrac{3}{2},\tfrac{3}{2}\|$ | $-X$ | 0 | $Y$ | 0 | 0 | 0 | 0 | 0 | 0 | 0 | 0 | 0 | 0 | 0 | 0 | 0 |
| $\langle\tfrac{3}{2},\tfrac{1}{2}\|$ | 0 | $-X$ | 0 | 0 | $Y$ | 0 | 0 | 0 | 0 | 0 | 0 | 0 | 0 | 0 | 0 | 0 |
| $\langle\tfrac{1}{2},\tfrac{3}{2}\|$ | $Y^*$ | 0 | $-Z$ | 0 | 0 | $T$ | 0 | 0 | 0 | 0 | 0 | 0 | 0 | 0 | 0 | 0 |
| $\langle\tfrac{3}{2},-\tfrac{1}{2}\|$ | 0 | 0 | 0 | $-X$ | 0 | 0 | 0 | $Y$ | 0 | 0 | 0 | 0 | 0 | 0 | 0 | 0 |
| $\langle\tfrac{1}{2},\tfrac{1}{2}\|$ | 0 | $Y^*$ | 0 | 0 | $-Z$ | 0 | 0 | 0 | $T$ | 0 | 0 | 0 | 0 | 0 | 0 | 0 |
| $\langle-\tfrac{1}{2},\tfrac{3}{2}\|$ | 0 | 0 | $T^*$ | 0 | 0 | $Z$ | 0 | 0 | 0 | $Y$ | 0 | 0 | 0 | 0 | 0 | 0 |
| $\langle\tfrac{3}{2},-\tfrac{3}{2}\|$ | 0 | 0 | 0 | 0 | 0 | 0 | $-X$ | 0 | 0 | 0 | $Y$ | 0 | 0 | 0 | 0 | 0 |
| $\langle\tfrac{1}{2},-\tfrac{1}{2}\|$ | 0 | 0 | 0 | $Y^*$ | 0 | 0 | 0 | $-Z$ | 0 | 0 | 0 | $T$ | 0 | 0 | 0 | 0 |
| $\langle-\tfrac{1}{2},\tfrac{1}{2}\|$ | 0 | 0 | 0 | 0 | $T^*$ | 0 | 0 | 0 | $Z$ | 0 | 0 | 0 | $Y$ | 0 | 0 | 0 |
| $\langle-\tfrac{3}{2},\tfrac{3}{2}\|$ | 0 | 0 | 0 | 0 | 0 | $Y^*$ | 0 | 0 | 0 | $X$ | 0 | 0 | 0 | 0 | 0 | 0 |
| $\langle\tfrac{1}{2},-\tfrac{3}{2}\|$ | 0 | 0 | 0 | 0 | 0 | 0 | $Y^*$ | 0 | 0 | 0 | $-Z$ | 0 | 0 | $T$ | 0 | 0 |
| $\langle-\tfrac{1}{2},-\tfrac{1}{2}\|$ | 0 | 0 | 0 | 0 | 0 | 0 | 0 | $T^*$ | 0 | 0 | 0 | $Z$ | 0 | 0 | $Y$ | 0 |
| $\langle-\tfrac{3}{2},\tfrac{1}{2}\|$ | 0 | 0 | 0 | 0 | 0 | 0 | 0 | 0 | $Y^*$ | 0 | 0 | 0 | $X$ | 0 | 0 | 0 |
| $\langle-\tfrac{1}{2},-\tfrac{3}{2}\|$ | 0 | 0 | 0 | 0 | 0 | 0 | 0 | 0 | 0 | 0 | $T^*$ | 0 | 0 | $Z$ | 0 | $Y$ |
| $\langle-\tfrac{3}{2},-\tfrac{1}{2}\|$ | 0 | 0 | 0 | 0 | 0 | 0 | 0 | 0 | 0 | 0 | 0 | $Y^*$ | 0 | 0 | $X$ | 0 |
| $\langle-\tfrac{3}{2},-\tfrac{3}{2}\|$ | 0 | 0 | 0 | 0 | 0 | 0 | 0 | 0 | 0 | 0 | 0 | 0 | 0 | $Y^*$ | 0 | $X$ |

[a] Legend:
$X = -\tfrac{3}{2} \mu_B \left(B_x g^a_{xz} + B_y g^a_{yz} + B_z g^a_{zz}\right);\ Y = \tfrac{\sqrt{3}}{2} \mu_B \left[\left(B_x g^a_{xx} + B_y g^a_{yx} + B_z g^a_{zx}\right) - i\left(B_x g^a_{xy} + B_y g^a_{yy} + B_z g^a_{zy}\right)\right]$
$Z = -\tfrac{1}{2} \mu_B \left(B_x g^a_{xz} + B_y g^a_{yz} + B_z g^a_{zz}\right);\ T = \mu_B \left[\left(B_x g^a_{xx} + B_y g^a_{yx} + B_z g^a_{zx}\right) - i\left(B_x g^a_{xy} + B_y g^a_{yy} + B_z g^a_{zy}\right)\right]$; the asterix stands for complex conjugate.

Table S21 Analytical matrix elements of $\hat{H}^{zee} = \mu_B \vec{B} \hat{g}_b \hat{S}_b$ expressed in the $|M_{Sa}, M_{Sb}\rangle$ uncoupled-spin basis.[a]

| $\mu_B \vec{B} \hat{g}_b \hat{S}_b$ | $\|\tfrac{3}{2},\tfrac{3}{2}\rangle$ | $\|\tfrac{3}{2},\tfrac{1}{2}\rangle$ | $\|\tfrac{1}{2},\tfrac{3}{2}\rangle$ | $\|\tfrac{3}{2},-\tfrac{1}{2}\rangle$ | $\|\tfrac{1}{2},\tfrac{1}{2}\rangle$ | $\|-\tfrac{1}{2},\tfrac{3}{2}\rangle$ | $\|\tfrac{3}{2},-\tfrac{3}{2}\rangle$ | $\|\tfrac{1}{2},-\tfrac{1}{2}\rangle$ | $\|-\tfrac{1}{2},\tfrac{1}{2}\rangle$ | $\|-\tfrac{3}{2},\tfrac{3}{2}\rangle$ | $\|\tfrac{1}{2},-\tfrac{3}{2}\rangle$ | $\|-\tfrac{1}{2},-\tfrac{1}{2}\rangle$ | $\|-\tfrac{3}{2},\tfrac{1}{2}\rangle$ | $\|-\tfrac{1}{2},-\tfrac{3}{2}\rangle$ | $\|-\tfrac{3}{2},-\tfrac{1}{2}\rangle$ | $\|-\tfrac{3}{2},-\tfrac{3}{2}\rangle$ |
|---|---|---|---|---|---|---|---|---|---|---|---|---|---|---|---|---|
| $\langle\tfrac{3}{2},\tfrac{3}{2}\|$ | $-X$ | $Y$ | 0 | 0 | 0 | 0 | 0 | 0 | 0 | 0 | 0 | 0 | 0 | 0 | 0 | 0 |
| $\langle\tfrac{3}{2},\tfrac{1}{2}\|$ | $Y^*$ | $-Z$ | 0 | $T$ | 0 | 0 | 0 | 0 | 0 | 0 | 0 | 0 | 0 | 0 | 0 | 0 |
| $\langle\tfrac{1}{2},\tfrac{3}{2}\|$ | 0 | 0 | $-X$ | 0 | $Y$ | 0 | 0 | 0 | 0 | 0 | 0 | 0 | 0 | 0 | 0 | 0 |
| $\langle\tfrac{3}{2},-\tfrac{1}{2}\|$ | 0 | $T^*$ | 0 | $Z$ | 0 | 0 | $Y$ | 0 | 0 | 0 | 0 | 0 | 0 | 0 | 0 | 0 |
| $\langle\tfrac{1}{2},\tfrac{1}{2}\|$ | 0 | 0 | $Y^*$ | 0 | $-Z$ | 0 | 0 | $T$ | 0 | 0 | 0 | 0 | 0 | 0 | 0 | 0 |
| $\langle-\tfrac{1}{2},\tfrac{3}{2}\|$ | 0 | 0 | 0 | 0 | 0 | $-X$ | 0 | 0 | $Y$ | 0 | 0 | 0 | 0 | 0 | 0 | 0 |
| $\langle\tfrac{3}{2},-\tfrac{3}{2}\|$ | 0 | 0 | 0 | $Y^*$ | 0 | 0 | $X$ | 0 | 0 | 0 | 0 | 0 | 0 | 0 | 0 | 0 |
| $\langle\tfrac{1}{2},-\tfrac{1}{2}\|$ | 0 | 0 | 0 | 0 | $T^*$ | 0 | 0 | $Z$ | 0 | 0 | $Y$ | 0 | 0 | 0 | 0 | 0 |
| $\langle-\tfrac{1}{2},\tfrac{1}{2}\|$ | 0 | 0 | 0 | 0 | 0 | $Y^*$ | 0 | 0 | $-Z$ | 0 | 0 | $T$ | 0 | 0 | 0 | 0 |
| $\langle-\tfrac{3}{2},\tfrac{3}{2}\|$ | 0 | 0 | 0 | 0 | 0 | 0 | 0 | 0 | 0 | $-X$ | 0 | 0 | $Y$ | 0 | 0 | 0 |
| $\langle\tfrac{1}{2},-\tfrac{3}{2}\|$ | 0 | 0 | 0 | 0 | 0 | 0 | $Y^*$ | 0 | 0 | $X$ | 0 | 0 | 0 | 0 | 0 | 0 |
| $\langle-\tfrac{1}{2},-\tfrac{1}{2}\|$ | 0 | 0 | 0 | 0 | 0 | 0 | 0 | $T^*$ | 0 | 0 | $Z$ | 0 | $Y$ | 0 | 0 | 0 |
| $\langle-\tfrac{3}{2},\tfrac{1}{2}\|$ | 0 | 0 | 0 | 0 | 0 | 0 | 0 | 0 | $Y^*$ | 0 | 0 | $-Z$ | 0 | $T$ | 0 | 0 |
| $\langle-\tfrac{1}{2},-\tfrac{3}{2}\|$ | 0 | 0 | 0 | 0 | 0 | 0 | 0 | 0 | 0 | 0 | $Y^*$ | 0 | $X$ | 0 | 0 | 0 |
| $\langle-\tfrac{3}{2},-\tfrac{1}{2}\|$ | 0 | 0 | 0 | 0 | 0 | 0 | 0 | 0 | 0 | 0 | 0 | $T^*$ | 0 | $Z$ | $Y$ | 0 |
| $\langle-\tfrac{3}{2},-\tfrac{3}{2}\|$ | 0 | 0 | 0 | 0 | 0 | 0 | 0 | 0 | 0 | 0 | 0 | 0 | 0 | $Y^*$ | 0 | $X$ |

[a] Legend:
$X = -\tfrac{3}{2} \mu_B \left(B_x g^b_{xz} + B_y g^b_{yz} + B_z g^b_{zz}\right);\ Y = \tfrac{\sqrt{3}}{2} \mu_B \left[\left(B_x g^b_{xx} + B_y g^b_{yx} + B_z g^b_{zx}\right) - i\left(B_x g^b_{xy} + B_y g^b_{yy} + B_z g^b_{zy}\right)\right]$
$Z = -\tfrac{1}{2} \mu_B \left(B_x g^b_{xz} + B_y g^b_{yz} + B_z g^b_{zz}\right);\ T = \mu_B \left[\left(B_x g^b_{xx} + B_y g^b_{yx} + B_z g^b_{zx}\right) - i\left(B_x g^b_{xy} + B_y g^b_{yy} + B_z g^b_{zy}\right)\right]$; the asterix stands for complex conjugate.

Table S22 SOCI energies, in cm$^{-1}$, corresponding to the 16 low-lying spin-orbit states of complex 2, calculated with various approaches.

| State idx | SO-CASSCF | SO-NEVPT2 | model//SO-NEVPT2 |
|---|---|---|---|
| $\psi_1$ | 0.0 | 0.0 | 0.0 |
| $\psi_2$ | 1.9 | 1.9 | 2.1 |
| $\psi_3$ | 3.9 | 3.6 | 3.9 |
| $\psi_4$ | 8.7 | 7.7 | 8.1 |
| $\psi_5$ | 214.6 | 177.0 | 178.7 |
| $\psi_6$ | 215.2 | 177.7 | 179.3 |
| $\psi_7$ | 218.9 | 180.6 | 182.2 |
| $\psi_8$ | 220.2 | 182.2 | 183.7 |
| $\psi_9$ | 238.0 | 206.4 | 206.6 |
| $\psi_{10}$ | 238.3 | 206.9 | 207.1 |
| $\psi_{11}$ | 241.4 | 209.6 | 210.4 |
| $\psi_{12}$ | 243.8 | 212.2 | 212.9 |
| $\psi_{13}$ | 448.8 | 379.7 | 382.3 |
| $\psi_{14}$ | 449.0 | 380.0 | 382.4 |
| $\psi_{15}$ | 457.8 | 388.4 | 390.5 |
| $\psi_{16}$ | 458.2 | 389.0 | 391.2 |

Table S23 Contribution of the $|S,M_S\rangle$ components of the low-lying $S = 3, 2, 1, 0$ states to the wavefunctions of the 16 low-energy spin-orbit states, obtained in SO-NEVPT2 calculations on complex 2 in an arbitrary xyz frame.$^a$

| SOCI state | $|3,3\rangle$ | $|3,2\rangle$ | $|3,1\rangle$ | $|3,0\rangle$ | $|3,-1\rangle$ | $|3,-2\rangle$ | $|3,-3\rangle$ | $|2,2\rangle$ | $|2,1\rangle$ | $|2,0\rangle$ | $|2,-1\rangle$ | $|2,-2\rangle$ | $|1,1\rangle$ | $|1,0\rangle$ | $|1,-1\rangle$ | $|0,0\rangle$ |
|---|---|---|---|---|---|---|---|---|---|---|---|---|---|---|---|---|
| $\psi_1$ | 0.1 | 0.04 | 0.04 | 0.07 | 0.04 | 0.04 | 0.1 | 0.01 | 0.02 | 0 | 0.02 | 0.01 | 0.02 | 0 | 0.02 | 0 |
| $\psi_2$ | 0.05 | 0.06 | 0.05 | 0.04 | 0.05 | 0.06 | 0.05 | 0.01 | 0 | 0.05 | 0 | 0.01 | 0.03 | 0.06 | 0.03 | 0 |
| $\psi_3$ | 0.07 | 0.02 | 0.02 | 0.01 | 0.02 | 0.02 | 0.07 | 0.02 | 0.01 | 0.02 | 0.01 | 0.02 | 0.07 | 0.04 | 0.07 | 0.02 |
| $\psi_4$ | 0 | 0 | 0.01 | 0.01 | 0.01 | 0 | 0 | 0.09 | 0.02 | 0.01 | 0.02 | 0.09 | 0.01 | 0.06 | 0.01 | 0.17 |
| $\psi_5$ | 0.02 | 0.15 | 0.02 | 0.05 | 0.02 | 0.15 | 0.02 | 0.02 | 0.02 | 0.12 | 0.02 | 0.02 | 0.01 | 0 | 0.01 | 0 |
| $\psi_6$ | 0.06 | 0.01 | 0.1 | 0.05 | 0.1 | 0.01 | 0.06 | 0.02 | 0.08 | 0 | 0.08 | 0.02 | 0 | 0.02 | 0 | 0.02 |
| $\psi_7$ | 0.04 | 0.04 | 0.02 | 0 | 0.02 | 0.04 | 0.04 | 0.02 | 0.08 | 0.04 | 0.08 | 0.02 | 0.09 | 0 | 0.09 | 0.01 |
| $\psi_8$ | 0.01 | 0.01 | 0.02 | 0.03 | 0.02 | 0.01 | 0.01 | 0.1 | 0.04 | 0.01 | 0.04 | 0.1 | 0.03 | 0.14 | 0.03 | 0.05 |
| $\psi_9$ | 0.06 | 0.09 | 0 | 0.06 | 0 | 0.09 | 0.06 | 0.03 | 0.05 | 0.13 | 0.05 | 0.03 | 0 | 0 | 0 | 0 |
| $\psi_{10}$ | 0.06 | 0.05 | 0.07 | 0.01 | 0.07 | 0.05 | 0.06 | 0.02 | 0.09 | 0.08 | 0.09 | 0.02 | 0 | 0.01 | 0 | 0 |
| $\psi_{11}$ | 0.04 | 0.02 | 0.03 | 0 | 0.03 | 0.02 | 0.04 | 0.03 | 0.11 | 0.04 | 0.11 | 0.03 | 0.05 | 0.01 | 0.05 | 0.05 |
| $\psi_{12}$ | 0.01 | 0 | 0.01 | 0.01 | 0.01 | 0 | 0.01 | 0.09 | 0.03 | 0.02 | 0.03 | 0.09 | 0.06 | 0.23 | 0.06 | 0.02 |
| $\psi_{13}$ | 0.05 | 0.02 | 0.27 | 0.02 | 0.27 | 0.02 | 0.05 | 0.03 | 0.02 | 0.01 | 0.02 | 0.03 | 0 | 0 | 0 | 0 |
| $\psi_{14}$ | 0.02 | 0.14 | 0.02 | 0.35 | 0.02 | 0.14 | 0.02 | 0.02 | 0.03 | 0.01 | 0.03 | 0.02 | 0 | 0 | 0 | 0 |
| $\psi_{15}$ | 0.02 | 0.01 | 0.01 | 0 | 0.01 | 0.01 | 0.02 | 0.11 | 0.02 | 0.09 | 0.02 | 0.11 | 0.04 | 0.05 | 0.04 | 0.26 |
| $\psi_{16}$ | 0.01 | 0.01 | 0.01 | 0.01 | 0.01 | 0.01 | 0.01 | 0.04 | 0.03 | 0.02 | 0.03 | 0.04 | 0.27 | 0 | 0.27 | 0.05 |

$^a$The corresponding state energies are printed in Table S22.

Table S24 Contribution of the $|S,M_S\rangle$ components of the low-lying $S = 3, 2, 1, 0$ states to the wavefunctions of the 16 low-energy spin-orbit states, obtained in SO-NEVPT2 calculations on complex 2 in the molecular PAF.

| SOCI state | $|3,3\rangle$ | $|3,2\rangle$ | $|3,1\rangle$ | $|3,0\rangle$ | $|3,-1\rangle$ | $|3,-2\rangle$ | $|3,-3\rangle$ | $|2,2\rangle$ | $|2,1\rangle$ | $|2,0\rangle$ | $|2,-1\rangle$ | $|2,-2\rangle$ | $|1,1\rangle$ | $|1,0\rangle$ | $|1,-1\rangle$ | $|0,0\rangle$ |
|---|---|---|---|---|---|---|---|---|---|---|---|---|---|---|---|---|
| $\psi_1$ | 0 | 0 | 0 | 0.42 | 0 | 0 | 0 | 0.01 | 0.02 | 0 | 0.02 | 0.01 | 0 | 0.04 | 0 | 0 |
| $\psi_2$ | 0 | 0 | 0.17 | 0 | 0.17 | 0 | 0 | 0.02 | 0.01 | 0.01 | 0.01 | 0.02 | 0.05 | 0 | 0.05 | 0 |
| $\psi_3$ | 0 | 0 | 0.12 | 0 | 0.12 | 0 | 0 | 0.02 | 0.01 | 0.01 | 0.01 | 0.02 | 0.09 | 0 | 0.09 | 0.01 |
| $\psi_4$ | 0 | 0.01 | 0 | 0 | 0 | 0.01 | 0 | 0 | 0 | 0.23 | 0 | 0 | 0.04 | 0 | 0.04 | 0.17 |
| $\psi_5$ | 0.01 | 0.19 | 0.01 | 0.01 | 0.01 | 0.19 | 0.01 | 0.09 | 0 | 0 | 0 | 0.09 | 0 | 0.01 | 0 | 0 |
| $\psi_6$ | 0.02 | 0.16 | 0.02 | 0 | 0.02 | 0.16 | 0.02 | 0.09 | 0.01 | 0 | 0.01 | 0.09 | 0.01 | 0 | 0.01 | 0.02 |
| $\psi_7$ | 0.01 | 0.01 | 0.07 | 0.01 | 0.07 | 0.01 | 0.01 | 0.01 | 0.11 | 0 | 0.11 | 0.01 | 0.08 | 0.04 | 0.08 | 0.01 |
| $\psi_8$ | 0.01 | 0 | 0.03 | 0 | 0.03 | 0 | 0.01 | 0.01 | 0.14 | 0 | 0.14 | 0.01 | 0.09 | 0 | 0.09 | 0.05 |
| $\psi_9$ | 0.02 | 0.12 | 0.03 | 0.02 | 0.03 | 0.12 | 0.02 | 0.14 | 0.01 | 0 | 0.01 | 0.14 | 0 | 0 | 0 | 0 |
| $\psi_{10}$ | 0.02 | 0.11 | 0.03 | 0.02 | 0.03 | 0.11 | 0.02 | 0.13 | 0.02 | 0.01 | 0.02 | 0.13 | 0.01 | 0 | 0.01 | 0 |
| $\psi_{11}$ | 0 | 0.02 | 0.07 | 0 | 0.07 | 0.02 | 0 | 0.04 | 0.11 | 0 | 0.11 | 0.04 | 0.05 | 0.02 | 0.05 | 0.05 |
| $\psi_{12}$ | 0 | 0 | 0.02 | 0 | 0.02 | 0 | 0 | 0 | 0.13 | 0 | 0.13 | 0 | 0.16 | 0.02 | 0.16 | 0.02 |
| $\psi_{13}$ | 0.35 | 0 | 0 | 0 | 0 | 0 | 0.35 | 0.03 | 0.02 | 0 | 0.02 | 0.03 | 0 | 0 | 0 | 0 |
| $\psi_{14}$ | 0.35 | 0 | 0 | 0 | 0 | 0 | 0.35 | 0.03 | 0.02 | 0 | 0.02 | 0.03 | 0 | 0 | 0 | 0 |
| $\psi_{15}$ | 0 | 0.02 | 0.01 | 0.01 | 0.01 | 0.02 | 0 | 0 | 0 | 0.34 | 0 | 0 | 0.01 | 0.1 | 0.01 | 0.26 |
| $\psi_{16}$ | 0 | 0 | 0 | 0.06 | 0 | 0 | 0 | 0.02 | 0.03 | 0.06 | 0.03 | 0.02 | 0.01 | 0.52 | 0.01 | 0.05 |

$^a$The corresponding state energies are printed in Table S22.

Table S25 Complex 2: Numerical matrix elements (cm$^{-1}$) of $\hat{H}^{MS} = J\hat{S}_a\hat{S}_b + \hat{S}_a\bar{\bar{D}}_a\hat{S}_a + \hat{S}_b\bar{\bar{D}}_b\hat{S}_b$ expressed in the $|M_{Sa}, M_{Sb}\rangle$ uncoupled-spin basis, obtained with the spin-free NEVPT2 $J = -2.85$ cm$^{-1}$ and the calculated local anisotropy tensors shown in Equation 10 of the main manuscript.

| $\hat{H}^{MS}$ | $\|\tfrac{3}{2},\tfrac{3}{2}\rangle$ | $\|\tfrac{3}{2},\tfrac{1}{2}\rangle$ | $\|\tfrac{1}{2},\tfrac{3}{2}\rangle$ | $\|\tfrac{3}{2},-\tfrac{1}{2}\rangle$ | $\|\tfrac{1}{2},\tfrac{1}{2}\rangle$ | $\|-\tfrac{1}{2},\tfrac{3}{2}\rangle$ | $\|\tfrac{3}{2},-\tfrac{3}{2}\rangle$ | $\|\tfrac{1}{2},-\tfrac{1}{2}\rangle$ | $\|-\tfrac{1}{2},\tfrac{1}{2}\rangle$ | $\|-\tfrac{3}{2},\tfrac{3}{2}\rangle$ | $\|\tfrac{1}{2},-\tfrac{3}{2}\rangle$ | $\|-\tfrac{1}{2},-\tfrac{1}{2}\rangle$ | $\|-\tfrac{3}{2},\tfrac{1}{2}\rangle$ | $\|-\tfrac{1}{2},-\tfrac{3}{2}\rangle$ | $\|-\tfrac{3}{2},-\tfrac{1}{2}\rangle$ | $\|-\tfrac{3}{2},-\tfrac{3}{2}\rangle$ |
|---|---|---|---|---|---|---|---|---|---|---|---|---|---|---|---|---|
| $\langle\tfrac{3}{2},\tfrac{3}{2}\|$ | 155.9 | $28.3 + 25.3i$ | $-27.5 - 24.7i$ | $45.1 + 10.8i$ | $0 - 11.9 - 12.5i$ | 0 | 0 | 0 | 0 | 0 | 0 | 0 | 0 | 0 | 0 | 0 |
| $\langle\tfrac{3}{2},\tfrac{1}{2}\|$ | $28.3 - 25.3i$ | $-6.8$ | $-4.3$ | $0 - 27.5 - 24.7i$ | $45.1 + 10.8i$ | $0 - 11.9 - 12.5i$ | 0 | 0 | 0 | 0 | 0 | 0 | 0 | 0 | 0 | 0 |
| $\langle\tfrac{1}{2},\tfrac{3}{2}\|$ | $-27.5 + 24.7i$ | $-4.3$ | $2.5$ | $28.3 + 25.3i$ | $45.1 + 10.8i$ | 0 | $0 - 11.9 - 12.5i$ | 0 | 0 | 0 | 0 | 0 | 0 | 0 | 0 | 0 |
| $\langle\tfrac{3}{2},-\tfrac{1}{2}\|$ | $45.1 - 10.8i$ | $0 - 27.5 + 24.7i$ | 0 | $-2.5$ | $-4.9$ | $0 - 28.3 - 25.3i$ | $27.5 + 24.7i$ | $0 - 11.9 - 12.5i$ | 0 | 0 | 0 | 0 | 0 | 0 | 0 | 0 |
| $\langle\tfrac{1}{2},\tfrac{1}{2}\|$ | $0 - 27.5 + 24.7i$ | $28.3 - 25.3i$ | $2.5$ | $-4.9$ | $-163.0$ | $-4.9$ | 0 | $-5.7$ | $28.3 + 25.3i$ | 0 | $45.1 + 10.8i$ | $0 - 11.9 - 12.5i$ | 0 | 0 | 0 | 0 |
| $\langle-\tfrac{1}{2},\tfrac{3}{2}\|$ | $0 - 11.9 + 12.5i$ | 0 | $0 - 28.3 - 25.3i$ | $-4.9$ | $-4.9$ | $6.8$ | 0 | $-161.6$ | $-5.7$ | 0 | $45.1 + 10.8i$ | 0 | $0 - 11.9 - 12.5i$ | 0 | 0 | 0 |
| $\langle\tfrac{3}{2},-\tfrac{3}{2}\|$ | 0 | $45.1 - 10.8i$ | $0 - 28.3 + 25.3i$ | 0 | 0 | 0 | 168.7 | $-4.3$ | 0 | 0 | $0 - 27.5 - 24.7i$ | 0 | $0 - 11.9 - 12.5i$ | 0 | 0 | 0 |
| $\langle\tfrac{1}{2},-\tfrac{1}{2}\|$ | 0 | $0$ | $45.1 - 10.8i - 27.5 + 24.7i$ | $0 - 28.3 + 25.3i$ | $-5.7$ | $-161.6$ | $-4.3$ | 0 | 0 | $28.3 - 25.3i$ | 0 | $27.5 + 24.7i$ | $0 - 11.9 - 12.5i$ | 0 | 0 | 0 |
| $\langle-\tfrac{1}{2},\tfrac{1}{2}\|$ | 0 | $0 - 11.9 - 12.5i$ | 0 | $27.5 - 24.7i$ | $28.3 + 25.3i$ | $-5.7$ | 0 | 0 | $-4.3$ | $168.7$ | 0 | $0 - 28.3 - 25.3i$ | 0 | $45.1 + 10.8i$ | 0 | 0 |
| $\langle-\tfrac{3}{2},\tfrac{3}{2}\|$ | 0 | 0 | $0 - 11.9 + 12.5i$ | 0 | 0 | $0 - 27.5 + 24.7i - 28.3 + 25.3i$ | 0 | 0 | $0 - 28.3 + 25.3i - 27.5 + 24.7i$ | 0 | 0 | 0 | $45.1 + 10.8i$ | 0 | 0 | 0 |
| $\langle\tfrac{1}{2},-\tfrac{3}{2}\|$ | 0 | 0 | 0 | $45.1 - 10.8i$ | 0 | 0 | $0 - 27.5 - 24.7i$ | $0 - 28.3 - 25.3i$ | 0 | 0 | $6.8$ | $-4.9$ | 0 | $0 - 11.9 - 12.5i$ | 0 | 0 |
| $\langle-\tfrac{1}{2},-\tfrac{1}{2}\|$ | 0 | 0 | 0 | 0 | $0 - 11.9 + 12.5i$ | 0 | $0 - 28.3 - 25.3i$ | 0 | 0 | 0 | $-4.9$ | $-163.0$ | $-4.9$ | $-4.9 - 28.3 - 25.3i$ | $27.5 + 24.7i$ | 0 |
| $\langle-\tfrac{3}{2},\tfrac{1}{2}\|$ | 0 | 0 | 0 | 0 | 0 | $0 - 11.9 + 12.5i$ | 0 | $27.5 - 24.7i$ | $28.3 - 25.3i$ | 0 | 0 | $-4.9$ | 0 | $-2.5$ | 0 | $45.1 + 10.8i$ |
| $\langle-\tfrac{1}{2},-\tfrac{3}{2}\|$ | 0 | 0 | 0 | 0 | 0 | 0 | 0 | $45.1 - 10.8i$ | $0 - 28.3 + 25.3i$ | 0 | $0 - 28.3 + 25.3i$ | $-4.9$ | 0 | $2.5$ | $-4.3$ | $27.5 + 24.7i$ |
| $\langle-\tfrac{3}{2},-\tfrac{1}{2}\|$ | 0 | 0 | 0 | 0 | 0 | 0 | 0 | 0 | $45.1 - 10.8i$ | $0 - 11.9 + 12.5i$ | 0 | $27.5 - 24.7i$ | 0 | $-4.3$ | $-6.8$ | $-28.3 - 25.3i$ |
| $\langle-\tfrac{3}{2},-\tfrac{3}{2}\|$ | 0 | 0 | 0 | 0 | 0 | 0 | 0 | 0 | 0 | 0 | $0 - 11.9 + 12.5i$ | 0 | $45.1 - 10.8i$ | $27.5 - 24.7i - 28.3 + 25.3i$ | | 155.9 |

Table S26 Complex 2: Numerical matrix elements (cm$^{-1}$) of $\hat{H}^{MS}=J\hat{S}_a\hat{S}_b+\hat{S}_a\bar{\bar{D}}_a\hat{S}_a+\hat{S}_b\bar{\bar{D}}_b\hat{S}_b$ expressed in the $|S,M_S\rangle$ coupled-spin basis, obtained with the spin-free NEVPT2 $J=-2.85$ cm$^{-1}$ and the the calculated local anisotropy tensors shown in Eq. 10 of the main manuscript.

| $\hat{H}^{MS}$ | $|3,3\rangle$ | $|3,2\rangle$ | $|3,1\rangle$ | $|3,0\rangle$ | $|3,-1\rangle$ | $|3,-2\rangle$ | $|3,-3\rangle$ | $|2,2\rangle$ | $|2,1\rangle$ | $|2,0\rangle$ | $|2,-1\rangle$ | $|2,-2\rangle$ | $|1,1\rangle$ | $|1,0\rangle$ | $|1,-1\rangle$ | $|0,0\rangle$ |
|---|---|---|---|---|---|---|---|---|---|---|---|---|---|---|---|---|
| $\langle 3,3|$ | 155.9 | 0.6+0.4i | 14.8−0.8i | 0 | 0 | 0 | 0 | 39.4+35.3i | 40.2+16.5i | 0 | 0 | 0 | 18.2−0.9i | 0 | 0 | 0 |
| $\langle 3,2|$ | 0.6−0.4i | −6.4 | 0.4+0.3i | 21.0−1.1i | 0 | 0 | 0 | −4.6 | 0 | 40.2+16.5i | 0 | 0 | −0.4−0.2i | 10.5−0.5i | 0 | 0 |
| $\langle 3,1|$ | 14.8+0.8i | 0.4−0.3i | −103.8 | 0.2+0.1i | 23.0−0.1i | 0 | 0 | 0−30.5+27.4i | −2.9−24.9−22.3i | 31.2+12.7i | 0 | 0 | 79.5 | −0.3−0.2i | 4.7−0.2i | 0 |
| $\langle 3,0|$ | 0 | 21.0+1.1i | 0.2−0.1i | −136.3 | −0.2−0.1i | 21.0−1.1i | 0 | −18.0+7.4i−35.2+31.6i | −35.2−31.6i | 0−24.9−22.3i | 2.9−30.5−31.6i | 18.0+7.4i | 0.2−0.1i | 97.4 | −0.2−0.1i | 0 |
| $\langle 3,-1|$ | 0 | 23.0+1.2i | −0.2−0.1i | −0.2+0.1i | −103.8 | −0.4−0.3i | 14.8−0.8i | 0−31.2+12.7i−24.9+22.3i | 14.8−0.8i | 0−40.2+16.5i | 2.9−30.5−27.4i | 18.0+7.4i | 4.7+0.2i | 0.3−0.2i | 79.5 | 0 |
| $\langle 3,-2|$ | 0 | 0 | 0 | 21.0+1.1i | −0.4+0.3i | −6.4 | −0.6−0.4i | 0 | 0−40.2+16.5i | 0 | 0 | 4.6 | 0 | 10.5+0.5i | 0.4−0.2i | 0 |
| $\langle 3,-3|$ | 0 | 0 | 0 | 0 | 14.8+0.8i | −0.6+0.4i | 155.9 | 0 | 0 | 0−40.2+16.5i | 39.4−35.3i | −35.3i | 0 | 0 | 18.2+0.9i | 0 |
| $\langle 2,2|$ | 39.4−35.3i | −4.6−30.5−27.4i | −18.0−7.4i | 0 | 0 | 0 | 0 | 2.1 | 0 | 0 | 0 | 0 | 24.9+22.3i | 36.0+14.7i | 0 23.5−1.2i | 0 |
| $\langle 2,1|$ | 40.2−16.5i | 0 | −2.9−35.2−24.9+22.3i−31.2−12.7i | 0 | 0 | 0 | 0 | 2.1 | 0 | 0 | 0 | −3.6−17.6−15.8i | 25.5+10.4i−0.6−0.4i | 0 |
| $\langle 2,0|$ | 40.2−16.5i−24.9+22.3i | 0−24.9−22.3i−40.2−16.5i | 0 | 0 | 0 | 0 | 2.1 | 0 | 0 | 0−30.5+27.4i | 0−30.5−27.4i | 162.3 |
| $\langle 2,-1|$ | 0 | 0 | 31.2−12.7i−35.2+31.6i | 2.9 | 0−40.2−16.5i | 0 | 0 | 0 | 2.1 | 0 | 0−25.5+10.4i−17.6+15.8i | 3.6 0.6−0.4i |
| $\langle 2,-2|$ | 0 | 0 | 0 | 18.0−7.4i−30.5+27.4i | 4.6 | 39.4+35.3i | 0 | 0 | 0 | 2.1 | 0−36.0+14.7i | 24.9−22.3i 23.5+1.2i |
| $\langle 1,1|$ | 18.2+0.9i | −0.4+0.2i | 0 | 0.2+0.1i | 4.7−0.2i | 0 | 0 | 24.9−22.3i | −3.6−30.5−27.4i−25.5−10.4i | 0 | 0−25.5+10.4i−17.6+15.8i | −57.1 | −0.4−0.3i−23.0+1.2i | 0 |
| $\langle 1,0|$ | 0 | 10.5+0.5i | −0.3+0.2i | 97.4 | 0.3+0.2i | 10.5−0.5i | 0 | 36.0−14.7i−17.6+15.8i | 0−17.6−15.8i−36.0−14.7i | −0.4+0.3i | −23.0−1.2i | 137.7 | 0.4+0.3i | 0 |
| $\langle 1,-1|$ | 0 | 0 | 4.7+0.2i | −0.2+0.1i | 79.5 | 0.4+0.2i | 18.2−0.9i | 0 | 25.5−10.4i−30.5+27.4i | 3.6 | 24.9+22.3i | 0 | 0.4−0.3i | −57.1 | 0 |
| $\langle 0,0|$ | 0 | 0 | 0 | 0 | 0 | 0 | 0 | 23.5+1.2i | −0.6+0.4i | 162.3 | 0.6+0.4i | 23.5−1.2i | 0 | 0 | 0 | 10.7 |

Table S27 Complex 2: Matrix elements, in cm$^{-1}$, of $\hat{\mathcal{H}}^{\text{eff}}$ expressed in the $|S, M_S\rangle$ coupled-spin basis, obtained from a SO-NEVPT2 calculation on complex 2 in the molecular PAF.

| $\hat{\mathcal{H}}^{\text{MS}}$ | $|3,3\rangle$ | $|3,2\rangle$ | $|3,1\rangle$ | $|3,0\rangle$ | $|3,-1\rangle$ | $|3,-2\rangle$ | $|3,-3\rangle$ | $|2,2\rangle$ | $|2,1\rangle$ | $|2,0\rangle$ | $|2,-1\rangle$ | $|2,-2\rangle$ | $|1,1\rangle$ | $|1,0\rangle$ | $|1,-1\rangle$ | $|0,0\rangle$ |
|---|---|---|---|---|---|---|---|---|---|---|---|---|---|---|---|---|
| $\langle 3,3|$ | 156.3 | 0.2−0.3i | 14.3+0.1i | 0.2i | 0 | 0 | 0 | 38.0+35.9i | 39.9+15.7i | 0.1 | −0.1i | 0 | 18.5+1.4i | −0.1i | −0.1i | 0.1 |
| $\langle 3,2|$ | 0.2+0.3i | −5.4 | 0.1−0.1i | 20.5−0.4i | 0 | 0 | 0 | −5.9+0.3i | 0.4 | 39.8+15.7i | −0.1 | 0 | 0.1i | 10.5−0.6i | 0 | 0 |
| $\langle 3,1|$ | 14.3−0.1i | 0.1+0.1i | −102.2 | 0 | 22.5−0.5i | −0.1i | 0 | 0−30.2+27.6i | −3.6−0.1i−34.9−32.0i | 30.9+12.2i | −0.1 | −0.1+0.1i | 10.5+0.6i | 0.2 |
| $\langle 3,0|$ | −0.2i | 20.5+0.4i | 0 | −134.4 | 0 | 20.5−0.4i | 0 | −17.9+7.1i−34.9+32.0i | −0.1−34.9−22.7i | −0.1−34.9−32.0i | 17.9+7.1i | 0 | 0.1i | 97.1 | −0.1 | 0 |
| $\langle 3,−1|$ | 0 | 0 | 22.5+0.5i | 0 | −102.2 | −0.1+0.1i | 14.3+0.1i | −30.9+12.2i−24.8+22.7i | −0.1−30.9−12.2i | 3.6+0.1i−30.2−27.6i | 0.4 | 79.3 | 0.1 | −0.2 |
| $\langle 3,−2|$ | 0 | 0 | 0.1i | 20.5+0.4i | −0.1−0.1i | −5.4 | −0.2+0.3i | 0 | −0.1−39.9+15.7i | −24.8+22.7i | 5.9+0.3i | 0.1i | 10.5+0.6i | 0 |
| $\langle 3,−3|$ | 0 | 0 | 0 | 0.2i | 14.3−0.1i | −0.2−0.3i | 156.3 | 0 | 0 | 0.1−39.9−15.7i | 0.2−0.3i | 38.0−35.9i | 0.1i | −0.1i | 18.5−1.4i | 0.1 |
| $\langle 2,2|$ | 38.0−35.9i | −5.9−0.3i−30.2−27.6i | −17.9−7.1i | 0 | 0 | 0 | 1.9 | 0.1 | 2.2 | 2.3 | −0.1i | 24.3−22.5i | 0.1+0.1i | 35.8+14.1i | 0 23.4−1.3i |
| $\langle 2,1|$ | 39.9−15.7i | 0.4 | −3.6−0.1i−34.9−32.0i | 0.1−24.8−22.7i | −0.1 | −39.9−15.7i | 0.1 | 2.2 | 0.2−0.3i | 0.2−0.2i | −0.1 | 35.8−14.1i−17.5−16.1i | −4.1−17.5−16.1i | 25.3+10.0i−0.8+0.2i |
| $\langle 2,0|$ | 0.1 39.8−15.7i | 0.4 −3.6+0.1i−34.9−32.0i | 0.1 39.8−22.7i | 0.1−24.8−22.7i | −39.9−15.7i | 0.1 | 0.2+0.3i | −0.1i | 2.3 | −0.2−0.1i | 0 | −17.5−16.1i−35.8−14.1i | 0 −30.3−27.8i | 161.5 |
| $\langle 2,−1|$ | 0.1i | −0.1 | 30.9−12.2i−34.9+32.0i | 0.1+27.8i | 3.6−0.1i | 0.4−39.9−15.7i | −0.1 | −0.1i | 0.2+0.3i | −0.2−0.1i | 2.2 | −0.1−25.3+10.0i−17.5+16.1i | 4.1 | 0.8+0.2i |
| $\langle 2,−2|$ | 0 | 0 | −0.1 | 17.9−7.1i−30.2+27.6i | 5.9−0.3i | 0 | 38.0+35.9i | 0.1−0.1i | 0.1i | −0.2−0.3i | −0.1 | 1.9 | 0−35.8+14.1i | 24.3−22.5i+23.4+1.3i |
| $\langle 1,1|$ | 18.5+1.4i | −0.1i | 0 | 0.1 | 4.7−0.2i | 0 | −0.1i | 24.3−22.5i | −4.1−16.1i−17.5+16.1i | 0 | −17.5−16.1i−35.8−14.1i | 0 | −58.4 | −0.8−0.2i | 0.8−0.2i | 0.1 |
| $\langle 1,0|$ | 0.1i 10.5+0.6i | −0.1−0.1i | 97.1 | 0.1−0.1i | 10.5+0.6i | 0.1i | 35.8−14.1i−17.5+16.1i | 25.3−10.0i−30.3−27.8i | 4.1 | 0 | −0.8−0.2i | 135.3 | 0.8+0.2i | 0 |
| $\langle 1,−1|$ | 0.1i | 0 | 4.7+0.2i | −0.1 | 79.3 | −0.1i | 18.5−1.4i | 0 25.3−10.0i−30.3+27.8i | 4.1 | 24.3+22.5i | −22.9−1.2i | −58.4 | −0.1 |
| $\langle 0,0|$ | 0.1 | 0 | 0 | 0.2 | 0 | 0 | 0.1 | 23.4+1.3i | 0.8−0.2i | 161.5 | 0.8−0.2i | 23.4−1.3i | −0.1 | 0 | −0.1 | 8.0 |

Table S28 Complex 2: Matrix elements, in cm$^{-1}$, of $\hat{\mathcal{H}}^{\text{eff}}$ (Table S27) expressed in the uncoupled-spin basis, $|M_{Sa}, M_{Sb}\rangle$.

| $\hat{\mathcal{H}}^{\text{MS}}$ | $\left|\tfrac{3}{2},\tfrac{3}{2}\right\rangle$ | $\left|\tfrac{3}{2},\tfrac{1}{2}\right\rangle$ | $\left|\tfrac{1}{2},\tfrac{3}{2}\right\rangle$ | $\left|\tfrac{1}{2},\tfrac{1}{2}\right\rangle$ | $\left|\tfrac{3}{2},-\tfrac{1}{2}\right\rangle$ | $\left|-\tfrac{1}{2},\tfrac{3}{2}\right\rangle$ | $\left|\tfrac{1}{2},-\tfrac{1}{2}\right\rangle$ | $\left|-\tfrac{1}{2},\tfrac{1}{2}\right\rangle$ | $\left|\tfrac{3}{2},-\tfrac{3}{2}\right\rangle$ | $\left|-\tfrac{3}{2},\tfrac{3}{2}\right\rangle$ | $\left|-\tfrac{1}{2},-\tfrac{1}{2}\right\rangle$ | $\left|\tfrac{1}{2},-\tfrac{3}{2}\right\rangle$ | $\left|-\tfrac{3}{2},\tfrac{1}{2}\right\rangle$ | $\left|-\tfrac{1}{2},-\tfrac{3}{2}\right\rangle$ | $\left|-\tfrac{3}{2},-\tfrac{1}{2}\right\rangle$ | $\left|-\tfrac{3}{2},-\tfrac{3}{2}\right\rangle$ |
|---|---|---|---|---|---|---|---|---|---|---|---|---|---|---|---|---|
| $\left\langle\tfrac{3}{2},\tfrac{3}{2}\right|$ | 156.3 | 27.0+25.2i | −26.7−25.6i | 44.8+10.4i | −0.6−0.9i−11.7−11.8i | 0.1 | −0.1 | 0 | 0 | 0 | 0 | 0 | 0 | 0 | 0 | 0 |
| $\left\langle\tfrac{3}{2},\tfrac{1}{2}\right|$ | 27.0−25.2i | −7.6 | −3.6−0.3i | 0.1−27.4−25.3i | 0.1−27.4−25.3i | −0.4 | −0.2−0.2i−28.3−25.2i | −0.1+0.2i | 0 | 0 | 0 | 0.1 | 0 | 0 | 0 |
| $\left\langle\tfrac{1}{2},\tfrac{3}{2}\right|$ | −26.7+25.6i | −3.6+0.3i | 4.1 | 27.5−25.0i | 0.3 | −3.5 | 44.5+10.7i−27.4−25.4i | −0.1+0.3i−11.7−12.1i | 0 | 0 | 0.1+0.1i | −0.1+0.1i | −0.1−0.1i | 0.1 |
| $\left\langle\tfrac{1}{2},\tfrac{1}{2}\right|$ | 44.8−10.4i | 0.1 | 0.3 | −3.5 | −4.1−0.1i | −4.1+0.1i | 44.5−10.7i−27.4−25.4i | −0.2+0.2i | 0 | 0 | −11.8−11.8i | −0.1−0.1i | 0.1+0.1i | 0 |
| $\left\langle\tfrac{3}{2},-\tfrac{1}{2}\right|$ | −0.6+0.9i−27.4−25.3i | 27.5+25.0i | −4.1−0.1i | −162.4 | −3.8−0.1i | 28.1+25.3i | −0.2−0.1i | −3.0 | 0.3 | 44.6+10.5i | −0.2+0.4i−11.8−11.8i | 0 | 0.4−11.7−11.8i |
| $\left\langle-\tfrac{1}{2},\tfrac{3}{2}\right|$ | −11.7+11.8i | −0.4 | −0.2 | −3.8+0.1i | 0.4 | −0.3+0.1i | −4.7+0.1i | −0.2 | 7.4 | 0 | −0.2−0.4i−11.8+11.8i | 0 | 0 0.1+0.1i | 0.2i | 0.1 |
| $\left\langle\tfrac{1}{2},-\tfrac{1}{2}\right|$ | 0.1 | 44.8−10.2i | −0.2−0.2i−28.4+25.2i | −27.3+25.4i | 28.1−25.3i | −0.2 | −162.4 | 0.4 | 0.1 | −3.8−0.1i | 0−27.0−25.5i | 0.1 | −4.1−0.1i−11.9−11.6i | −0.2−0.1i |
| $\left\langle-\tfrac{1}{2},\tfrac{1}{2}\right|$ | −0.1 | −0.2−0.2i+44.5−10.7i | 44.5+10.7i | −0.1+0.3i | −4.7−0.1i | 0 | 0.4 | −161.0 | −3.0 | −0.2−28.0−25.2i | 0−27.5−25.0i | 44.5+10.7i | −0.2+0.2i |
| $\left\langle\tfrac{3}{2},-\tfrac{3}{2}\right|$ | −0.2−11.9+11.6i | −0.1−0.3i−11.8+11.8i | 0 | −0.3+0.1i | 0.1 | 0.2 | −3.0 | 166.6 | −0.2 | −3.0 | 0.1 | 0 | −0.2+0.2i | 44.8+10.2i | −0.1 |
| $\left\langle-\tfrac{3}{2},\tfrac{3}{2}\right|$ | 0 | −11.9+11.6i | 44.5−10.7i | −4.7+0.1i | −161.0 | −3.0 | 166.6 | 0.2 | 7.4 | 0−27.5+25.0i | 0 | 44.6+10.2i | 0 |
| $\left\langle-\tfrac{1}{2},-\tfrac{1}{2}\right|$ | 0 | −0.2−11.7+12.0i | −0.1−0.3i | 0 | 0.2 | 27.0+25.5i | −0.2−0.2i−27.2+25.5i | −28.1+25.2i | −0.4 | 0.1 | 0.4 | −0.1−0.1i−27.5−25.0i | −0.1−27.5−25.0i | 27.4+25.3i | −0.6+0.9i |
| $\left\langle\tfrac{1}{2},-\tfrac{3}{2}\right|$ | 0.1i | 0 | 0 | 0 | 0 | 0 | 44.6+10.5i | −0.2−0.4i−10.5i | 0.1i | 0 | 0.1i | −25.4i | 28.3−25.2i | −0.3 | −0.1 | −0.2i |
| $\left\langle-\tfrac{3}{2},\tfrac{1}{2}\right|$ | 0 | −0.1−0.1i | 0.1+0.1i | 0 | −0.1 | −0.3+0.1i | 0 | 0.1 | 0.2 | −0.2 | 0 | −0.2 | −3.5 | −0.3 | −0.1 |
| $\left\langle-\tfrac{1}{2},-\tfrac{3}{2}\right|$ | 0 | 0 | 0.1i | 0.1i | 0 | 0.2i | 0.1 | 27.1+25.4i | −4.7+0.1i | 44.6+10.5i | −4.1+0.1i | −4.1+0.1i | −0.3 | 4.1 | −3.6−0.3i | −0.1 |
| $\left\langle-\tfrac{3}{2},-\tfrac{1}{2}\right|$ | 0 | 0 | 0 | 0.1 | 0 | 0 | 0.1+0.1i | −0.2−0.5i | 28.1+25.0i | −0.3 | −0.1 | 0.4 | −3.6+0.3i | −7.6−25.6i |
| $\left\langle-\tfrac{3}{2},-\tfrac{3}{2}\right|$ | 0 | 0 | 0 | 0 | 0 | 0 | 0.2i | 0.1i | 0.1i | 44.8−10.4i | −0.3−0.4i | 26.7−25.6i−27.0−25.2i | 26.7−25.6i−27.0−25.2i | 156.3 |

Table S29 Complex 2: Matrix elements of $\hat{H}^{\mathrm{MS}} - \hat{H}^{\mathrm{eff}}$ (cm$^{-1}$) expressed in the $|M_{Sa}, M_{Sb}\rangle$ uncoupled-spin basis. $\hat{H}^{\mathrm{MS}} = J\hat{S}_a\hat{S}_b + \hat{S}_a\bar{\bar{D}}_a\hat{S}_a + \hat{S}_b\bar{\bar{D}}_b\hat{S}_b$ and $\hat{H}^{\mathrm{eff}}$ are taken from Tables S25 and S28 respectively.

| Diff. | $\|\tfrac{3}{2},\tfrac{3}{2}\rangle$ | $\|\tfrac{3}{2},\tfrac{1}{2}\rangle$ | $\|\tfrac{1}{2},\tfrac{3}{2}\rangle$ | $\|\tfrac{3}{2},-\tfrac{1}{2}\rangle$ | $\|\tfrac{3}{2},-\tfrac{1}{2}\rangle$ | $\|-\tfrac{1}{2},\tfrac{3}{2}\rangle$ | $\|\tfrac{1}{2},\tfrac{1}{2}\rangle$ | $\|-\tfrac{1}{2},\tfrac{3}{2}\rangle$ | $\|\tfrac{3}{2},-\tfrac{3}{2}\rangle$ | $\|\tfrac{1}{2},-\tfrac{1}{2}\rangle$ | $\|-\tfrac{1}{2},\tfrac{1}{2}\rangle$ | $\|-\tfrac{3}{2},\tfrac{3}{2}\rangle$ | $\|\tfrac{1}{2},-\tfrac{3}{2}\rangle$ | $\|-\tfrac{1}{2},-\tfrac{1}{2}\rangle$ | $\|-\tfrac{3}{2},\tfrac{1}{2}\rangle$ | $\|-\tfrac{1}{2},-\tfrac{3}{2}\rangle$ | $\|-\tfrac{3}{2},-\tfrac{1}{2}\rangle$ | $\|-\tfrac{3}{2},-\tfrac{3}{2}\rangle$ |
|---|---|---|---|---|---|---|---|---|---|---|---|---|---|---|---|---|---|---|

(Full numerical matrix omitted here due to density; see source.)

Table S30 Complex 2: Matrix elements of $\hat{H}^{\mathrm{MS}} - \hat{H}^{\mathrm{eff}}$ (cm$^{-1}$) expressed in the $|S, M_S\rangle$ coupled-spin basis. $\hat{H}^{\mathrm{MS}} = J\hat{S}_a\hat{S}_b + \hat{S}_a\bar{\bar{D}}_a\hat{S}_a + \hat{S}_b\bar{\bar{D}}_b\hat{S}_b$ and $\hat{H}^{\mathrm{eff}}$ are taken from Tables S26 and S27 respectively.

Table S31 Analytical matrix elements of $\hat{S}_a \bar{\bar{D}}_{ab} \hat{S}_b$ for dicobalt(II) complexes expressed, in the molecular PAF, in the $|S, M_S\rangle$ coupled spin-basis spanned by the $S = 3, 2, 1, 0$ spin functions.

| $\langle \hat{S}_a \bar{\bar{D}}_{ab} \hat{S}_b \rangle$ | $|3,3\rangle$ | $|3,2\rangle$ | $|3,1\rangle$ | $|3,0\rangle$ | $|3,-1\rangle$ | $|3,-2\rangle$ | $|3,-3\rangle$ | $|2,2\rangle$ | $|2,1\rangle$ | $|2,0\rangle$ | $|2,-1\rangle$ | $|2,-2\rangle$ | $|1,1\rangle$ | $|1,0\rangle$ | $|1,-1\rangle$ | $|0,0\rangle$ |
|---|---|---|---|---|---|---|---|---|---|---|---|---|---|---|---|---|
| $\langle 3,3|$ | $\frac{3D_{ab}}{2}$ | 0 | $\frac{3\sqrt{15}E_{ab}}{10}$ | 0 | 0 | 0 | 0 | 0 | 0 | 0 | 0 | 0 | $-\frac{3\sqrt{10}E_{ab}}{10}$ | 0 | 0 | 0 |
| $\langle 3,2|$ | 0 | 0 | 0 | $\frac{3\sqrt{30}E_{ab}}{10}$ | 0 | 0 | 0 | 0 | 0 | 0 | 0 | 0 | 0 | $-\frac{\sqrt{30}E_{ab}}{10}$ | $\frac{\sqrt{6}E_{ab}}{10}$ | 0 |
| $\langle 3,1|$ | $\frac{3\sqrt{15}E_{ab}}{10}$ | 0 | $-\frac{9D_{ab}}{10}$ | 0 | $\frac{9E_{ab}}{5}$ | 0 | 0 | 0 | 0 | 0 | 0 | 0 | $-\frac{\sqrt{6}D_{ab}}{5}$ | 0 | 0 | 0 |
| $\langle 3,0|$ | 0 | $\frac{3\sqrt{30}E_{ab}}{10}$ | 0 | $-\frac{3D_{ab}}{5} + \frac{9E_{ab}}{5}$ | 0 | $\frac{3\sqrt{30}E_{ab}}{10}$ | 0 | 0 | 0 | 0 | 0 | 0 | 0 | $-\frac{4D_{ab}}{5} - \frac{3E_{ab}}{5}$ | 0 | 0 |
| $\langle 3,-1|$ | 0 | 0 | $\frac{9E_{ab}}{5}$ | 0 | $-\frac{9D_{ab}}{10}$ | 0 | $\frac{3\sqrt{15}E_{ab}}{10}$ | 0 | 0 | 0 | 0 | 0 | $-\frac{\sqrt{6}E_{ab}}{10}$ | $\frac{\sqrt{30}E_{ab}}{10}$ | $-\frac{\sqrt{6}D_{ab}}{5}$ | 0 |
| $\langle 3,-2|$ | 0 | 0 | 0 | $\frac{3\sqrt{30}E_{ab}}{10}$ | 0 | 0 | 0 | 0 | 0 | 0 | 0 | 0 | 0 | 0 | $-\frac{3\sqrt{10}E_{ab}}{10}$ | 0 |
| $\langle 3,-3|$ | 0 | 0 | 0 | 0 | $\frac{3\sqrt{15}E_{ab}}{10}$ | 0 | $\frac{3D_{ab}}{2}$ | 0 | 0 | 0 | 0 | 0 | 0 | 0 | 0 | 0 |
| $\langle 2,2|$ | 0 | 0 | 0 | 0 | 0 | 0 | 0 | $D_{ab}$ | $-\frac{D_{ab}}{2}$ | $\frac{\sqrt{6}E_{ab}}{2}$ | $\frac{3E_{ab}}{2}$ | 0 | 0 | 0 | 0 | $-\frac{\sqrt{6}E_{ab}}{2}$ |
| $\langle 2,1|$ | 0 | 0 | 0 | 0 | 0 | 0 | 0 | 0 | $\frac{3E_{ab}}{2}$ | 0 | 0 | $\frac{\sqrt{6}E_{ab}}{2}$ | 0 | 0 | 0 | 0 |
| $\langle 2,0|$ | 0 | 0 | 0 | 0 | 0 | 0 | 0 | $\frac{\sqrt{6}E_{ab}}{2}$ | 0 | $\frac{4D_{ab}}{3} - E_{ab}$ | 0 | $\frac{\sqrt{6}E_{ab}}{2}$ | 0 | 0 | 0 | $-\frac{2D_{ab}}{3} + E_{ab}$ |
| $\langle 2,-1|$ | 0 | 0 | 0 | 0 | 0 | 0 | 0 | 0 | 0 | 0 | $\frac{D_{ab}}{2}$ | 0 | 0 | 0 | 0 | 0 |
| $\langle 2,-2|$ | 0 | 0 | 0 | 0 | 0 | 0 | 0 | 0 | 0 | $\frac{\sqrt{6}E_{ab}}{2}$ | 0 | $D_{ab}$ | 0 | 0 | 0 | 0 |
| $\langle 1,1|$ | $-\frac{3\sqrt{10}E_{ab}}{10}$ | 0 | $-\frac{\sqrt{6}D_{ab}}{5}$ | 0 | $-\frac{\sqrt{6}E_{ab}}{10}$ | 0 | 0 | 0 | 0 | 0 | 0 | 0 | $\frac{17D_{ab}}{30}$ | 0 | $\frac{17E_{ab}}{10}$ | 0 |
| $\langle 1,0|$ | 0 | $-\frac{\sqrt{30}E_{ab}}{10}$ | 0 | $-\frac{4D_{ab}}{5} - \frac{3E_{ab}}{5}$ | 0 | $-\frac{\sqrt{30}E_{ab}}{10}$ | 0 | 0 | 0 | 0 | 0 | 0 | 0 | $-\frac{16D_{ab}}{15} + \frac{E_{ab}}{5}$ | 0 | 0 |
| $\langle 1,-1|$ | 0 | 0 | $-\frac{\sqrt{6}E_{ab}}{10}$ | 0 | $-\frac{\sqrt{6}D_{ab}}{5}$ | 0 | $-\frac{3\sqrt{10}E_{ab}}{10}$ | 0 | 0 | 0 | 0 | 0 | $\frac{17E_{ab}}{10}$ | 0 | $\frac{17D_{ab}}{30}$ | 0 |
| $\langle 0,0|$ | 0 | 0 | 0 | 0 | 0 | 0 | 0 | $-\frac{\sqrt{6}E_{ab}}{2}$ | 0 | $-\frac{2D_{ab}}{3} + E_{ab}$ | 0 | $-\frac{\sqrt{6}E_{ab}}{2}$ | 0 | 0 | 0 | $\frac{D_{ab}}{3} - E_{ab}$ |

Table S32 Complex 2: Numerical matrix elements (cm$^{-1}$) of $\hat{H}^{MS} = J\hat{S}_a\hat{S}_b + \left[\hat{S}_a\bar{\bar{D}}_{ab}\hat{S}_a + \hat{S}_b\bar{\bar{D}}_b\hat{S}_b\right]_{calc} + \hat{S}_a\bar{\bar{D}}_{ab}\hat{S}_b$, expressed in the $|S, M_S\rangle$ coupled-spin basis, obtained with calculated local anisotropies and $J$, $D_{ab}$, $E_{ab}$ listed in Equation 10 and Table 8, respectively, of the main manuscript.

| $\hat{H}^{MS}$ | $|3,3\rangle$ | $|3,2\rangle$ | $|3,1\rangle$ | $|3,0\rangle$ | $|3,-1\rangle$ | $|3,-2\rangle$ | $|3,-3\rangle$ | $|2,2\rangle$ | $|2,1\rangle$ | $|2,0\rangle$ | $|2,-1\rangle$ | $|2,-2\rangle$ | $|1,1\rangle$ | $|1,0\rangle$ | $|1,-1\rangle$ | $|0,0\rangle$ |
|---|---|---|---|---|---|---|---|---|---|---|---|---|---|---|---|---|
| $\langle 3,3|$ | 157.1 | 0.6+0.4i | 14.7−0.8i | 0 | 0 | 0 | 0 | 39.4+35.3i | 40.2+16.5i | 0 | 0 | 0 | 18.3−0.9i | 0 | 0 | 0 |
| $\langle 3,2|$ | 0.6−0.4i | −5.1 | 0.4+0.3i | 20.8−1.1i | 0 | 0 | 0 | −4.6 | 0 | 40.2+16.5i | 0 | 0 | −0.4−0.2i | 10.6−0.5i | 0 | 0 |
| $\langle 3,1|$ | 14.7+0.8i | 0.4−0.3i | −102.4 | 0.2+0.1i | 22.8−1.2i | 0 | 0 | −30.5+27.4i | −2.9−24.9−22.3i | 31.2+12.7i | 0 | 0 | 79.6 | −0.3−0.2i | 4.7−0.2i | 0 |
| $\langle 3,0|$ | 0 | 20.8+1.1i | 0.2−0.1i | −135.2 | −0.2−0.1i | 20.8−1.1i | 0 | −18.0+7.4i | −35.2−31.6i | 0 | −35.2−31.6i | 18.0+7.4i | 0.2−0.1i | 97.5 | −0.2−0.1i | 0 |
| $\langle 3,-1|$ | 0 | 0 | 22.8+1.2i | −0.2+0.1i | −102.4 | −0.2−0.3i | 14.7−0.8i | 0 | −31.2+12.7i−24.9+22.3i | 0 | 2.9−30.5−31.6i | 18.0+7.4i | 4.7+0.2i | 0.3−0.2i | 79.6 | 0 |
| $\langle 3,-2|$ | 0 | 0 | 0 | 20.8+1.1i | −0.4+0.3i | −5.1 | −0.6−0.4i | 0 | 0 | −40.2+16.5i | −0.2 | 4.6 | 0 | 10.6+0.5i | 0.4−0.2i | 0 |
| $\langle 3,-3|$ | 0 | 0 | 0 | 0 | 14.7+0.8i | −0.6+0.4i | 157.1 | 0 | 0 | 0 | 0 | −35.3i | 0 | 0 | 18.3+0.9i | 0 |
| $\langle 2,2|$ | 39.4−35.3i | −4.6−30.5−27.4i | −18.0−7.4i | 0 | 0 | 0 | 0 | 1.6 | 0 | −0.2 | 0 | 0 | 24.9−22.3i | 36.0+14.7i | 0 | 23.6−1.2i |
| $\langle 2,1|$ | 40.2−16.5i | 0 | −2.9−35.2−31.6i−31.2−12.7i | 0 | −24.9+22.3i | 0 | 0 | 0 | 1.7 | 0 | −0.2 | 0 | −3.6−17.6−15.8i | 25.5+10.4i | −0.6−0.4i |
| $\langle 2,0|$ | 0 | 40.2−16.5i−24.9+22.3i | 0 | 0 | −24.9−22.3i−40.2−16.5i | 0 | −0.2 | 0 | 1.9 | 0 | 0 | −30.5−27.4i | 0 | −30.5−27.4i | 162.2 |
| $\langle 2,-1|$ | 0 | 0 | 31.2−12.7i−35.2+31.6i | 0 | 2.9 | 0 | −40.2−16.5i | 0 | −0.2 | 0 | 1.7 | 0 | −25.5+10.4i−17.6+15.8i | 3.6 | 0.6−0.4i |
| $\langle 2,-2|$ | 0 | 0 | 0 | 18.0−7.4i−30.5+27.4i | 0 | 4.6 | 39.4−35.3i | 0 | 0 | −0.2 | 0 | 1.6 | 0 | −36.0+14.7i | 24.9−22.3i | 23.6+1.2i |
| $\langle 1,1|$ | 18.3+0.9i | −0.4+0.2i | 79.6 | 0.2+0.1i | 4.7−0.2i | 0 | 0 | 24.9−22.3i | −3.6−30.5+27.4i | 0 | −25.5−10.4i−17.6−15.8i | 0 | −58.7 | −0.4−0.3i | −23.2+1.2i | 0 |
| $\langle 1,0|$ | 0 | 10.6+0.5i | −0.3+0.2i | 97.5 | 0.3+0.2i | 10.6−0.5i | 0 | 36.0−14.7i−17.6+15.8i | 0 | 25.5−10.4i−30.5+27.4i | −36.0−14.7i | 136.2 | −0.4+0.3i |
| $\langle 1,-1|$ | 0 | 0 | 4.7+0.2i | −0.2+0.1i | 79.6 | 0.4+0.2i | 18.3−0.9i | 0 | 3.6 | 0 | 24.9+22.3i | −23.2−1.2i | 0.4+0.3i | −58.7 | 0 |
| $\langle 0,0|$ | 0 | 0 | 0 | 0 | 0 | 0 | 0 | 23.6+1.2i | −0.6+0.4i | 162.2 | 0.6+0.4i | 23.6−1.2i | 0 | 0 | 0 | 8.7 |

Table S33 Complex 2: Matrix elements of $\hat{H}^{MS} - \hat{H}^{eff}$ (cm$^{-1}$) expressed in the $|S, M_S\rangle$ coupled-spin basis. $\hat{H}^{MS}$ and $\hat{H}^{eff}$ matrices are taken from Tables S32 and S27 respectively.

| Diff. | $|3,3\rangle$ | $|3,2\rangle$ | $|3,1\rangle$ | $|3,0\rangle$ | $|3,-1\rangle$ | $|3,-2\rangle$ | $|3,-3\rangle$ | $|2,2\rangle$ | $|2,1\rangle$ | $|2,0\rangle$ | $|2,-1\rangle$ | $|2,-2\rangle$ | $|1,1\rangle$ | $|1,0\rangle$ | $|1,-1\rangle$ | $|0,0\rangle$ |
|---|---|---|---|---|---|---|---|---|---|---|---|---|---|---|---|---|
| $\langle 3,3|$ | 0.8 | 0.4+0.6i | 0.4−0.8i | −0.2i | 0 | 0 | 0 | 1.4−0.6i | 0.3+0.7i | −0.1 | 0.1i | 0 | −0.2+0.4i | 0.1i | 0.1i | −0.1 |
| $\langle 3,2|$ | 0.4−0.6i | 0.2 | 0.4+0.4i | 0.3−0.7i | −0.1i | 0 | 0 | 1.2−0.3i | −0.4 | 0.4+0.7i | 0.1 | 0 | −0.3−0.4i | 0.1+0.1i | 0 | 0 |
| $\langle 3,1|$ | 0.4+0.8i | 0.4−0.4i | −0.2 | 0.2+0.1i | 0.2−0.6i | 0.1i | 0 | −0.3−0.2i | 0.7−0.1i | −0.1+0.3i | 0.3+0.5i | 0.1 | 0.3 | −0.2−0.3i | 0.1 | −0.2 |
| $\langle 3,0|$ | 0.2i | 0.3+0.7i | 0.2−0.1i | −0.7 | −0.2−0.1i | 0.3−0.7i | 0.2i | −0.1+0.2i | −0.3−0.4i | 0.1i | −0.3+0.4i | 0.1+0.2i | 0.1−0.1i | 0.4 | −0.1−0.1i | 0 |
| $\langle 3,-1|$ | 0 | 0.1i | 0.2+0.6i | −0.2−0.1i | −0.2 | −0.4−0.4i | 0.4−0.8i | 0.1 | −0.3+0.5i | −0.1−0.3i | −0.7−0.1i | −0.3+0.2i | 0.1 | 0.2−0.3i | 0.3 | 0.2 |
| $\langle 3,-2|$ | 0 | 0 | −0.1i | 0.3+0.7i | −0.4+0.4i | 0.2 | −0.4−0.6i | 0.1 | 0.1 | −0.4+0.7i | −0.4 | −1.2−0.3i | 0 | 0.1−0.1i | 0.3−0.4i | 0 |
| $\langle 3,-3|$ | 0 | 0 | 0 | −0.2i | 0.4+0.8i | −0.4+0.6i | 0.8 | 0 | 0.1i | −0.1 | −0.3+0.7i | 1.4+0.6i | −0.1i | 0.1i | −0.2−0.4i | −0.1 |
| $\langle 2,2|$ | 1.4+0.6i | 1.2+0.3i | −0.3+0.2i | −0.1−0.2i | 0.1 | 0.1 | 0 | −0.3 | −0.1 | −0.1 | −0.1−0.1i | −0.1−0.1i | 0.7−0.1i | 0.2+0.6i | 0 | 0.2+0.1i |
| $\langle 2,1|$ | 0.3−0.7i | −0.4 | 0.7+0.1i | −0.3+0.4i | −0.3−0.5i | 0.1 | −0.1i | −0.1 | −0.5 | 0.2−0.1i | −0.3+0.3i | 0.1i | 0.5 | −0.1+0.3i | 0.2+0.4i | 0.2−0.5i |
| $\langle 2,0|$ | −0.1 | 0.4−0.7i | −0.1−0.3i | −0.1i | −0.1+0.3i | −0.4−0.7i | −0.1 | −0.3−0.3i | 0.2+0.1i | −0.4 | −0.2+0.1i | 0.1+0.3i | −0.1+0.3i | 0 | 0.1 | 0.8 |
| $\langle 2,-1|$ | −0.1i | 0.1 | 0.3−0.5i | −0.3−0.4i | −0.7+0.1i | −0.4 | −0.3−0.7i | 0.1i | −0.3−0.3i | 0.2+0.1i | −0.5 | 0.1 | −0.1−0.3i | −0.1−0.3i | −0.5 | −0.2−0.5i |
| $\langle 2,-2|$ | 0 | 0 | 0.1 | 0.1−0.2i | −0.3−0.2i | −1.2+0.3i | 1.4−0.6i | −0.1−0.1i | 0.5 | 0.3i | 0.1 | −0.3 | 0 | −0.2+0.6i | 0.7+0.1i | 0.2−0.1i |
| $\langle 1,1|$ | −0.2−0.4i | −0.3+0.4i | 0.3 | 0.1+0.1i | 0.1 | 0 | 0.1i | 0.7+0.1i | −0.1−0.3i | −0.2+0.4i | 0.1 | −0.3 | −0.3 | 0.4−0.4i | −0.4 | −0.1 |
| $\langle 1,0|$ | −0.1i | 0.1−0.1i | −0.2+0.3i | 0.4 | 0.2+0.3i | 0.1+0.1i | −0.1i | 0.2−0.6i | 0.5 | 0 | −0.1+0.3i | −0.2−0.6i | 0.4+0.4i | 0.9 | −0.4+0.4i | 0.1 |
| $\langle 1,-1|$ | −0.1i | 0 | 0.1 | −0.1+0.1i | 0.3 | 0.3+0.4i | −0.2−0.4i | 0 | 0.2−0.4i | −0.2−0.4i | −0.5 | 0.7−0.1i | −0.4 | −0.4−0.4i | −0.3 | 0.1 |
| $\langle 0,0|$ | −0.1 | 0 | −0.2 | 0 | −0.2 | −0.1 | −0.1 | 0.2−0.1i | 0.2+0.5i | 0.8 | −0.2+0.5i | 0.2+0.1i | 0.1 | 0 | 0.1 | 0.7 |

## S2 xyz coordinates

### S2.1 Complex 1: $[Co_2Cl_6]^{2-}$

Arbitrary frame:

| | | | |
|---|---|---|---|
| Co | -0.000682 | 0.004123 | -1.595481 |
| Co | 0.000682 | -0.004123 | 1.595481 |
| Cl | 0.000000 | -1.717325 | 0.000000 |
| Cl | 1.895916 | -0.178834 | -2.743890 |
| Cl | -1.861051 | 0.153232 | -2.805935 |
| Cl | 0.000000 | 1.717325 | -0.000000 |
| Cl | -1.895916 | 0.178834 | 2.743890 |
| Cl | 1.861051 | -0.153232 | 2.805935 |

Molecular PAF:

| | | | |
|---|---|---|---|
| Co | -1.587716 | -0.136461 | 0.078181 |
| Co | 1.587716 | 0.136461 | -0.078181 |
| Cl | -0.048059 | -0.351600 | -1.680260 |
| Cl | -2.557145 | -2.122392 | 0.334479 |
| Cl | -2.963284 | 1.603233 | -0.094113 |
| Cl | 0.048059 | 0.351600 | 1.680260 |
| Cl | 2.557145 | 2.122392 | -0.334479 |
| Cl | 2.963284 | -1.603233 | 0.094113 |

### S2.2 Complex 2: $[Co_2(L)_2(acac)_2(H_2O)]$

Arbitrary frame:

| | | | |
|---|---|---|---|
| O | 0.113492 | -1.374674 | -0.077237 |
| Co | 0.016448 | 0.115158 | -1.554198 |
| Co | -0.018680 | -0.109455 | 1.565701 |
| O | -0.319718 | 1.222363 | 3.089591 |
| O | 0.000432 | 1.389143 | 0.007798 |
| O | -2.112259 | -0.119467 | 1.251910 |
| H | -2.566740 | -0.907787 | 1.597783 |
| H | -2.243365 | -0.141475 | 0.253473 |
| O | 1.975431 | 0.158534 | 1.798366 |
| O | -2.030940 | -0.098165 | -1.451936 |
| O | -0.237416 | 1.408603 | -3.105331 |
| O | 0.092614 | -1.469756 | -2.913571 |
| N | -0.059152 | -1.808842 | 2.759532 |
| N | 2.083351 | 0.416352 | -1.629002 |
| C | 0.564069 | 1.931802 | 3.678787 |
| C | -0.770256 | -3.319332 | 0.982473 |
| C | -0.434915 | -2.952446 | 2.355083 |
| H | -0.503954 | -3.793547 | 3.070450 |
| C | 0.701539 | 2.498988 | 0.027175 |
| C | 1.966996 | 2.602957 | -0.618920 |
| C | 0.220071 | 3.632736 | 0.700156 |
| C | -0.449995 | -2.550761 | -0.158652 |
| C | -0.735973 | -3.153432 | -1.414724 |
| C | -1.375494 | -4.565359 | 0.853066 |
| H | -1.588155 | -5.130032 | 1.772610 |
| C | -2.860978 | 0.246368 | -2.355766 |
| C | 2.551072 | 1.010673 | 2.523319 |
| C | 1.923858 | 1.844741 | 3.440702 |
| H | 2.571187 | 2.521007 | 4.010060 |
| C | -1.099940 | 3.609857 | 1.317907 |

```
H    -1.663625    2.652511    1.197804
C    -1.708827   -5.126159   -0.377668
C    -2.546186    0.999344   -3.445803
C     2.595889    1.523487   -1.299456
H     3.663564    1.712413   -1.547700
C    -1.344872   -4.411890   -1.486602
H    -1.533774   -4.828295   -2.487513
C    -0.376233   -2.582105   -2.676728
H    -0.562981   -3.254792   -3.552925
H    -1.449124   -2.652236    5.980214
C    -0.539125   -3.258345    5.878647
C     0.318757   -1.638219    4.180843
H    -0.416620   -0.936334    4.617335
H     1.277453   -1.085958    4.164811
C    -0.423979   -4.417566    6.623957
C     0.050577    2.904875    4.691326
H    -0.581399    3.661968    4.189807
H     0.856567    3.420210    5.238207
H    -0.602258    2.383176    5.414714
C     0.448015   -2.875103    5.013415
C     3.682513   -1.741578   -0.110669
H     2.622296   -1.899261    0.113777
C    -4.277203   -0.224867   -2.160618
H    -4.296355   -1.327852   -2.081622
H    -4.949842    0.085316   -2.976063
H    -4.676485    0.171190   -1.207669
C     4.045327    1.105327    2.340558
H     4.532029    0.213702    2.772093
H     4.474943    2.004226    2.812365
H     4.283448    1.109595    1.262796
C    -2.371102   -6.465188   -0.459840
H    -1.812382   -7.232655    0.107574
H    -2.452770   -6.816577   -1.502735
H    -3.394815   -6.441024   -0.039746
C     1.565054   -3.674520    4.932264
H     2.384000   -3.391249    4.262072
C     3.016039   -0.576199   -2.242488
H     3.506234   -0.118504   -3.124101
H     2.401409   -1.411023   -2.617514
C     4.049676   -1.063389   -1.274319
C     4.655030   -2.200927    0.768833
C     0.692801   -5.209407    6.516260
H     0.761142   -6.132143    7.106360
F    -1.726798   -6.112369    7.557215
C    -1.502240   -4.816766    7.563022
F    -2.660718   -4.242148    7.305607
F    -1.239495   -4.511708    8.808129
C     4.279066   -2.887951    2.004205
F     3.056259   -3.437514    1.948283
F     4.244695   -2.079687    3.052563
F     5.088540   -3.868775    2.352848
C     1.682326   -4.836019    5.697083
H     2.592398   -5.442117    5.601930
C     5.401009   -0.853778   -1.529788
H     5.703744   -0.326985   -2.444567
C     6.001439   -1.977793    0.491513
```

| | | | |
|---|---|---|---|
| H | 6.773111 | -2.346500 | 1.177712 |
| C | 6.342486 | -1.293225 | -0.645868 |
| H | 7.405444 | -1.116734 | -0.859792 |
| O | -1.596394 | 4.553136 | 1.898839 |
| C | 2.673551 | 3.798467 | -0.529406 |
| H | 3.658107 | 3.843182 | -1.020668 |
| C | -1.296880 | 1.537853 | -3.760791 |
| C | 0.983257 | 4.809942 | 0.748140 |
| H | 0.541497 | 5.657760 | 1.291951 |
| H | -3.357897 | 1.220261 | -4.147388 |
| C | 2.211621 | 4.914082 | 0.137340 |
| C | 3.008678 | 6.200682 | 0.185976 |
| H | 3.094622 | 6.583589 | 1.219155 |
| H | 2.525519 | 6.995864 | -0.413030 |
| H | 4.030009 | 6.062093 | -0.209204 |
| C | -1.164294 | 2.360590 | -5.013161 |
| H | -0.769726 | 3.360285 | -4.755399 |
| H | -2.110876 | 2.482281 | -5.564556 |
| H | -0.419100 | 1.889569 | -5.680001 |

Molecular PAF:

| | | | |
|---|---|---|---|
| O | -0.195445 | 0.946496 | -0.987178 |
| Co | 1.328819 | 0.688948 | 0.434342 |
| Co | -1.338206 | -0.697582 | -0.431663 |
| O | -2.451548 | -2.243338 | 0.315139 |
| O | 0.313972 | -0.979562 | 0.933627 |
| O | -2.133707 | 0.526377 | 1.101550 |
| H | -2.832311 | 1.136828 | 0.806708 |
| H | -1.373539 | 1.108085 | 1.415093 |
| O | -0.465787 | -2.011719 | -1.702237 |
| O | 0.163317 | 1.823982 | 1.699673 |
| O | 2.790580 | 0.685822 | 1.850925 |
| O | 2.133582 | 2.443892 | -0.364518 |
| N | -2.744314 | -0.081519 | -1.830989 |
| N | 2.501547 | -0.531414 | -0.791965 |
| C | -2.333421 | -3.483884 | 0.034736 |
| C | -1.971153 | 2.228824 | -1.929614 |
| C | -2.860477 | 1.114221 | -2.242473 |
| H | -3.684750 | 1.381620 | -2.930252 |
| C | 0.906817 | -2.123501 | 1.185861 |
| C | 2.106065 | -2.513634 | 0.523118 |
| C | 0.365383 | -3.012813 | 2.127175 |
| C | -0.682595 | 2.098343 | -1.365674 |
| C | 0.080207 | 3.295176 | -1.273820 |
| C | -2.455499 | 3.474945 | -2.314625 |
| H | -3.458374 | 3.518720 | -2.764031 |
| C | 0.577196 | 2.454761 | 2.726999 |
| C | -0.582990 | -3.264445 | -1.702965 |
| C | -1.470399 | -3.991909 | -0.919410 |
| H | -1.462497 | -5.079529 | -1.051927 |
| C | -0.819049 | -2.637812 | 2.889450 |
| H | -1.223661 | -1.620092 | 2.667612 |
| C | -1.727990 | 4.653663 | -2.167418 |
| C | 1.816869 | 2.312421 | 3.272156 |
| C | 2.740553 | -1.733283 | -0.483047 |
| H | 3.528532 | -2.282141 | -1.044378 |

```
C    -0.456745    4.523076   -1.678274
H     0.185450    5.412222   -1.589365
C     1.443818    3.343732   -0.841745
H     1.924183    4.349305   -0.956536
H    -6.320278   -0.398151   -2.189255
C    -5.917117   -0.382362   -3.210526
C    -3.697991   -1.104240   -2.317397
H    -4.269973   -1.443366   -1.433215
H    -3.074473   -1.969321   -2.612706
C    -6.746892    0.000185   -4.248635
C    -3.210732   -4.414167    0.809771
H    -2.936942   -4.374771    1.881029
H    -3.141305   -5.457847    0.463101
H    -4.262083   -4.079531    0.743937
C    -4.611215   -0.717992   -3.438379
C     1.545371   -0.585860   -3.724895
H     0.788164   -0.050801   -3.142100
C    -0.407209    3.404881    3.354738
H    -0.736903    4.149528    2.606415
H     0.004439    3.935849    4.227794
H    -1.310407    2.851791    3.675242
C     0.343587   -3.995838   -2.641943
H     0.023881   -3.832187   -3.685544
H     0.374380   -5.080734   -2.448961
H     1.361820   -3.579777   -2.550378
C    -2.301830    5.970314   -2.586343
H    -2.669770    5.942295   -3.628920
H    -1.555681    6.780541   -2.518757
H    -3.161620    6.261248   -1.952964
C    -4.165339   -0.681495   -4.739645
H    -3.129581   -0.959433   -4.963315
C     3.253258    0.000545   -1.968137
H     4.339673   -0.127597   -1.794070
H     3.063475    1.085653   -2.011370
C     2.855469   -0.665341   -3.249335
C     1.197038   -1.195963   -4.923541
C    -6.277299    0.044784   -5.538326
H    -6.946905    0.362677   -6.347586
F    -8.570788    1.382523   -4.699427
C    -8.163839    0.356182   -3.985239
F    -8.400528    0.668024   -2.726170
F    -8.997824   -0.614419   -4.258713
C    -0.179243   -1.141732   -5.415886
F    -0.875254   -0.108821   -4.916820
F    -0.883058   -2.216764   -5.095767
F    -0.287932   -1.037340   -6.725888
C    -5.010811   -0.308192   -5.785400
H    -4.612951   -0.295659   -6.808234
C     3.797098   -1.368702   -3.993593
H     4.832640   -1.433861   -3.633890
C     2.157498   -1.895386   -5.650065
H     1.889844   -2.370683   -6.601246
C     3.434042   -1.979278   -5.158151
H     4.188204   -2.534199   -5.732688
O    -1.335302   -3.339707    3.734761
C     2.663008   -3.755599    0.812625
```

```
H        3.578208    -4.039446     0.269921
C        2.832489     1.459560     2.834983
C        0.981206    -4.249800     2.374549
H        0.501399    -4.893766     3.126061
H        2.043097     2.919466     4.155456
C        2.132426    -4.638948     1.729525
C        2.789984    -5.970223     2.026724
H        2.061345    -6.799882     1.979678
H        3.228673    -5.984099     3.042561
H        3.601453    -6.192078     1.312028
C        4.131630     1.441748     3.593153
H        4.346216     0.410981     3.929289
H        4.142009     2.111548     4.468437
H        4.953560     1.729122     2.912199
```



\end{document}